\documentclass[11pt,a4paper]{article}

\usepackage[dvipdfmx]{graphicx}
\usepackage{graphicx}
\usepackage{subcaption}
\usepackage{float}
\usepackage{color}
\usepackage{amsmath,amssymb}
\usepackage{fdsymbol}
\usepackage{bbold}
\usepackage{t1enc}
\usepackage{placeins}
\usepackage{tablefootnote}
\usepackage{enumitem}
\usepackage{hyperref}
\usepackage[title]{appendix}
\usepackage{soul}
\usepackage{makecell}

\usepackage{color}
\usepackage[dvipsnames]{xcolor}

\DeclareMathSymbol{\shm}{\mathbin}{AMSa}{"39}

\newcommand{\logpt}{$\ln p_{\mathrm{T}}$ }
\newcommand{\pt}{$p_{\mathrm{T}}$ }
\newcommand{\ppt}{p_{\mathrm{T}} }
\newcommand{\none}{$\tilde \chi_1^0$}

\newcommand{\cone}{$\tilde \chi_1^\pm$}
\newcommand{\wino}{$\tilde W$}
\newcommand{\hino}{$\tilde h$}
\newcommand{\bino}{$\tilde B$}

\newcommand{\lhc}{$L=300~\mathrm{fb}^{-1}$}
\newcommand{\hllhc}{$L=3000~\mathrm{fb}^{-1}$}

\begin{document}

\begin{center}
\begin{Large}
{\bf Machine Learning Electroweakino Production}
\end{Large}

\vspace{0.5cm}
\renewcommand*{\thefootnote}{\fnsymbol{footnote}}
\begin{large}
Rafa\l{} Mase\l{}ek$^{(\rm a, b)}$, 
Mihoko M.~Nojiri$^{(\rm c,d, e)}$
and
Kazuki Sakurai$^{(\rm b)}$
 \\[1mm]
\end{large}
\setcounter{footnote}{0}
\vspace{3mm}

\begin{small}

$^{(\rm a)}${\em
Laboratoire de Physique Subatomique et de Cosmologie (LPSC), Université Grenoble-Alpes,
CNRS/IN2P3, 53 Avenue des Martyrs, F-38026 Grenoble, France}

\vskip 0.15cm

$^{(\rm b)}${
Institute of Theoretical Physics, Faculty of Physics,\\
University of Warsaw, Pasteura 5, 02-093, Warsaw, Poland 
}

\vskip 0.15cm

$^{(\rm c)}${
Theory Center, IPNS, KEK, 1-1 Oho,
Tsukuba, Ibaraki 305-0801, Japan}

\vskip 0.15cm

$^{(\rm d)}${
The Graduate University of Advanced Studies (Sokendai),
\\  1-1 Oho, Tsukuba, Ibaraki 305-0801, Japan}

\vskip 0.15cm

$^{(\rm e)}${
Kavli IPMU (WPI), University of Tokyo,
5-1-5 Kashiwanoha, Kashiwa, Chiba 277-8583, Japan}

\end{small}

\end{center}
\vspace{5em}
\begin{abstract}
The system of light electroweakinos and heavy squarks gives rise to one of the most challenging signatures to detect at the LHC.  
It consists of missing transverse energy recoiled against a few hadronic jets originating either from QCD radiation or squark decays.
The analysis generally suffers from the large irreducible $Z + {\rm jets}$ ($Z \to \nu \bar \nu$) background.
In this study, we explore Machine Learning (ML) methods for efficient signal/background discrimination.
Our best attempt uses both reconstructed (jets, missing transverse energy, etc.) and low-level (particle-flow) objects.
We find that the discrimination performance improves as the $p_\mathrm{T}$ threshold for soft particles is lowered from 10 GeV to 1 GeV, at the expense of larger systematic uncertainty.
In many cases, the ML method provides a factor two enhancement in $S/\sqrt{S+B}$ from a simple kinematical selection. 
The sensitivity on the squark-elecroweakino mass plane is derived with this method, assuming the Run-3 and HL-LHC luminosities. 
Moreover, we investigate the relations between input features and the network's classification performance to reveal the physical information used in the background/signal discrimination process.

\end{abstract}
\newpage
\tableofcontents
\newpage
\section{Introduction}

The nature of dark matter has remained a longstanding mystery. 
Despite the plethora of experiments aimed at directly detecting dark matter in underground laboratories and indirectly through secondary particles from dark matter annihilation or decay, no definitive evidence has yet been found.
Another promising avenue is to produce dark matter candidates at the Large Hadron Collider (LHC) and study them.
At the LHC, the production of dark matter will be detected as the excess of missing energy events.  

One of the most promising candidates of dark matter is the lightest neutralino ($\tilde \chi_1^0$) in supersymmetric theories.
The system of heavy squarks $(\tilde q)$ and light electroweakinos $(\tilde \chi)$ is probably the simplest and most studied case for the dark matter production at the LHC.
The current literature mainly focuses on a combination of moderately heavy squarks, $1 ~{\rm TeV} \lesssim m_{\tilde q} \lesssim 2 ~{\rm TeV}$ with a light Bino-like neutralino ($\tilde \chi_1^0$) \cite{ATLAS:2020syg,CMS:2019ybf}.
In this case, the production is dominated by squark pairs, $pp \to \tilde q \tilde q$,\footnote{$\tilde q$ denotes both squarks and anti-squarks.} followed by the squark decay $\tilde q \to q \tilde \chi_1^0$, which gives rise to final states with two randomly distributed high $p_\mathrm{T}$ jets and large missing energy ($E_\mathrm{T}^\mathrm{miss} = | {\bf p}_T^{\rm miss} |$) originating from two $\tilde \chi_1^0$s.
A similar signature also arises from the direct production of Wino- or Higgsino-like electroweakinos, $pp \to \tilde \chi \tilde \chi + {\rm jets}$, assuming the mass splitting between the electroweak multiplet is small enough so that decays among the multiplet are unresolvable.\footnote{When the mass splitting between the electroweak 
multiplet is moderately small, one can improve the sensitivity by 
exploiting soft-leptons coming from decays of heavier states in the 
multiplet \cite{ATLAS:2019lng, ATLAS:2021moa, CMS:2021edw}. 
For even smaller mass splitting the chargino's lifetime might be long enough to produce detectable disappearing track signature \cite{ATLAS:2022rme, CMS:2020atg}.
We, however, do not consider such cases in our study.
}
Unlike the squark pair production case, the origin of high $p_\mathrm{T}$ jets is QCD radiation, similar to the background processes.   

In both cases, the final state consists of no isolated leptons and large missing transverse energy recoiled against a small number of high $p_\mathrm{T}$ jets. 
Such a signature is (misleadingly) referred to as {\it monojet} channel, as the event selection usually allows more than one, but up to four, high $p_\mathrm{T}$ jets.   
The signal/background discrimination in the monojet channel is particularly challenging due to the overwhelmingly large irreducible background of $pp \to Z + {\rm jets}$, followed by $Z \to \nu \bar \nu$ \cite{ATLAS:2021kxv,CMS:2021far}.
In the high-level analysis, based on detector objects, the events are characterised only by a couple of jets and ${\bf p}_\mathrm{T}^{\rm miss}$, out of which not many useful kinematical variables can be constructed.\footnote{Notable examples of such kinamatical variables are $m_{T2}$ \cite{Lester:1999tx, Barr:2003rg} and $\alpha_T$ \cite{Randall:2008rw} variables.}
The jet tagging technique is not particularly helpful here because the jets in the signal processes originate from single light-flavour quarks or gluons, as in the background.
For the $pp \to \tilde \chi \tilde \chi + {\rm jets}$ process, in particular, the jets are from the QCD radiation, and their $p_\mathrm{T}$ distribution does not peak at the mass scale of the produced supersymmetric particles.
Despite these challenges, enhancing signal/background discrimination is the only way to improve current sensitivity, at least until a more powerful collider with higher collision energy is constructed and becomes operational.

Although the background mimics the signal very well in human eyes, there may be a subtle difference that has not been captured by the traditional analyses.
Those non-trivial characteristics might be detected by modern machine learning (ML) methods. 
In fact, with ML we can use the low-level data, consisting of reconstructed particles and charged tracks, to characterise the events.
A typical monojet channel has more than 100 reconstructed low-level objects on average.
The low-level data, therefore, has a much higher expressibility of events compared to the high-level data with only a few jets and ${\bf p}_T^{\rm miss}$.

In this work, we investigate how ML techniques can help to enchance the signal/background discrimination in the monojet channel. 
The ML application to the monojet channel is not new. 
In \cite{Khosa:2019kxd}, the classification of the signal models in the monojet channel has been studied using high-level variables. 
Ref.\ \cite{Lv:2022pme} explored the application of jet substructure techniques to look for the direct Higgsino production using a Convolutional Neural Network (CNN).

In this paper, two ML methods are tested and compared:
One is the state-of-the-art ParticleNet \cite{Qu:2019gqs} method based on Graph Neural Networks (GNN), and the other is more traditional XGBoost \cite{DBLP:journals/corr/ChenG16} with Boosted Decision Trees (BDT).
Our analysis uses both the high-level and low-level variables as inputs. 
Our hybrid approach tries to compromise the strengths and weaknesses of low-level and high-level inputs.

Although low-level variables, in principle, contain more information than high-level ones, they are more prone to background mismodelling. 
{Notably, during LHC Run-3 and the High Luminosity LHC, substantial contamination from pileup events is anticipated in the low-momentum region.
However, there has been significant progress on pileup mitigation techniques. 
The traditional approach involves estimating the uniform pileup energy per unit area on an event-by-event basis \cite{Cacciari:2008gd, Cacciari:2008gd, Cacciari:2014jta, Soyez:2018opl}.
This estimated energy is then subtracted uniformly from the calorimeter towers of the event.
In contrast, recently adopted techniques focus on estimating the likelihood that a particle originates from the primary vertex (the hard scattering) \cite{Cacciari:2014gra, Bertolini:2014bba, ArjonaMartinez:2018eah}.
These methods operate on a per-particle basis, suppressing particles likely to originate from pileup interactions or weighting them proportionally to their likelihood.
The state-of-the-art technique, based on Graph Neural Networks, demonstrates impressive classification accuracy, achieving an area under the curve (AUC) of $\sim$96 \% for some benchmark processes \cite{ArjonaMartinez:2018eah}.
}

Despite such optimism, the effect of pileup in the low momentum region is potentially non-negligible and difficult to simulate. 
In our analysis with low-level variables, we therefore pay particular attention to the dependence of the $p_\mathrm{T}$ threshold for soft objects. 
We also perform several indirect checks to see to what extent the soft particle information is used in the signal/background classification.

Recently, squark and electroweakino searches have been revisited, and some new insights have been obtained.
Ref.\ \cite{Buanes:2022wgm} shows that the analysis designed for the squark search has an unexpectedly high sensitivity to the direct electroweakino production. 
The current best limit on the direct Wino- and Higgsino productions is obtained by recasting the squark search analysis \cite{Buanes:2022wgm}.  
In Ref.\ \cite{Lara:2022new}, it has been noted that if the lightest supersymmetric particle (LSP) is Wino-like and as light as $200$ GeV, the current squark mass limit ($m_{\tilde q} \gtrsim 1.9$ TeV \cite{ATLAS:2020syg}) forces the squark pair production to be a subdominant channel, making the direct Wino production, $pp \to \tilde \chi \tilde \chi + {\rm jets}$, and the associated squark-Wino production $pp \to \tilde q \tilde \chi$, the dominant ones.  
In fact, the current ATLAS and CMS analyses interpret the monojet analysis almost exclusively for the Bino-like neutralino, assuming the pair production of electroweakinos and squark-electroweakino associate production are subdominant.
We relax these assumptions in our analysis. 
In particular, we consider the Bino-, Wino- and Higgsino-like LSP scenarios separately and, for all cases, include all three relevant production processes: (i) squark pair, (ii) squark-electroweakino and (iii) electroweakino pair. 
We investigate to what extent these processes contribute to signal/background discrimination.

The rest of the paper is organised as follows. 
First, in Sec.\ \ref{sec:dataset}, we introduce our data set. 
We describe our SM background and signal benchmarks and provide details of the Monte Carlo simulation, preselection procedure and data preparation. 
Next, we describe in Sec.\ \ref{sec:gnn} the architecture and training 
process of our Graph Neural Network. Sec.\ \ref{sec:eval} contains a detailed evaluation of our approach. 
We start with an evaluation of an ensemble of networks 
trained on Wino-like and Higgsino-like signals. 
We compare our algorithm with Boosted Decision Trees and investigate the impact of the cut on particle \pt. 
In Sec.\ \ref{sec:limits}, we derive the limit for both Run-3 and HL-LHC on the squark vs electroweakino mass planes. 
Next, in Sec.\ \ref{sec:interp}, we interpret the results of our neural networks, aiming at understanding the importance of the used variables and their connection with underlying physics. 
Sec.\ \ref{sec:concl} is devoted to conclusions.

\section{Data set}\label{sec:dataset}

\subsection{Simulation}

In this study, we consider supersymmetric (SUSY) spectra with a light electroweakino, which is either Wino-like (\wino), Higgsino-like (\hino) or Bino-like (\bino),
and somewhat heavy but not decoupled light-flavour squarks, $(\tilde u, \tilde d, \tilde s, \tilde c)_{L/R}$.
In this setup, we want to consider the most challenging signal, which possesses no isolated leptons and extra ($b$-)jets. 
To this aim, we assume other SUSY particles are heavy enough so that their effect is not directly observable.
For simplicity, we assume all eight light-flavour squarks are mass-degenerate and collectively call them the squark ($\tilde q$). 

The Bino-like LSP gives rise to the unique on-shell state, $\tilde \chi_1^0$.
On the other hand, if the LSP is Wino- or Higgsino-like, 
the lightest states come with the electroweak multiplet:
($\tilde \chi_1^0$,$\tilde \chi_1^\pm$) for Wino
and 
($\tilde \chi_1^0$,$\tilde \chi_2^0$,$\tilde \chi_1^\pm$) for Higgsino,
where 
$\tilde \chi_{1(2)}^0$ and $\tilde \chi_1^\pm$
are the lightest (second lightest) neutralino
and the lightest chargino, respectively. 
In our benchmark scenarios, the mass splitting among the multiplet is generally 1 GeV or less, and their internal decays (e.g.\ \cone $\to$ \none) cannot be resolved because the decay products are too soft. 
We, therefore, call the set of light electroweakino states the electroweakino ($\tilde \chi$) or Wino/ Higgsino/ Bino, collectively.

\begin{figure}[t!]
    \centering
    \includegraphics[width=1\linewidth]{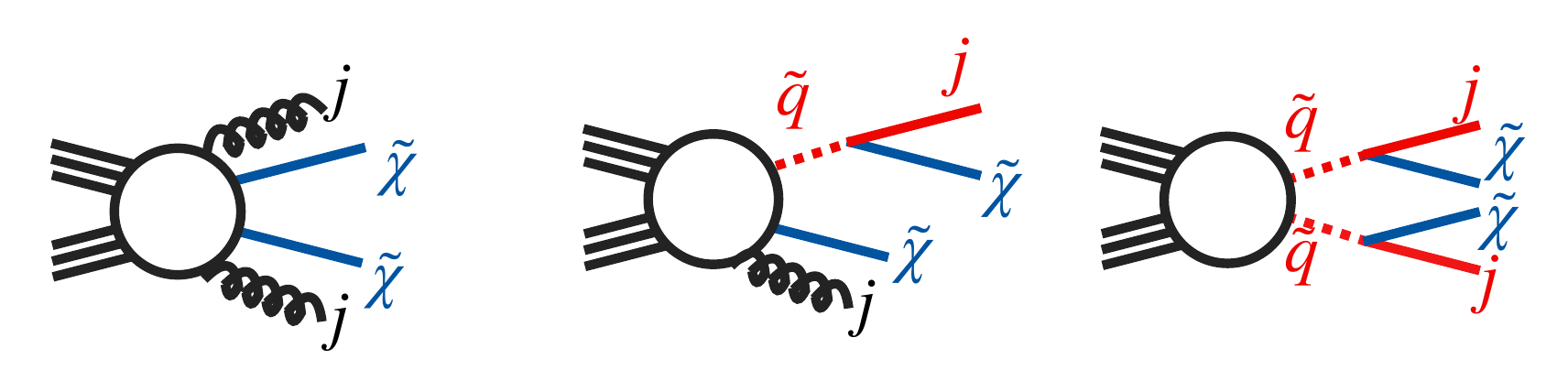}
    \caption{\small Three classes of processes contributing to signal samples: electroweakino pair production + ISR (left), associated squark-electroweakino production + ISR (center), and squark pair production (right). }
    \label{fig:diagrams}
\end{figure}

Our benchmark mass point used for the 
training of the artificial Neural Networks is $m_{\tilde \chi_1^0} = 300$ GeV, $m_{\tilde q} = 2.2$ TeV. 
In this scenario, three relevant SUSY production processes, depicted in Fig.\ \ref{fig:diagrams}, contribute to signal regions with high $p_T$ jets:
\begin{itemize}
    \item 0 $\tilde q$: electroweakino pair production associated with hard initial state radiation (ISR),
    \item  1 $\tilde q$: squark-electroweakino production,
    \item  2 $\tilde q$: squark-pair production.
\end{itemize}

In the 0 $\tilde q$ process, the source of high $p_T$ jets is only ISR, while for 1 $\tilde q$ and 2 $\tilde q$ processes, they may also originate from the squark decay, $\tilde q \to q \tilde \chi$.
The branching ratio of the latter decay is 100\,\% in our setup.

{
The dominant background in the monojet signal region, 
constituting between 60\% and 75\% of the total SM background \cite{ATLAS:2021kxv}, is the 
production of a $Z$ boson decaying to a pair of 
neutrinos, $Z \to \nu \bar \nu$, associated with ISR jets. 
The second most important background process is $W$ + jets (15-30\%), 
with unidentified leptons in the final state. Small contributions may arise from $Z\to l l + \mathrm{jets}~(l=e,\mu,\tau)$ ($<1\%$), multijet ($<1\%$), $t \bar t$ and single-top (0.6-3.5\%), 
and diboson $WW, WZ, ZZ$ (2-6\%) processes. Production of a single top quark associated with additional vector bosons is even more negligible.

In this study, we simulate only the $Z + {\rm jets}$ ($Z \to \nu \bar \nu$) dominant background 
process. The second biggest background, $W+\mathrm{jets}$, is taken into account by using a constant
correction factor estimated to be 1.3\footnote{We would like to stress that while these 
simplifications are justified for a phenomenological study, if our approach is to be used in 
experimental search, a detailed simulation of background is required for better precision and control.} (after the preselection described in Tab. \ref{tab:preselection}).
We omit the effects of other background processes completely due to their much smaller size. 

}

SUSY mass spectra are calculated using SuSpect \cite{Kneur_2023}. 
Throughout the training, the gluino mass is fixed at 10 TeV
and the ratio of the Higgs vacuum expectation is taken as $\tan \beta = v_u/v_d = 10$.
{ The squark mass is fixed at 2.2 TeV to evade exclusion limits.}
For the Wino-like case, the Higgsino mass parameter, $\mu$, is fixed at 1 TeV and Bino soft mass is set to  $M_1 = 3$ TeV, while the Wino soft mass parameter, $M_2$, is tuned such that the mass of the lightest neutralino is 300 GeV. 
When $\tilde \chi$ is Higgsino-like, 
Bino soft mass is again set to  $M_1 = 3$ TeV,
$M_2$ is set to 1 TeV, while $\mu$ is varied. 
For the Bino search, we use the model trained on Winos instead, as described in Sec.\ \ref{sec:cross}.
{
The rest of the particles are set to be very heavy.
These parameters are selected to realise simple low energy spectra in which only squarks and the lightest electroweakino contribute to the signal region.
Changing other parameters gives a rather mild effect on the result.
Increasing the gluino mass results in slightly larger squark-squark cross sections. 
The heavier electroweakino masses and $\tan\beta$ affect the splitting between the lightest electroweakino multiplet, which may impact the lepton veto efficiency. 
However, the leptonic branching ratio is, in any case, small, and this effect is rather mild.
}
Decays of SUSY particles are calculated with SUSY-HIT package \cite{djouadi2006decayssupersymmetricparticlesprogram}.

For evaluation of the algorithm and derivation of LHC sensitivity, we
study neutralino masses ranging between 200 GeV and 1.1 TeV, and squark masses varying from 2.0 TeV to 3.0 TeV. We generate samples of near-pure Winos, Higgsino and Binos.

Data used in this study is obtained using Monte Carlo methods. 
Events are generated by simulating $pp$ collisions at $\sqrt{s}=13$ TeV energy, 
mimicking the LHC setup. Parton-level events are generated 
at the leading order
using Madgraph-5 with the NNPDF 2.3 NLO parton distribution function \cite{BALL2013244}. 
For the BSM signal processes, we include up to two additional partons in the matrix element calculations, and exactly two for the $Z\to \nu \bar \nu$ background.
Parton showering and hadronisation are conducted with Pythia-8 \cite{Sjostrand:2014zea}. 
A simplified detector simulation is performed with Delphes-3 \cite{deFavereau:2013fsa} using the default detector card. 
Jet clustering is done with Fastjet \cite{Cacciari:2011ma} using the anti-$k_T$ algorithm. 
Exact versions of the software used in this study are listed in Table \ref{tab:software} in Appendix \ref{app:a}.

\subsection{Preselection}

In order to remove the majority of the background and force the Neural Network to focus on subtle 
differences between signal and background events, a preselection was applied to data. 
Table \ref{tab:preselection} contains a list of event selections. {These were inspired by the ATLAS search for squarks and gluinos in final states with jets and missing transverse momentum \cite{ATLAS:2020syg}. We have modified the ``SR2j-1600'' signal region selection from \cite{ATLAS:2020syg} by focusing on 2-4 jets, raising the cuts on jet \pt, and increasing the required size of  $E^\mathrm{miss}_\mathrm{T}$. The exact numbers come from optimisation for the $m_{\tilde W}=300~\mathrm{GeV}$ $m_{\tilde q}=2.2~\mathrm{TeV}$ mass point with the help of a Decision Tree algorithm.}
Events passing the selection are characterised by high \pt jets, no isolated leptons, and large missing transverse momentum, $E_T^{\rm miss}$. 
After the preselection, the effective cross section of the SM background is $6.6~\mathrm{fb}$.
The effective cross sections for the signal are $6.9\cdot 10^{-2}~\mathrm{fb}$, $5.3\cdot 10^{-2}~\mathrm{fb}$ and $5.1\cdot 10^{-2}~\mathrm{fb}$ for the Wino, Higgsino and Bino scenarios, respectively, with $m_{\tilde \chi_1^0}=300~\mathrm{GeV}$ and 
$m_{\tilde q}=2.2~\mathrm{TeV}$.
The event samples that passed cuts in Table \ref{tab:preselection} are publicly available on \cite{monojetdataset}.

Before the training, the SM data and that of the Wino/Higgsino signal for $m_{\tilde \chi_1^0}=300~{\rm GeV}$ $m_{\tilde q}=2.2~{\rm TeV}$ benchmark point are combined into a balanced data set consisting of 1.17M events,\footnote{The size of the training sample, 1.17M events, is much larger than the number of expected background, $\sim 20$K.
Our method, therefore, requires the generation of training samples with Monte Carlo simulation.
Low-level features are particularly prone to background mismodeling.
One of the reasons to use both low-level and high-level inputs is to make our results less dependent on the background modelling.} which is subsequently split into the training (70\%) and validation (30\%) sets. 
The splitting is done ten times with ten different values of the random seed in order to produce data for the ensemble training\footnote{An optimal approach would be to use Monte Carlo to generate ten independent data sets. However, it was not feasible with the available resources.}. In addition, completely independent test sets are generated for the SM background and all considered signal model points, each consisting of $\mathcal{O}(\rm 10k)$ events.

\begin{table}[t!]
    \centering
     {\small
    \begin{tabular}{r|ll|l}
     \hline 
    
        1 &  $N_{\mathrm{jets}}(p_T > 30~\mathrm{GeV})$ & $\geq 2$ & number of jets with $p_\mathrm{T}>30~\mathrm{GeV}$\\
        2 &  $p_{\mathrm{T}}(j_1)$ [GeV] & $>520$ & \pt of the leading jet\\
        3 &   $p_{\mathrm{T}}(j_2)$ [GeV] & $>320$ & \pt of the second jet\\
        4 &   $E_{\mathrm{T}}^{\mathrm{miss}}$ [GeV] & $> 820$ & Missing Transverse Energy (MET)\\
        5 & lepton veto & & no baseline lepton with $p_\mathrm{T}>7$ GeV and $|\eta|<2.7$ \\
        6 &   $|\eta(j_1)|$ & $<2$ & pseudorapidity of the leading jet\\
        7 &   $|\eta(j_2)|$ & $<2$ & pseudorapidity of the second jet\\
        8 &   $\Delta \phi \left[(j_1, j_2, (j_3)), p_\mathrm{T}^\mathrm{miss}\right]$ & $>0.8$ & azimuthal angle between $p_\mathrm{T}^\mathrm{miss}$ and \pt of the 3 first jets\\
        9 &   $\Delta \phi \left[(j_{\geq 4}), p_\mathrm{T}^\mathrm{miss}\right]$ & $>0.4$ & azimuthal angle between $p_\mathrm{T}^\mathrm{miss}$ and \pt of the subleading jets\\
        10 &   $E_\mathrm{T}^\mathrm{miss}/\sqrt{H_\mathrm{T}}$ [GeV$^{1/2}$] & $>16$ & ratio of MET and square root of scalar sum of jet $p_\mathrm{T}$s\\
        11 &  $m_\mathrm{eff}$ [GeV] & $> 1600$ & scalar sum of MET and  $p_\mathrm{T}$s of all jets with ${p_\mathrm{T}} > 50~\mathrm{GeV}$\\
        12 &  $N_{\mathrm{jets}}(p_T > 30~\mathrm{GeV})$ & $\leq 4$& number of jets with $p_\mathrm{T}>30~\mathrm{GeV}$\\
        \hline 
        
    \end{tabular}
    }
\caption{\label{tab:preselection}
\small Summary of data preselection applied before feeding it to Neural Network.}            
\end{table}

\subsection{Preparation}

We extract two types of features from the event data: high-level and low-level variables. 
The ordered list of high-level variables is given in Table \ref{tab:high-level-variables}. 
They include, for instance, the $p_T$, $\eta$ and mass of the individual jets, as well as the distance between two jets in their azimuthal angles. 
The number of jets is a variable itself, ranging in our analysis between 2 and 4. 
When the third or fourth jet is missing, we use zero values for the corresponding high-level variables.
All high-level variables are scaled before feeding into the neural network, i.e.\ we subtract the mean and divide by the standard deviation, where the mean and the standard deviation are calculated with the training set.

Low-level data are composed of ``particles,'' which are reconstructed objects using the energy flow algorithm \cite{CMS:2009nxa} and belong to one of the three categories: charged objects, neutral hadrons or photons. 
We use up to 250 particles per event, satisfying the default $p_T$ cut, $p_T > 1$ GeV.
We will later vary this $p_T$ cut and study how the result responds to it.    
We pad with zeros when the multiplicity is lower than the threshold. 
The low-level information is ordered by particle's \pt in descending order. 
The particle-level features are listed in Table \ref{tab:low-level-variables}. 
Following the original implementation in \cite{Qu:2019gqs}, these variables are not scaled.

\begin{table}[t!]
    \centering
    \begin{tabular}{r|ll}
    \hline
        1 &  $E_\mathrm{T}^\mathrm{miss}$ &  missing transverse energy\\
        2 &  $H_\mathrm{T}$&  scalar sum of \pt of objects\\
        3 &  $\eta(\mathrm{event})$ & $\eta$ coordinate of the event centroid\\
        4 &  $M(\mathrm{event})$ & invariant mass of the event particles\\
        5 &  MT2 & stransverse mass\\
        6 &  $p_\mathrm{T}(j_1)$ & \pt of the leading jet \\
        7 &  $\eta (j_1)$ & pseudorapidity of the leading jet\\
        8 &  $M(j_1)$& mass of the leading jet\\
        9 &  $\Delta \phi \left( j_1, \phi_\mathrm{event} \right)$& azimuthal angle difference\tablefootnote{The convention we use throughout the study is that $\Delta\phi \in (-\pi, \pi]$ }
 between the leading jet and event centroid\tablefootnote{The event centroid is calculated as a vectorial sum of four-momenta of all particles in the event.}\\
        10 &  $p_\mathrm{T}(j_2)$ & \pt of the second jet \\
        11 &  $\eta (j_2)$ & pseudorapidity of the second jet\\
        12 &  $M(j_2)$& mass of the second jet\\
        13 &  $\Delta \phi \left( j_2, \phi_\mathrm{event} \right)$& azimuthal angle difference between the second jet and event centroid\\
        14 &  $p_\mathrm{T}(j_3)$ & \pt of the third jet \\
        15 &  $\eta (j_3)$ & pseudorapidity of the third jet\\
        16 &  $M(j_3)$& mass of the third jet\\
        17 &  $\Delta \phi \left( j_3, \phi_\mathrm{event} \right)$& azimuthal angle difference between the third jet and event centroid\\
        18 &  $p_\mathrm{T}(j_4)$ & \pt of the fourth jet \\
        19 &  $\eta (j_4)$ & pseudorapidity of the fourth jet\\
        20 &  $M(j_4)$& mass of the fourth jet\\
        21 &  $\Delta \phi \left( j_4, \phi_\mathrm{event} \right)$& azimuthal angle difference between the fourth jet and event centroid\\
    \hline
    \end{tabular}
    \caption{\label{tab:high-level-variables}
    \small List of the high-level inputs to the Neural Network. The order is the same as the order of data fed into the NN model.}    
\end{table}

\begin{table}[t!]
    \centering
    \begin{tabular}{r|ll}
    \hline
         1&  $\log p_\mathrm{T}$ & natural logarithm of particle's \pt\\
         2&  $\log E$ & natural logarithm of particle's energy\\
         3&  $\Delta \eta $ & the absolute value of $\eta$ difference between a particle and the event centroid\\
         4&  $\Delta \phi$& azimuthal angle difference between a particle and the event centroid\\
         5&  origin &  \makecell{one-hot-encoded type of the eflow object: \\charged (0), photons (1) or neutral hadrons (2)} \\
    \hline
    \end{tabular}        
     \caption{\label{tab:low-level-variables}
     \small List of the low-level inputs.}   
\end{table}

\section{Graph Neural Network}\label{sec:gnn}

\subsection{Architecture}\label{sec:architecture}

The architecture proposed in this study is implemented in Tensorflow \cite{tensorflow2015-whitepaper} with Keras \cite{chollet2015keras} and consists of two parts: one with the high-level inputs and the other with the low-level particle information. 
The former is a Multilayer Perceptron Network consisting of six layers. 
The first layer has 256 neurons, and the remaining five layers each have 128 neurons. 
All layers have L2=0.001 regularisation and are followed by batch normalisation and ReLU activation function. 

The latter network is closely modelled after the ParticleNet-lite jet tagging architecture \cite{Qu:2019gqs}, which has been proven very effective for jet classification\footnote{ParticleNet has eventually been superseded by ParticleTransformer \cite{Qu:2022mxj} and some other more complex architectures. 
In our study, we tried using fully connected networks, convolutional networks applied to event images, a combination of the two, and a convolutional graph network based on ParticleNet. 
We have found that the latter performs best. 
Transformer-based networks are typically much more computationally demanding, and available resources were not sufficient to investigate that avenue. 
Nonetheless, we leave it for future exploration.}.
Similarly to the original ParticleNet, this network uses relative $\phi$ and $\eta$ coordinates to construct a graph using the $k$-nearest neighbours algorithm. 
In our study, however, we extend the usability of ParticleNet to event-level data, i.e.\ we include information about particles from multiple jets, and we calculate the coordinates with respect to event centroid rather than jet axis. 
Our architecture has two EdgeConv \cite{10.1145/3326362} blocks with $k=7$ nearest neighbours used. 
The first block uses the physical information of the neighbouring particles to construct their representations in a new space. 
The second block repeats the operation but starts in the abstract space.
These blocks are followed by a Global Average Pooling operation, a fully connected layer with 256 neurons, a high dropout with 50\,\% probability, which randomly drops connections between the neurons, and a final dense layer with 128 neurons. 
Both dense layers have the L2 regularisation, batch normalisation and ReLU activation function. 
We found that a high probability dropout is necessary to protect the network from overfitting and failing to generalise correct predictions to test data.

\begin{figure}[t!]
    \centering
    \includegraphics[width=0.5\linewidth]{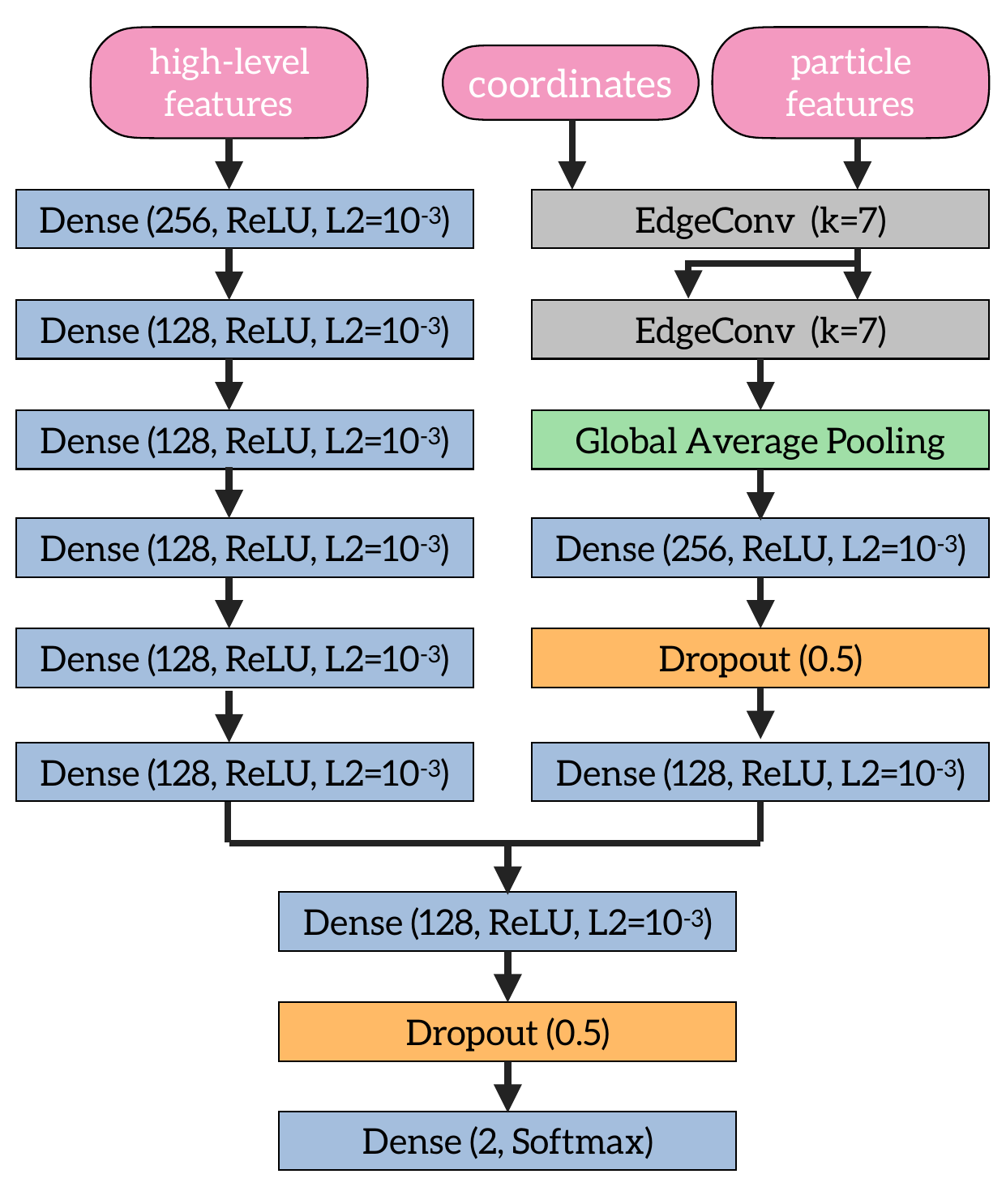}
    \caption{\small Neural Network architecture.}
    \label{fig:NN-architecture}
\end{figure}

Both networks (with high- and low-level inputs) analyse different data describing the same event in parallel. 
To combine their predictive power, the last layers of both networks are concatenated, followed by a dense layer with 128 neurons, a 50\,\%  dropout, and the final (output) dense layer. 
The output layer has only two neurons, one SM and the other BSM labels, with the sigmoid activation function. 
The details of the NN architecture are summarised in Fig.\ \ref{fig:NN-architecture}.
The full network has 211,318 parameters in total, 
out of which 207,724 are trainable. This is about 4 times less than the size of the training set. The hyperparameters of the 
architecture have been partially optimised. A more comprehensive and systematic hyperparameter
optimisation is recommended before applying the algorithm to experimental data.

\subsection{Training}

The weights are optimised with the Adam optimiser \cite{Kingma2014AdamAM} for the Binary Cross Entropy loss function. 
The gradient of each weight is individually clipped so that its norm is no higher than $5\cdot 10^{-4}$.
Adam is used together with Cosine Decay Learning Rate Scheduler \cite{loshchilov2017sgdr} and the linear warmup \cite{goyal2018accuratelargeminibatchsgd}. 
The learning scheduler starts with the learning rate 0 and linearly increases it for 8000 optimiser steps up to $5\cdot 10^{-4}$, then decreases it to $10^{-5}$ over 248000 steps.
Training data is split into minibatches with 1024 events each. 
The early stop mechanism finishes the training process if the loss calculated on the validation set does not improve for 50 epochs. 
At the end of the training, the best model's weights, in terms of the validation loss, are stored on disk.

The ensemble of ten\footnote{One may wonder if ten 
networks constitute a large enough sample for meaningful statistical 
analysis, since 30 data samples are commonly used in statistics as a 
minimal threshold. However, training large Neural Networks is 
computationally expensive and having 30 members of the ensemble is 
in practice unfeasible. Other studies in the field typically use less, e.g. in \cite{Hammad:2023sbd, Birk:2024knn} conclusions are drawn 
based on only 5 networks. Many other studies provide no uncertainty 
estimate at all. 
} networks is used to estimate the uncertainty of the results presented in this study. 
Each member of the ensemble corresponds to a different random seed number used to initialise the weights and to split data into training and validation sets. 
The uncertainty is quantified using a single standard deviation from the mean.
All trained models are publicly available in the GitHub repository of the project \cite{githubrepo}.


\section{Evaluation}\label{sec:eval}

\subsection{Wino-like LSP}\label{sec:wino-eval}

In this subsection, we discuss the result of machine learning classification for the Wino-like scenario with $m_{\tilde W}=300$ GeV and $m_{\tilde q}=2.2$ TeV.
The output of the neural network is the value, $s \in (0,1)$, called the NN score, which represents the machine's confidence that the event belongs to the BSM signal. 
It is straightforward to create a binary classifier by selecting some threshold value, $s_{\rm th}$, such that events with $s > s_{\rm th}$ are classified as the signal and otherwise the background. 
This classification is analogous to the traditional cut-and-count approach, where the signal region is defined as $s > s_{\rm th}$.
The threshold value can be optimised, for example, such that it maximises the naive significance, $S/\sqrt{S+B}$, where $S$ and $B$ are the signal and background yields, respectively.

\begin{figure}[t!]
    \centering
    \includegraphics[width=0.49\linewidth]{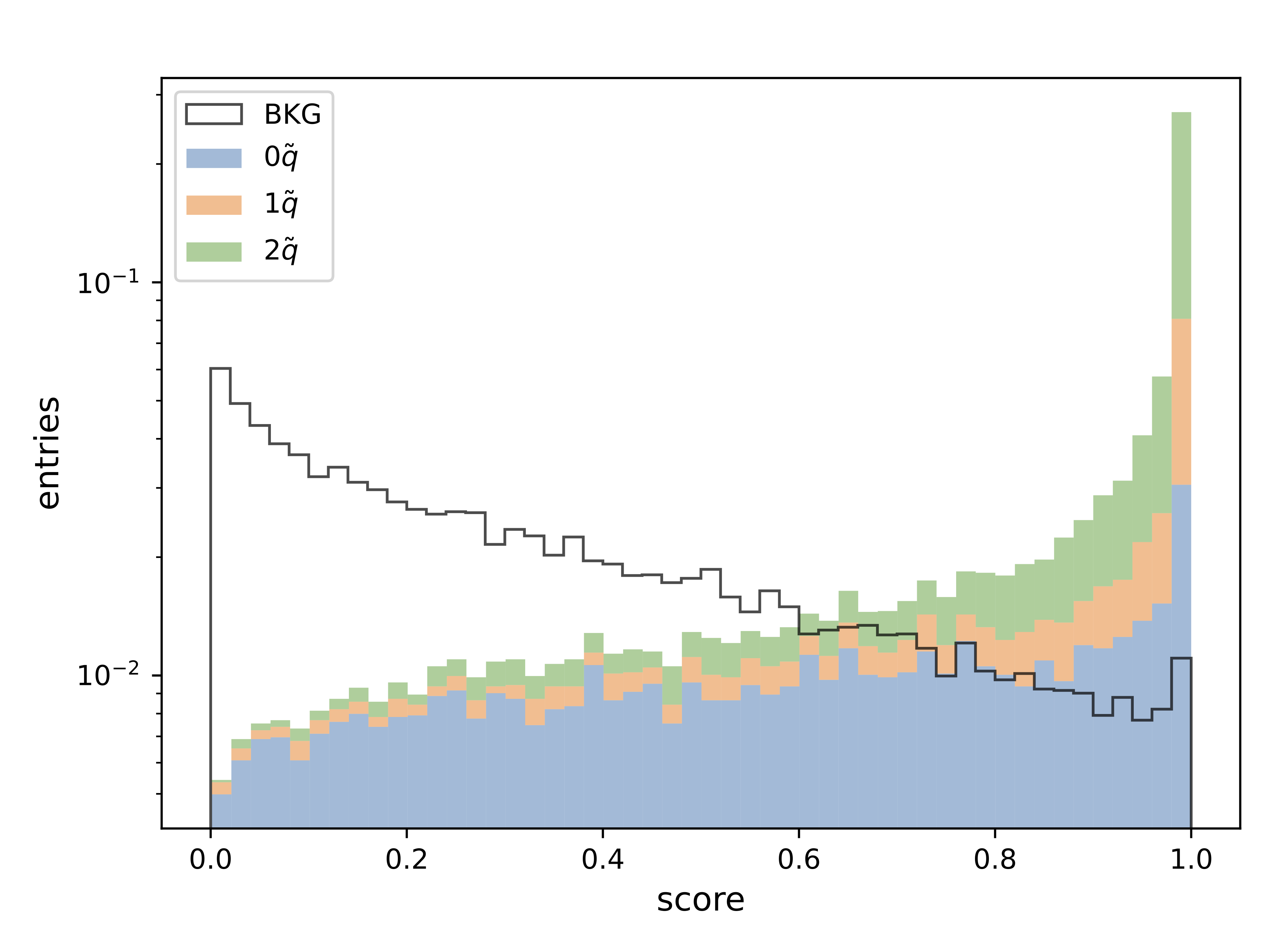}
     \includegraphics[width=0.49\linewidth]{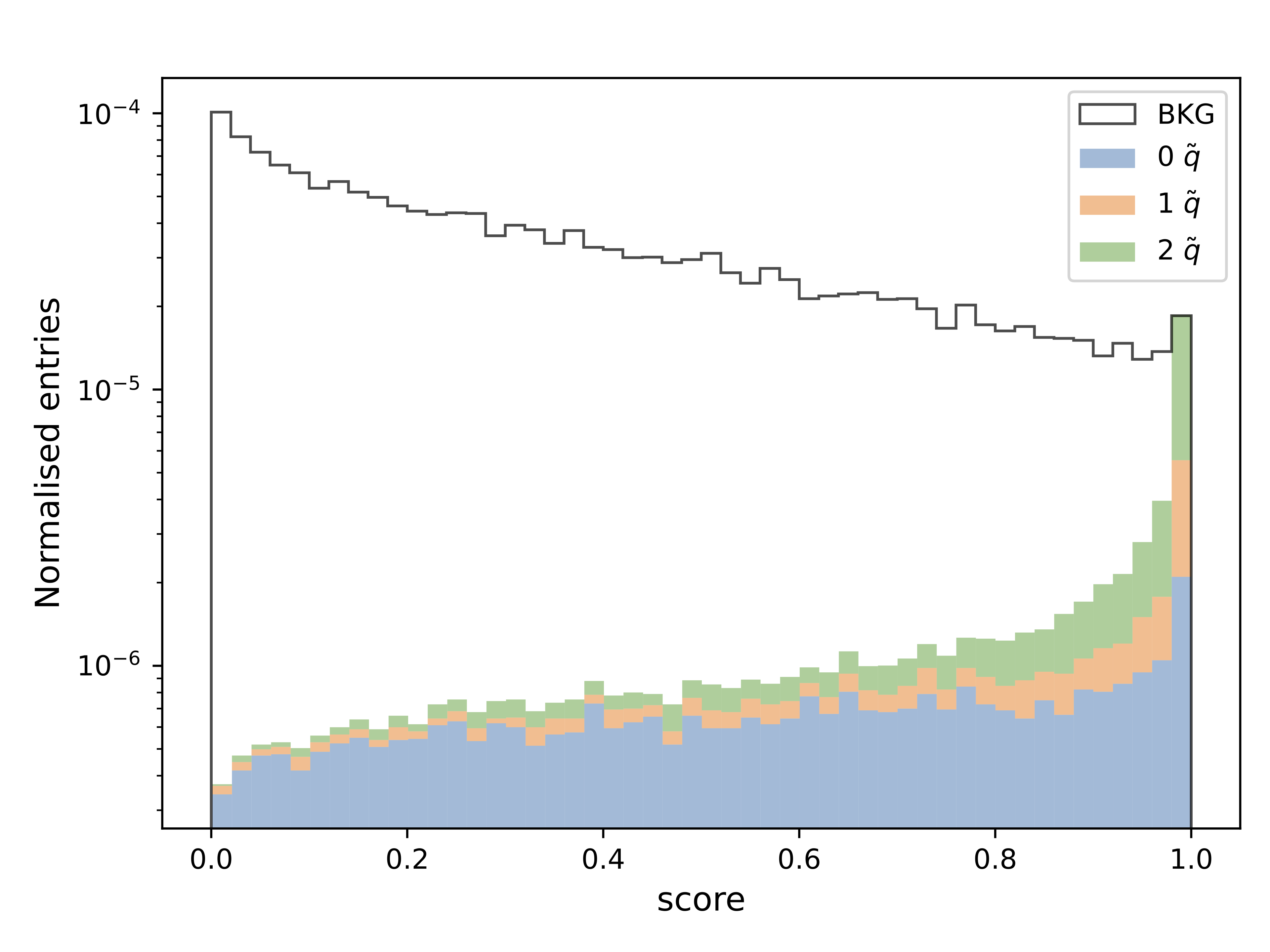}
    \caption{\small 
    The output of a neural network for the Wino-like scenario.
    The distributions are normalised to 1 in the left panel and to the value of the cross section in the right panel.
    The colours correspond to contributions from different classes of signals: 0 $\tilde q$ (blue), 1 $\tilde q$ (orange), 2 $\tilde q$ (green). The histogram for the SM background is shown as an unfilled thick black line.}
    \label{fig:wino-score}
\end{figure}

Fig.\ \ref{fig:wino-score} shows histograms of NN scores.
The signal and background samples on the left panel are normalised to 1, while on the right panel, they are normalised to their cross sections.
In the histograms, the signal sample is split into three classes: 0 $\tilde q$, 1 $\tilde q$ and 2 $\tilde q$, introduced in Sec.\ \ref{sec:dataset}.
On the left panel, we see that the network is very well able to separate the signal from the SM background. 
The score for the signal peaks at $s=1$, while the peak is at $s=0$ for the background.   
However, we observe that the peak for the background is smaller than for the signal, and the small second peak is visible at the wrong position, $s=1$, for the background. 
This indicates that the network finds some background events challenging to classify correctly.
Another observation is that the network assigns high scores for the events with one or two on-shell squarks.
This suggests that the network is confident that these events come from the BSM processes. 
This is not surprising because the jets from heavy squark 
decays are expected to be different from 
ISR jets. 
Due to the large mass splitting between the squark and electroweakino, quarks produced in squark decays tend to have large transverse momentum of the order of half of the squark mass, while ISR jets are expected to be, in general, less energetic.
On the right panel, we see that the background yield is typically two orders of magnitude larger than the signal yield, except for the region with a large $s$. 
The inclusion of all three processes in the analysis is crucial for an accurate estimate of the mass limits and sensitivities.   

\begin{figure}[t!]
    \centering
    \includegraphics[width=0.55\linewidth]{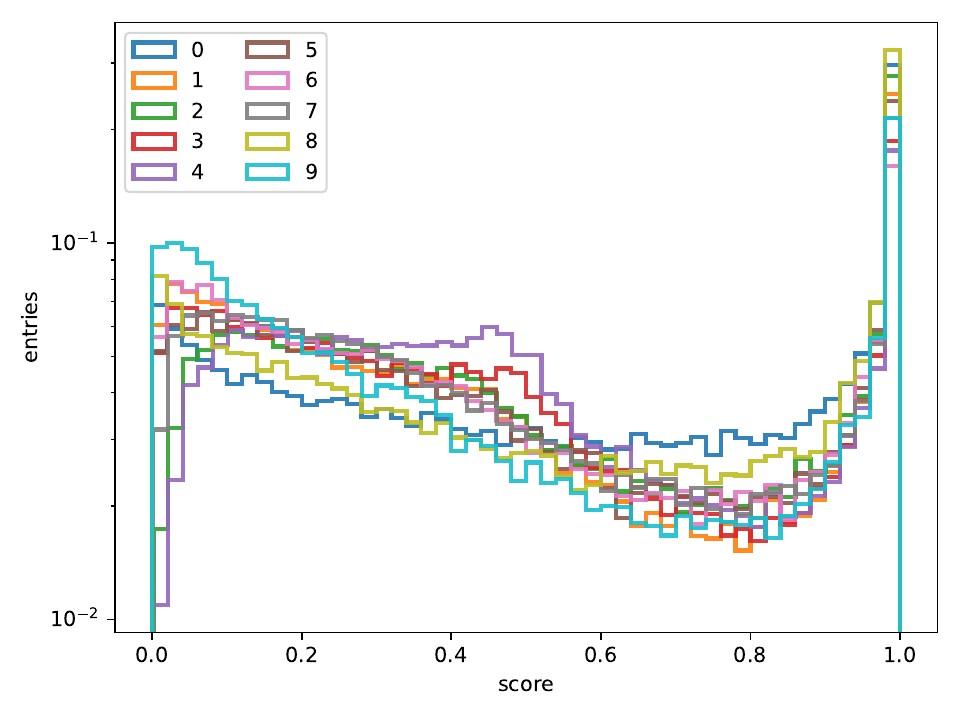}
    \caption{\small 
    The output of 10 neural networks trained to discriminate between the SM background and the signal of the Wino-like scenario. 
    Different colours indicate different networks and random seed values.
    Each histogram shows the output from the combined data (signal+background) and is normalised to one.  }
    \label{fig:wino-score-ensemble}
\end{figure}

To roughly estimate the statistical fluctuation in our NN classification, we show in Fig.\ \ref{fig:wino-score-ensemble} the score histograms for the combined signal and background events (unweighted) obtained by ten different networks trained with different training samples.    
One can see that all histograms peak at $s=1$ due to the signal events. 
Most of the histograms have a second, background-related peak around $s=0$. 
However, the height and exact location of this peak varies between the networks. 
Moreover, some networks, e.g.\ 3 and 4, exhibit a third peak around $\rm s=0.5$, which is totally absent for other members of the ensemble, e.g.\ 0 and 9. 
The presence of the third peak may suggest the existence of a subset of events whose characteristics the network failed to learn during the training phase. 
This could be due to the network converging to a suboptimal minimum influenced by these events, or it might be that these problematic events were underrepresented in the training dataset, preventing the network from adequately encountering and learning from them.
In the following sections, we will track how this random fluctuation propagates into physical quantities. 
In practical applications of ML, it is important to acknowledge and handle these fluctuations.

\begin{figure}[t!]
    \centering
     \includegraphics[width=0.448\linewidth]{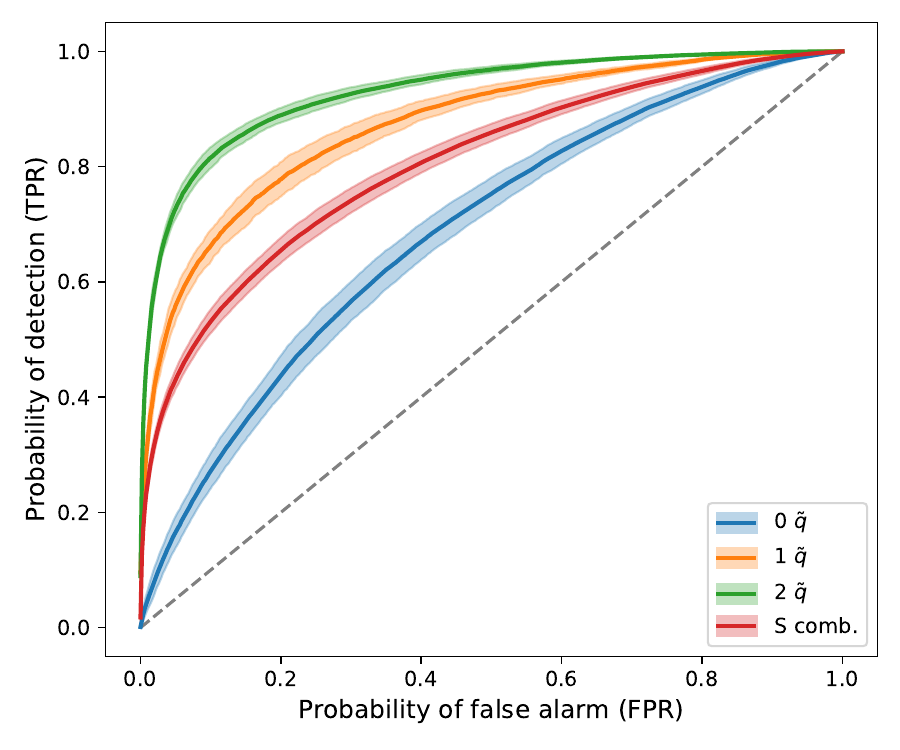}
    \includegraphics[width=0.5\linewidth]{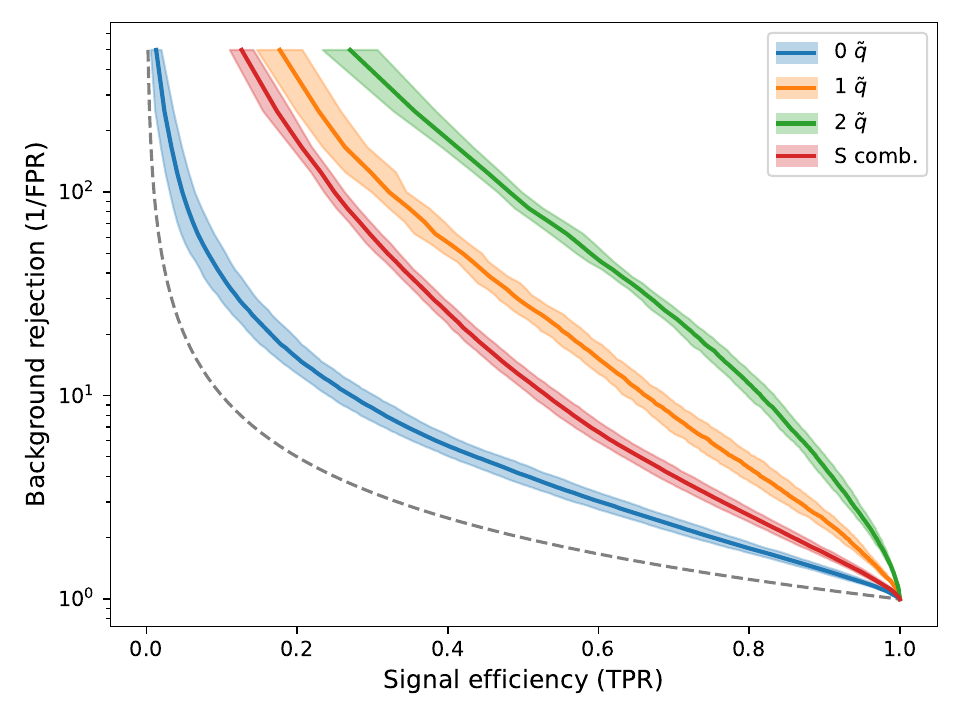}
    \caption{\small The ROC curves for the Wino-like neutralino with $m_{\tilde \chi_1^0} = 300$ GeV and $m_{\tilde q}=2.2$ TeV. 
    Different colours correspond to different classes of the signal: 
    0 $\tilde q$ (blue), 1 $\tilde q$ (orange), 2 $\tilde q$ (green) and combined (red). 
    The solid lines represent the mean result of an ensemble of 10 networks, and the bands correspond to $\pm 1 \sigma$ deviation from the mean. 
    }
    \label{fig:wino-roc}
\end{figure}

The left panel of Fig.\ \ref{fig:wino-roc} shows a Receiver Operator Characteristics (ROC) curve for our classifier. 
The thin bands around the curves represent the standard deviation evaluated from the ensemble of ten neural networks.
There are four curves in the plot corresponding to different classes of signal events. 
The red curve corresponds to the total signal combining all three classes, while the blue, orange and green curves represent 0 $\tilde q$, 1 $\tilde q$ and 2 $\tilde q$ processes, respectively. 
In evaluating the ROC curves for subclasses of signal events, the appropriate portion of the background sample is used to balance the size of the signal and background samples.
Comparing the ROC curves between the signal subclasses, we see that the best classification performance is achieved for events with two on-shell squarks. 
This is because those events typically possess two very high \pt jets originating from squark decays, which makes it easier to distinguish them from the background events.
The worst performance is seen for events with the 0 $\tilde q$ process, in which the high $p_T$ jets can come only from ISR as in the background.  
Notice, however, that the classification is much better than 
the random label assignment, meaning that the network can discriminate the 0 $\tilde q$ process from the SM background to some extent.  
The ROC curve for the combined signal lies between the 0 $\tilde q$ and 1 $\tilde q$ ones because the signal sample is dominated by the 0 $\tilde q$ process. 
The Area Under Curve (AUC) values for these ROC curves are listed in Table \ref{tab:wino-auc}. 

The right panel of Fig.\ \ref{fig:wino-roc} presents the same information as the ROC curves but in a different format. 
Unlike the left panel, which uses the False Positive Rate (FPR) and True Positive Rate (TPR) for the $x$ and $y$ axes, the right panel shows the trajectories in the (TPR, 1/FPR) plane when $s_{\rm th}$ is continuously varied. This representation is often useful because the $x$ and $y$ axes correspond to the signal efficiency and the background rejection, respectively. It illustrates the trade-off between signal loss and background rejection when choosing a threshold $s_{\rm th}$ for a desired level of background suppression.
One can see that the background rejection of 200 can be achieved, keeping the signal efficiency level around 0.2.
Note that these are the rejection and efficiency factors defined after the preselection cuts. 

\begin{table}[t!]
\centering
    \begin{tabular}{|c|c|c|}
        \hline
         Signal class &  mean AUC & standard deviation\\
         \hline
         0 $\tilde q$&    0.6866 &  0.0220 \\\hline
         1 $\tilde q$&    0.8718 &  0.0131\\\hline
         2 $\tilde q$&    0.9295 &  0.0065\\\hline
         combined    &   0.8039 &  0.0131 \\\hline         
    \end{tabular}
    \caption{\label{tab:wino-auc}
    \small The AUC for the ensemble of NN models trained for the Wino-like scenario with $m_{\tilde W} = 300$ GeV, $m_{\tilde q}=2.2$ TeV.
    }
\end{table}

\begin{figure}
    \centering
    \includegraphics[width=1\linewidth]{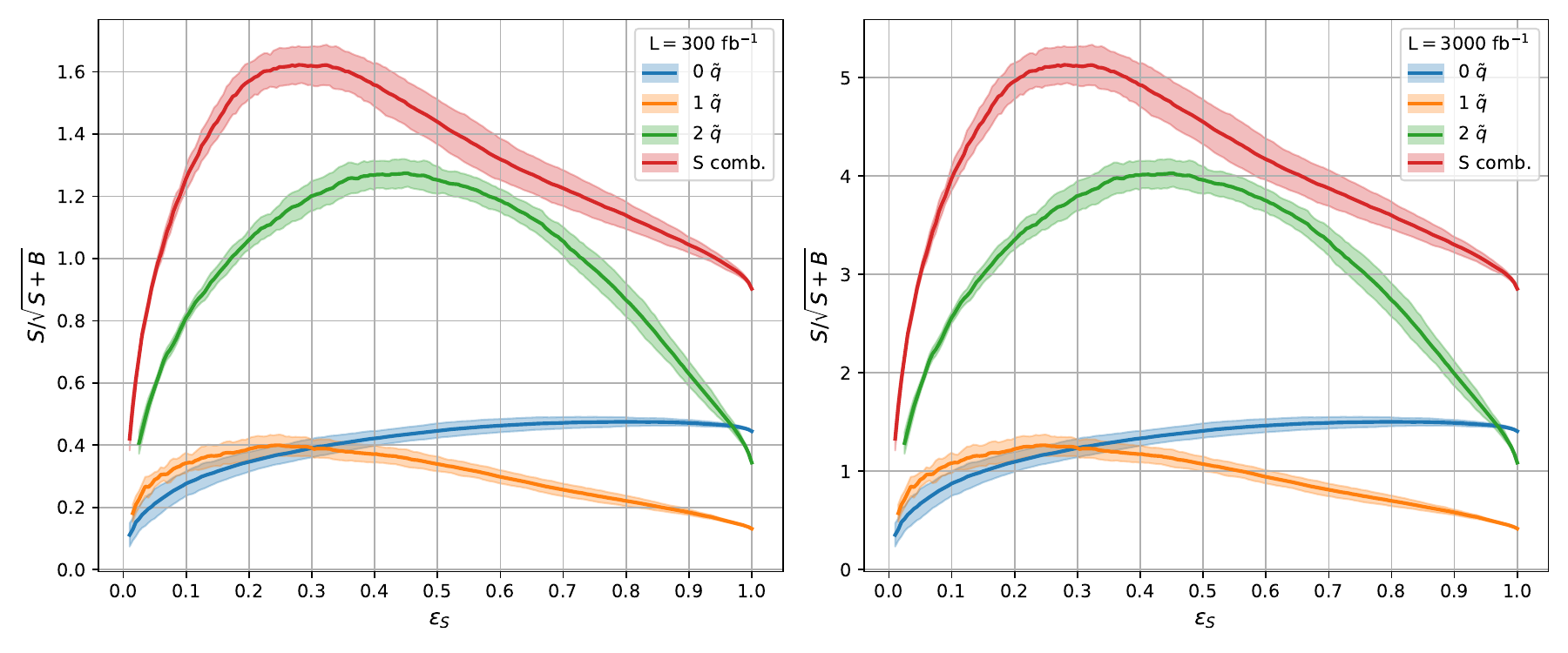}
    \caption{\small 
    The signal efficiency $\epsilon_S$ vs the naive significance $\rm S/\sqrt{S+B}$ for $L=300~\mathrm{fb}^{-1}$ (left) and  $L=3000~\mathrm{fb}^{-1}$ (right) for the Wino-like scenario. 
    The colours indicate the signal classes: 
    0 $\tilde q$ (blue), 1 $\tilde q$ (orange), 2 $\tilde q$ (green) and combined (red). 
    The solid lines represent the mean result of an ensemble of 10 networks and the bands correspond to $\pm 1 \sigma$ deviation from the mean. 
    }
    \label{fig:wino-eff-vs-sign}
\end{figure}

In order to assess the usefulness of our classification model for the SUSY search, we introduce a quantity $\rm Z \equiv S/\sqrt{S+B}$, referred to as \textit{naive significance}, which provides a good estimate of the statistical significance of the signal excess. 
In the above formula, $S$ and $B$ denote the signal and background yields after the preselection and the final selection with $s > s_{\rm th}$, respectively.
In Fig.\ \ref{fig:wino-eff-vs-sign} we plot the naive significance $Z$ as a function of the signal 
efficiency $\epsilon_S$, for $L = 300$ $\mathrm{fb}^{-1}$ (left) and $L=3000$ $\mathrm{fb}^{-1}$ (right). 
These plots differ only by a constant scaling with a factor $\sqrt{10}\simeq 3.16$.
In addition to the total significance (red), the contributions from three different signal processes, 0 $\tilde q$ (blue), 1 $\tilde q$ (orange) and 2 $\tilde q$ (green), are also shown. 
As can be seen, the largest significance $Z \sim 1.6$ (5.1)
can be achieved with $\epsilon_S \sim 0.3$ for 
$L = 300$ (3000) $\mathrm{fb}^{-1}$.
Around this region, the dominant contribution comes from the 2 $\tilde q$ process, while the other two production processes have smaller but still relevant impacts.

\subsection{Higgsino-like LSP}\label{sec:higgsino-eval}

\begin{figure}[t!]
    \centering
    \includegraphics[width=0.49\linewidth]{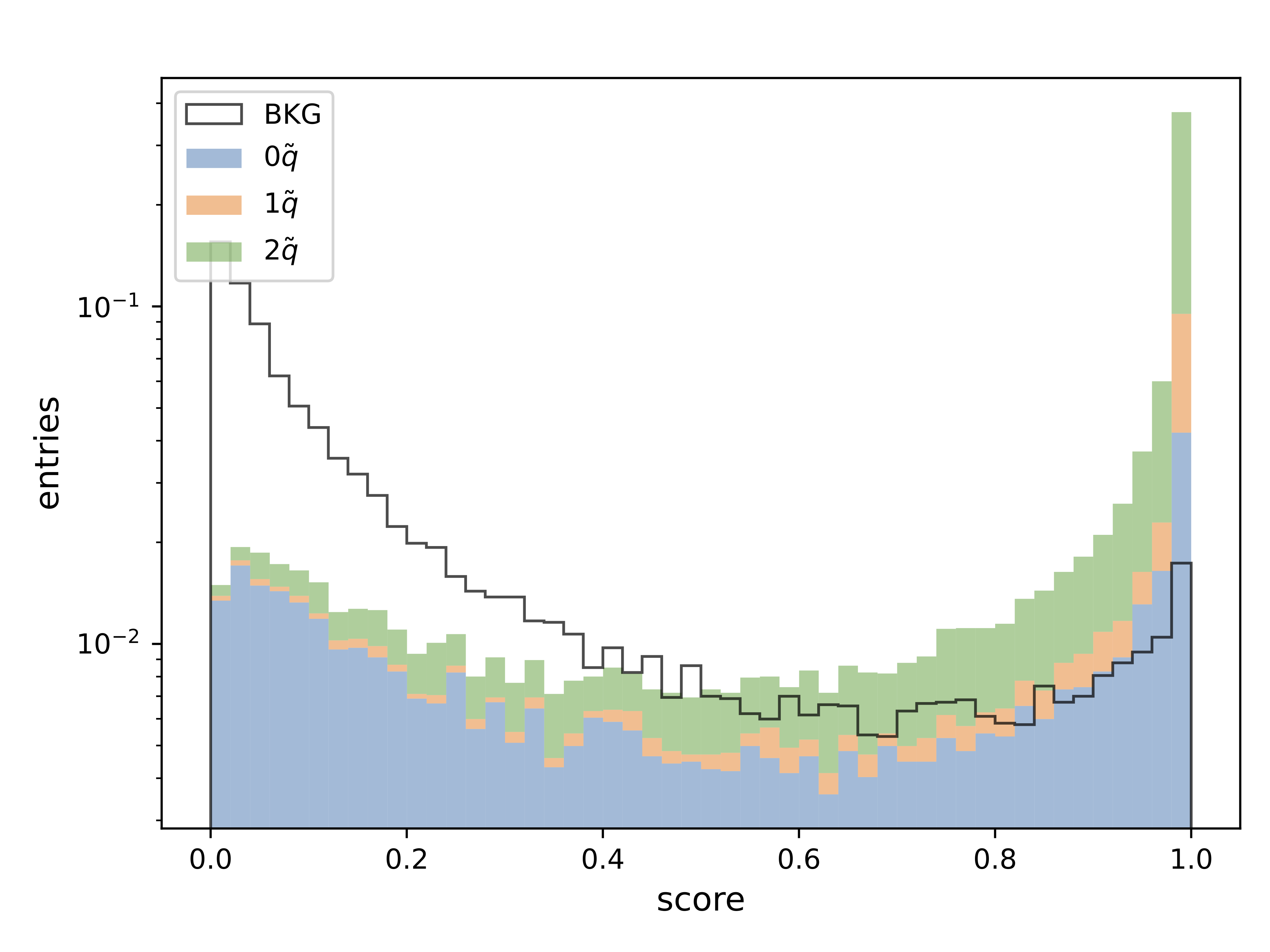}
     \includegraphics[width=0.49\linewidth]{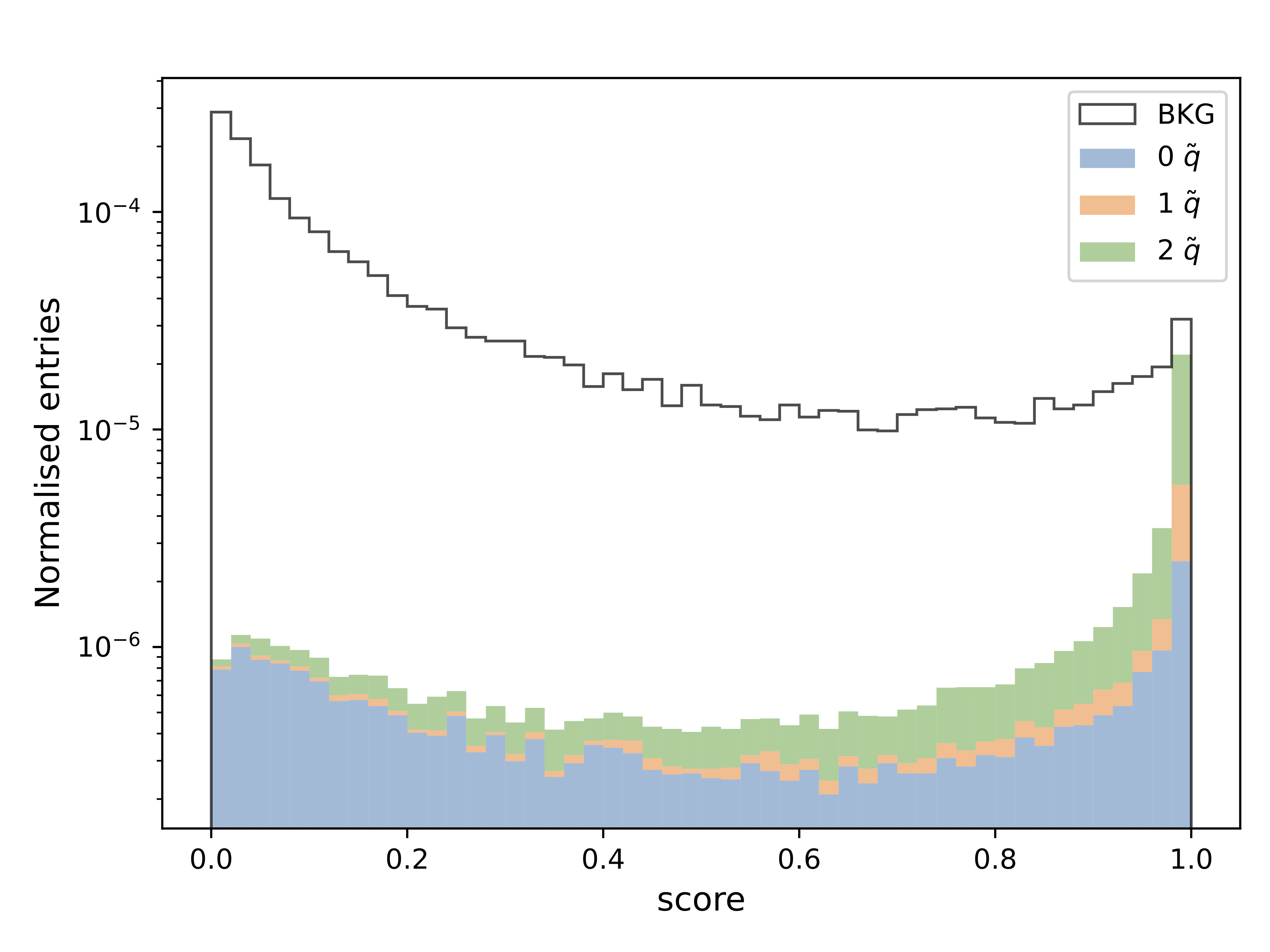}
    \caption{\small 
    The output of a neural network for the Higgsino-like scenario.
    The distributions are normalised to 1 in the left panel and to the corresponding cross section in the right panel.
    The colours indicate the signal classes: 0 $\tilde q$ (blue), 1 $\tilde q$ (orange), 2 $\tilde q$ (green). 
    The histogram for the SM background is shown as an unfilled thick black line.    }
    \label{fig:hino-score}
\end{figure}

We discuss the results for the Higgsino-like neutralino with $m_{\tilde h}=300$ GeV and $m_{\tilde q}=2.2$ TeV. 
The output of one of the ten NN models in the ensemble is shown in Fig.\ \ref{fig:hino-score}, 
where the left panel displays the normalised distributions of the NN score,
while the right panel presents the same distributions weighted with the cross sections of the corresponding processes.
The open histogram represents the score for the SM background, while the coloured histograms show the contributions from the three classes of the signal process: 
0 $\tilde q$ (blue), 1 $\tilde q$ (orange) and 2 $\tilde q$ (green).
Compared to the Wino case in Fig.\ \ref{fig:wino-score},
we see that the score distribution for Higgsinos has a U-shape, with two sharp peaks at $\rm s \simeq 0$ and $\rm s \simeq 1$, which usually indicates good discriminative power and stability of the classifier. 
Also, unlike the Wino case, the events with on-shell squarks are more evenly distributed. 
This suggests that the network focuses more on the 0 $\tilde q$ signal class.

\begin{figure}[t!]
    \centering
     \includegraphics[width=0.448\linewidth]{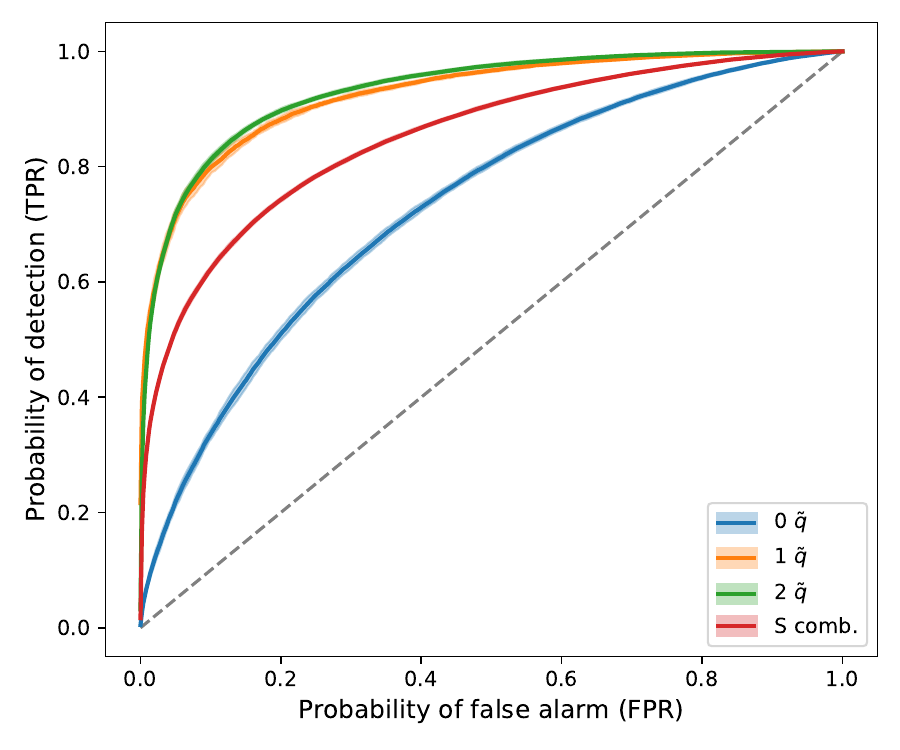}
    \includegraphics[width=0.5\linewidth]{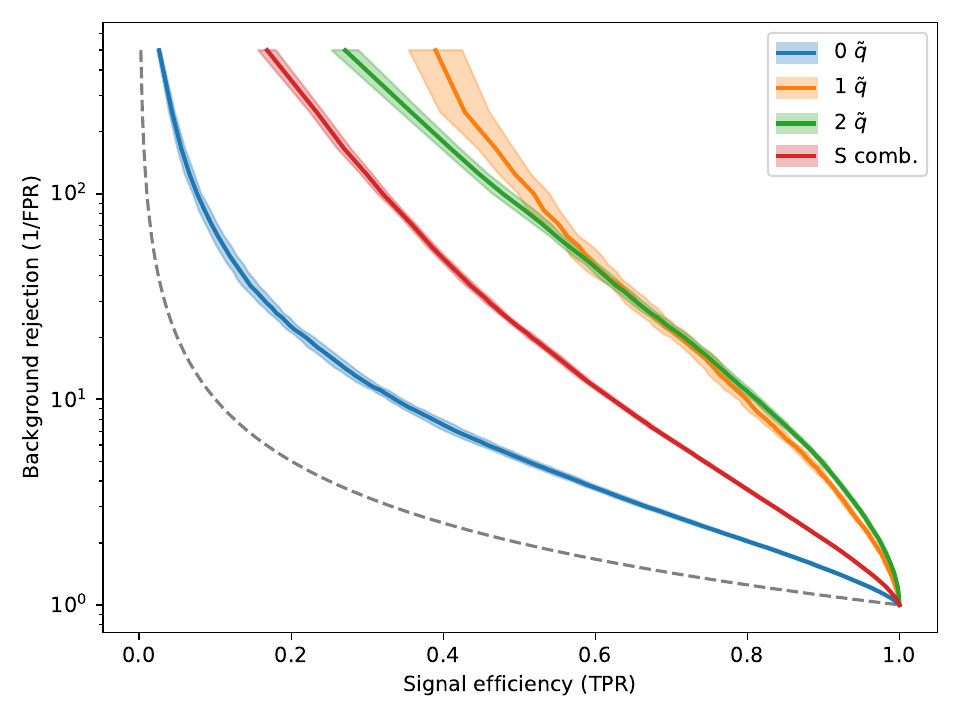}
    \caption{\small 
    The ROC curves for the Higgsino-like neutralino with $m_{\tilde \chi_1^0} = 300$ GeV and $m_{\tilde q}=2.2$ TeV. 
    The colours indicate the signal classes: 
    0 $\tilde q$ (blue), 1 $\tilde q$ (orange), 2 $\tilde q$ (green) and combined (red). 
    The solid lines represent the mean result of an ensemble of 10 networks, and the bands correspond to $\pm 1 \sigma$ deviation from the mean.    
    }
    \label{fig:hino-roc}
\end{figure}

The ROC curves for models trained on Higgsino-like samples are shown in Fig.\ \ref{fig:hino-roc}.
As in the Wino study, we estimate the fluctuation from the ten different network models trained with different samples, which is depicted as the widths of the curves.  
As can be seen, the curve's widths are much smaller compared to the Wino case, indicating that the predictions from different networks are largely consistent.
We also observe, similarly to Winos, that events with on-shell squarks are easier to distinguish from the background than the 0 $\tilde q$ production events. 
The discrimination performances for 1 $\tilde q$ and 2 $\tilde q$ signal classes are very similar, but 1 $\tilde q$ exhibits slightly better performance when the signal efficiency is lower than 0.5, as can be seen in the right panel. 
The mean AUC and the standard deviations for all signal classes are listed in Table \ref{tab:hino-auc}. 
The AUC for the combined signal is 0.8515(23).
Compared with Table \ref{tab:wino-auc}, we see that signal/background discrimination with GNN is better for Higgsinos than for Winos.

\begin{table}[t!]
 \centering
    \begin{tabular}{|c|c|c|}
        \hline
         Signal class &  mean AUC & standard deviation\\
         \hline
         0 $\tilde q$&   0.7292  &  0.0059    \\\hline
         1 $\tilde q$&   0.9265 &  0.0032    \\\hline
         2 $\tilde q$&   0.9330 &  0.0025     \\\hline
         combined    &   0.8515   &  0.0023       \\\hline
    \end{tabular}
    \caption{\small 
    The AUC for the ensemble of NN models trained for the Higgsino-like scenario with $m_{\tilde h} = 300$ GeV and $m_{\tilde q}=2.2$ TeV.
    }
    \label{tab:hino-auc}
\end{table}

\begin{figure}[t!]
    \centering
    \includegraphics[width=0.95\linewidth]{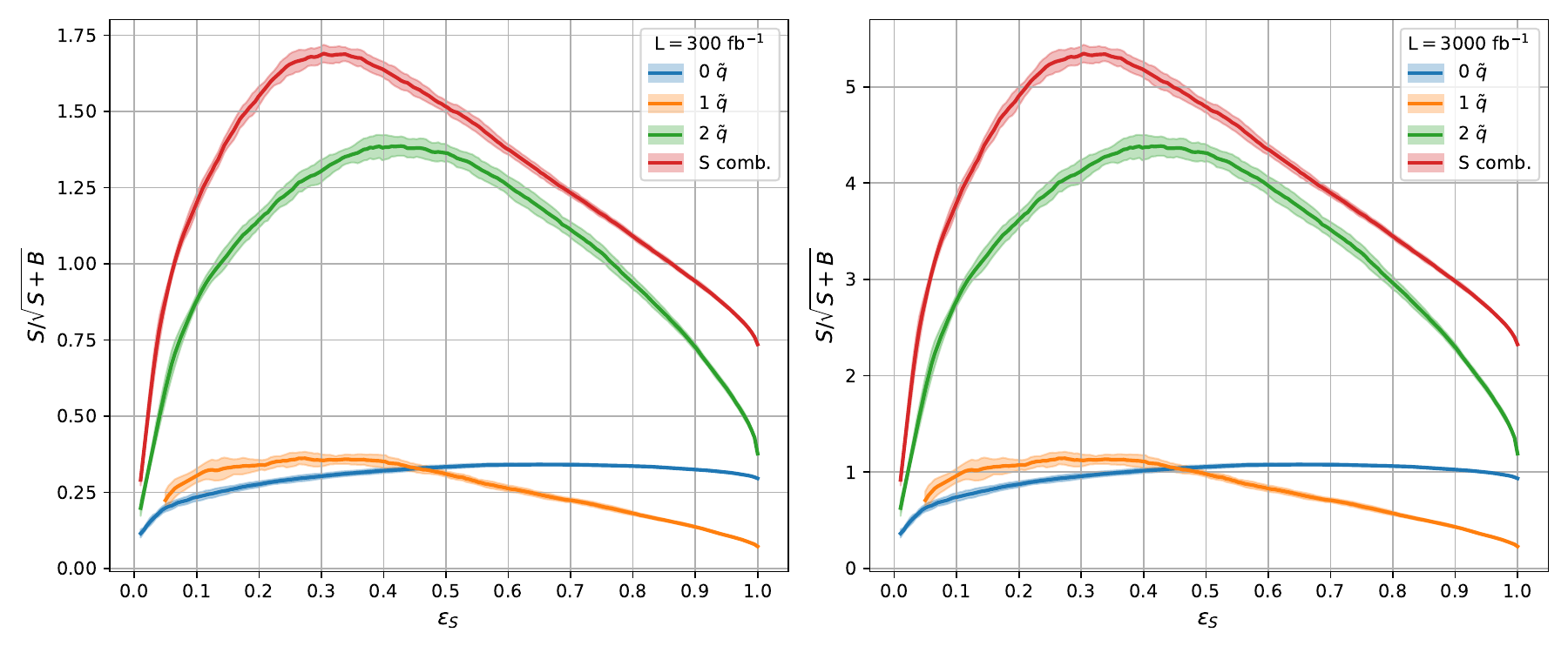}
    \caption{\small 
   The signal efficiency $\epsilon_S$ vs the naive significance $\rm S/\sqrt{S+B}$ for $L=300~\mathrm{fb}^{-1}$ (left) and  $L=3000~\mathrm{fb}^{-1}$ (right) for the Higgsino-like scenario. 
    The colours indicate the signal classes: 
    0 $\tilde q$ (blue), 1 $\tilde q$ (orange), 2 $\tilde q$ (green) and combined (red). 
    The solid lines represent the mean result of an ensemble of 10 networks and the bands correspond to $\pm 1 \sigma$ deviation from the mean.  
    }
    \label{fig:hino-eff-vs-sign}
\end{figure}

Fig.\ \ref{fig:hino-eff-vs-sign} shows the naive significance for different classes of the signal as a function of the signal efficiency. 
The left and right panels of Fig.\ \ref{fig:hino-eff-vs-sign} are for $L = 300$ fb$^{-1}$ (Run-3 LHC) and $L = 3000$ fb$^{-1}$ (HL-LHC).
The highest naive significance, reaching $Z \simeq 1.7$ for Run-3 LHC and $Z\simeq 5.4$ for HL-LHC, is achieved with $\epsilon_S \simeq 0.3$. 
Similarly to Winos, around this efficiency, the most relevant contribution to the total naive significance comes from the 2 $\tilde q$ events, with the two other signal classes being subdominant but non-negligible.

\subsection{Comparison with Boosted Decision Trees}

In this subsection, we compare the performances of our GNNs described in
Sec.\ \ref{sec:wino-eval} with that of the Boosted Decision Trees models implemented with the XGBoost library \cite{DBLP:journals/corr/ChenG16} for the Wino-like benchmark scenario with $m_{\tilde W}=300$ GeV and $m_{\tilde q}=2.2$ TeV.
Boosted Decision Trees (BDTs) are an ensemble learning technique that combines the predictions of multiple decision trees to improve accuracy and robustness. 
XGBoost (Extreme Gradient Boosting) \cite{DBLP:journals/corr/ChenG16} is an advanced implementation of BDTs that excels in speed and performance. 
It uses a gradient boosting framework to iteratively add trees, optimising for the residual errors of previous iterations.

The ensemble of BDTs is trained on the same data as the ensemble of GNNs\footnote{We also tried training BDTs on solely high-level or low-level data, but we obtained worse results.}. 
Low-level data for up to 250 particles, padded with zeros\footnote{The zero-padding is, strictly speaking, not neccessary, because ParticleNet is capable of handling a variable-sized input. However, our architecture is based on the official implementation of ParticleNet as provided in \cite{particlenetcode} by its authors. In this approach, a fixed-size input is used in order to have a simple implementation for an efficient batched training on GPU. Particle features are padded with zeros such that they always have the same length, and a \textit{mask} array is used to indicate if a position is occupied by a real particle or by a zero-padded value.}, is reshaped to the form of a 2D array, where the first dimension corresponds to events, and the second dimension contains concatenated variables, as listed in Table \ref{tab:low-level-variables}. 
High-level variables, listed in Table \ref{tab:high-level-variables}, are appended in front of the low-level data.
A subset of training data is used to find the optimal hyperparameters using the Optuna package \cite{DBLP:journals/corr/abs-1907-10902} and 100 trials. 
Values of the optimal hyperparameters are listed in Table \ref{tab:bdt-hyper} in Appendix \ref{sec:hyperparameters}. 
When the hyperparameters are set, ten models are trained using the same train-validation split as for the GNNs.

\begin{figure}[t!]
    \centering
    \includegraphics[width=0.6\linewidth]{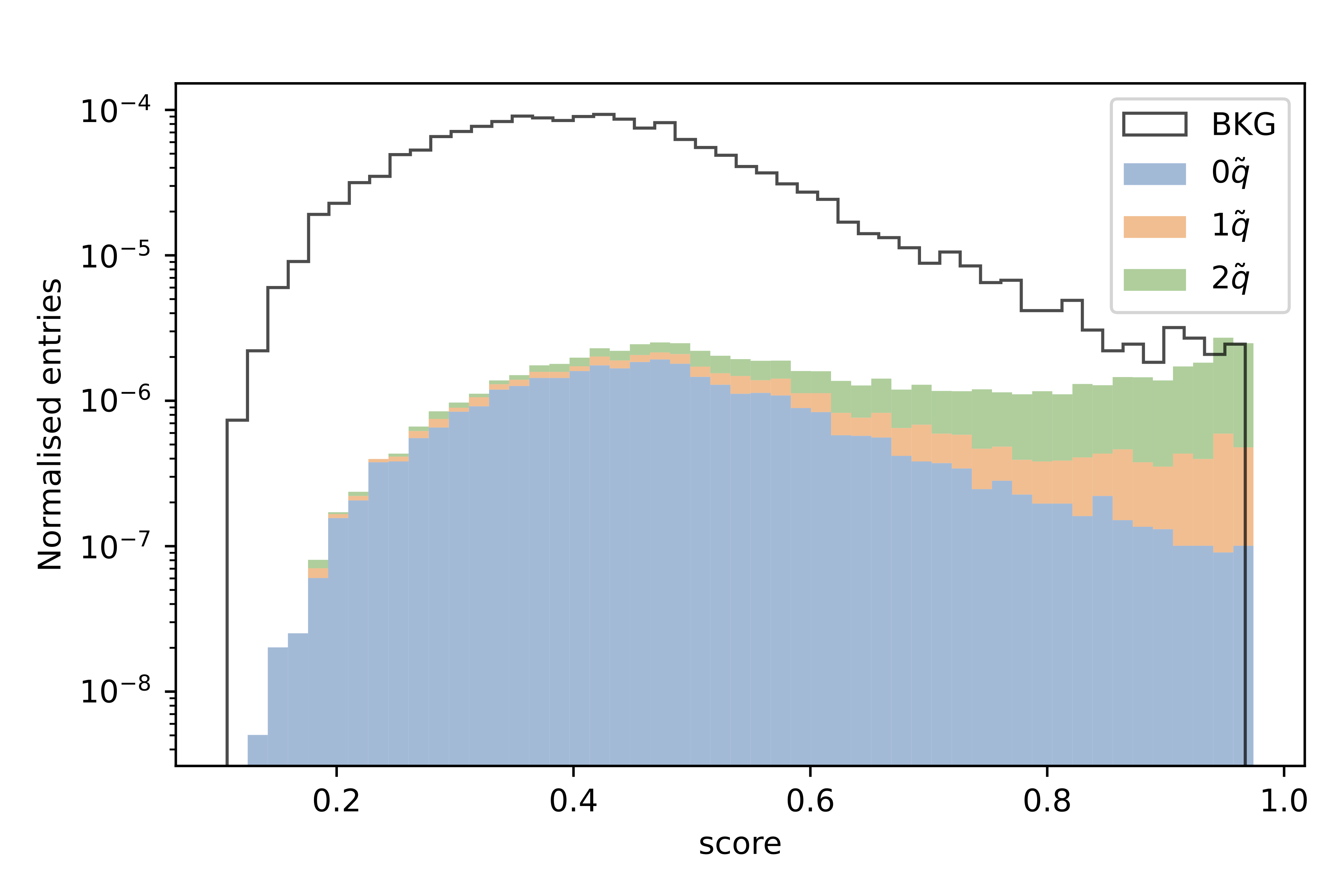}
    \caption{\small 
    The distributions of the BDT output normalised to the cross section of the corresponding processes. 
    }
    \label{fig:bdt-score}
\end{figure}

The output of a BDT model is shown in Fig.\ \ref{fig:bdt-score}. 
The figure depicts the score of one of the ten models weighted by the cross section.
One can see from the plot that, unlike the GNN output, there is no peak for $s \simeq 0$ for the background. 
The peak at $s \simeq 1$ for the signal is also very weak.
This suggests that the BDT is not able to distinguish well between the background and the signal. 
Nevertheless, most 1 $\tilde q$ and 2 $\tilde q$ events are assigned high score values, which implies that the algorithm can find differences between jets originating from squark decays and ISR.

\begin{figure}[t!]
    \centering
    \includegraphics[width=0.7\linewidth]{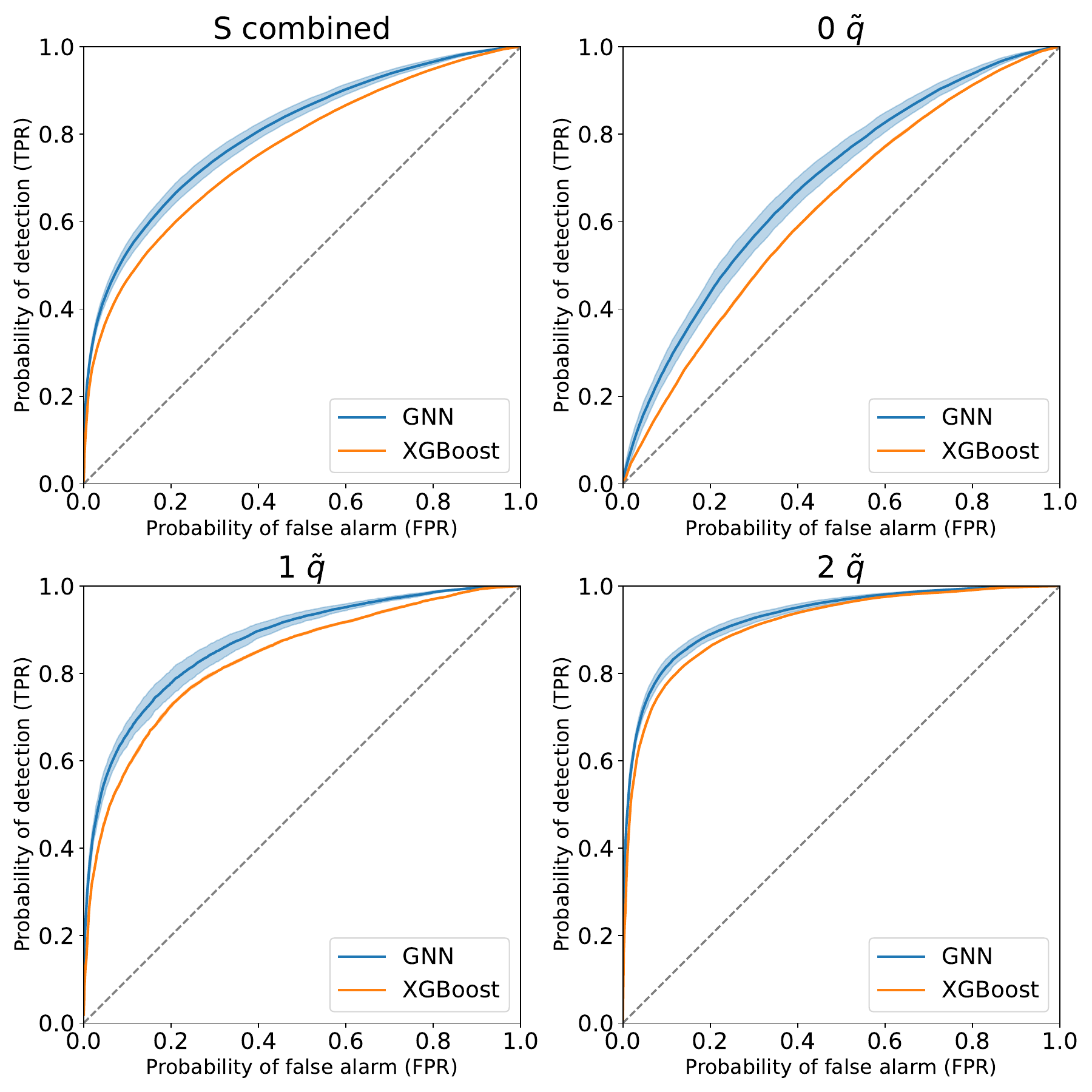}
    \caption{\small 
    The ROC curves for the Wino-like scenario with $m_{\tilde W}=300$ GeV and $m_{\tilde q}=2.2$ TeV obtained from Graph Neural Networks (blue) and BDT models (orange). 
    The plots are made for the combined signal (top left),
    0 $\tilde q$ (top right), 1 $\tilde q$ (bottom left)
    and 2 $\tilde q$ (bottom right).
    }
    \label{fig:bdt-roc-comparison}
\end{figure}

In Fig.\ \ref{fig:bdt-roc-comparison}, we directly compare the ROC curves of BDT models (orange) with those of GNNs (blue). 
The top-left plot depicts the combined signal sample, while the top-right, bottom-left and bottom-right plots are for 0 $\tilde q$, 1 $\tilde q$ and 2 $\tilde q$ signal classes, respectively. 
As can be seen, GNNs consistently perform better than the BDTs for all types of signals. Nevertheless, the largest difference is observed for the 0 $\tilde q$ signal class.  
On the other hand, the random fluctuation in the BDT classification is much smaller, leading to a much smaller spread of the curves compared to the GNNs results.

\begin{figure}[t!]
    \centering
    \includegraphics[width=0.9\linewidth]{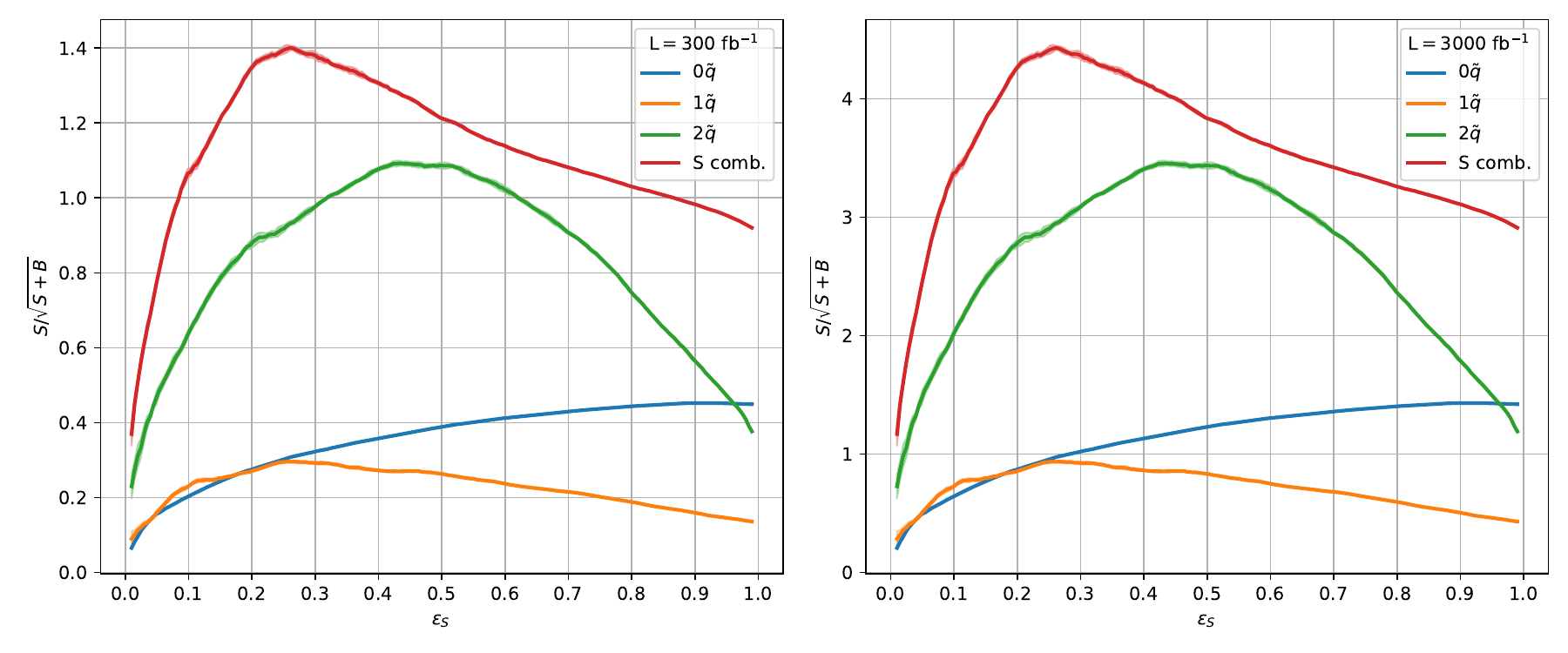}
    \caption{\small 
    The signal efficiency $\epsilon_S$ vs the naive significance $\rm S/\sqrt{S+B}$ for $L=300~\mathrm{fb}^{-1}$ (left) and  $L=3000~\mathrm{fb}^{-1}$ (right) obtained from BDT models trained on Wino-like neutralino sample with $m_{\tilde \chi_1^0} = 300$ GeV and $m_{\tilde q}=2.2$ TeV. 
    The colours indicate the signal classes: 
    0 $\tilde q$ (blue), 1 $\tilde q$ (orange), 2 $\tilde q$ (green) and combined (red). 
    The solid lines represent the mean result of an ensemble of 10 networks and the bands correspond to $\pm 1 \sigma$ deviation from the mean. 
    }
    \label{fig:bdt-eff-vs-sign}
\end{figure}

Fig.\ \ref{fig:bdt-eff-vs-sign} shows the trajectories of the signal efficiency and the naive significance when the score threshold $s_{\rm th}$ is varied. 
The panel on the left-hand side is for \lhc, while the panel on the right is for \hllhc. 
Interestingly, the shapes of all curves do not differ much from the GNN results shown in Fig.\ \ref{fig:wino-eff-vs-sign}. 
However, the uncertainty is much smaller, and the values of naive significance are significantly lower. 
For the end of Run-3 (\lhc), the maximal value of $\rm S/\sqrt{S+B}$ for the combined signal sample is 1.4 and corresponds to $\epsilon_S \approx 0.25$. For \hllhc, it is $Z_{\rm max} \approx 4.4$.

\subsection{Impact of the \pt cut}\label{sec:ptcut}

So far, we have used the default \pt cut, $\ppt > 1$ GeV, to select the particle flow objects that enter the low-level input data.   
This selection is quite relaxed on purpose since we expect that the low-energy activity might be helpful in discriminating between the signal and 
the background.
However, the exploitation of soft objects is not straightforward because they are less reliably simulated in the Monte Carlo tools and may also be affected by pileup events. 
Accurately simulating pileup events is notoriously difficult and beyond the scope of this paper.
We instead study the impact of the soft \pt threshold on the classification results for the Wino and Higgsino samples.
On top of the impressive performance of the state-of-the-art pileup mitigation techniques mentioned in the introduction, the pileup contamination will be significantly removed when the $p_T$ threshold is raised to 5 or 10 GeV.

\begin{figure}[t!]
    \centering
    \includegraphics[width=0.7\linewidth]{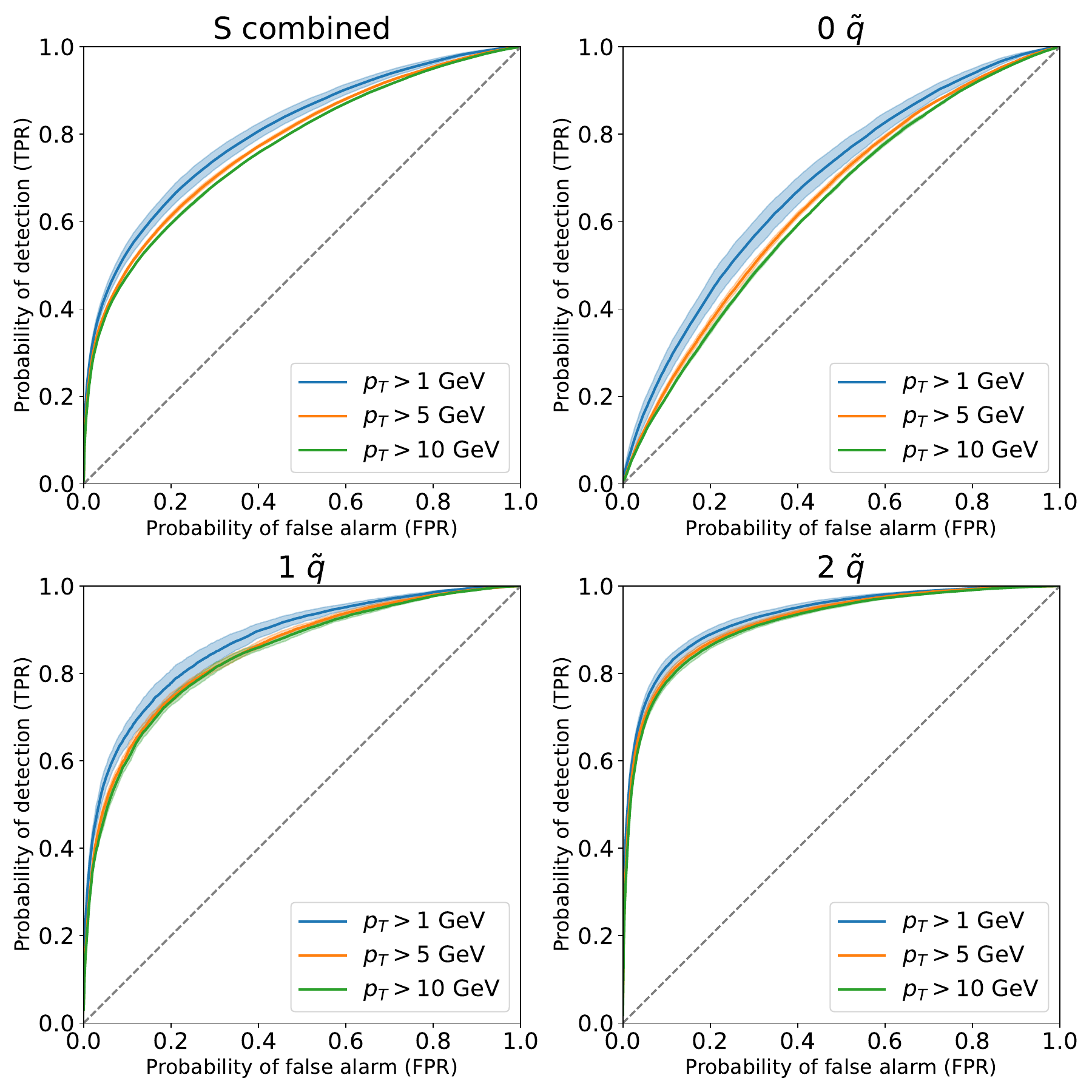}
    \caption{\small 
    The ROC curves for the Wino-like scenario with $m_{\tilde W}=300$ GeV and $m_{\tilde q}=2.2$ TeV. 
    The blue, orange and green curves correspond to the \pt threshold for soft particles of 1, 5 and 10 GeV, respectively.  
    The plots are for the combined signal (top left), 
    0 $\tilde q$ (top right), 
    1 $\tilde q$  (bottom left)
    and 2 $\tilde q$ (bottom right). }
    \label{fig:rocpt}
\end{figure}

Fig.\ \ref{fig:rocpt} shows the ROC curves for the combined signal (top-left) and three signal subclasses: 0 $\tilde q$ (top-right),
1 $\tilde q$ (bottom-left) and 2 $\tilde q$ (bottom-right) for the Wino sample.
Each panel shows three ROC curves corresponding to the different soft thresholds: $\ppt > 1$ GeV (blue), $\ppt > 5$ GeV (orange) and $\ppt > 10$ GeV (green). 
The bands of the curves indicate the fluctuation estimated with the ten different training samples. 
As can be seen, increasing \pt cut from 1 GeV leads to some decrease 
in classification performance for all classes of signal events.
The deterioration is more significant for the 0 $\tilde q$ subclass than the 1 $\tilde q$ and 2 $\tilde q$.
This is not surprising because the networks have to learn the difference in ISR between the signal and background for the 0 $\tilde q$ signal class.  
We also observe that varying the threshold from 1 to 5 GeV leads to noticeable deterioration, while varying it from 5 to 10 GeV does not significantly change the situation.
Moreover, the fluctuation observed in ten different networks is much smaller for the 5 and 10 GeV thresholds than for the 1 GeV one. 

\begin{figure}[t!]
    \centering
    \includegraphics[width=0.7\linewidth]{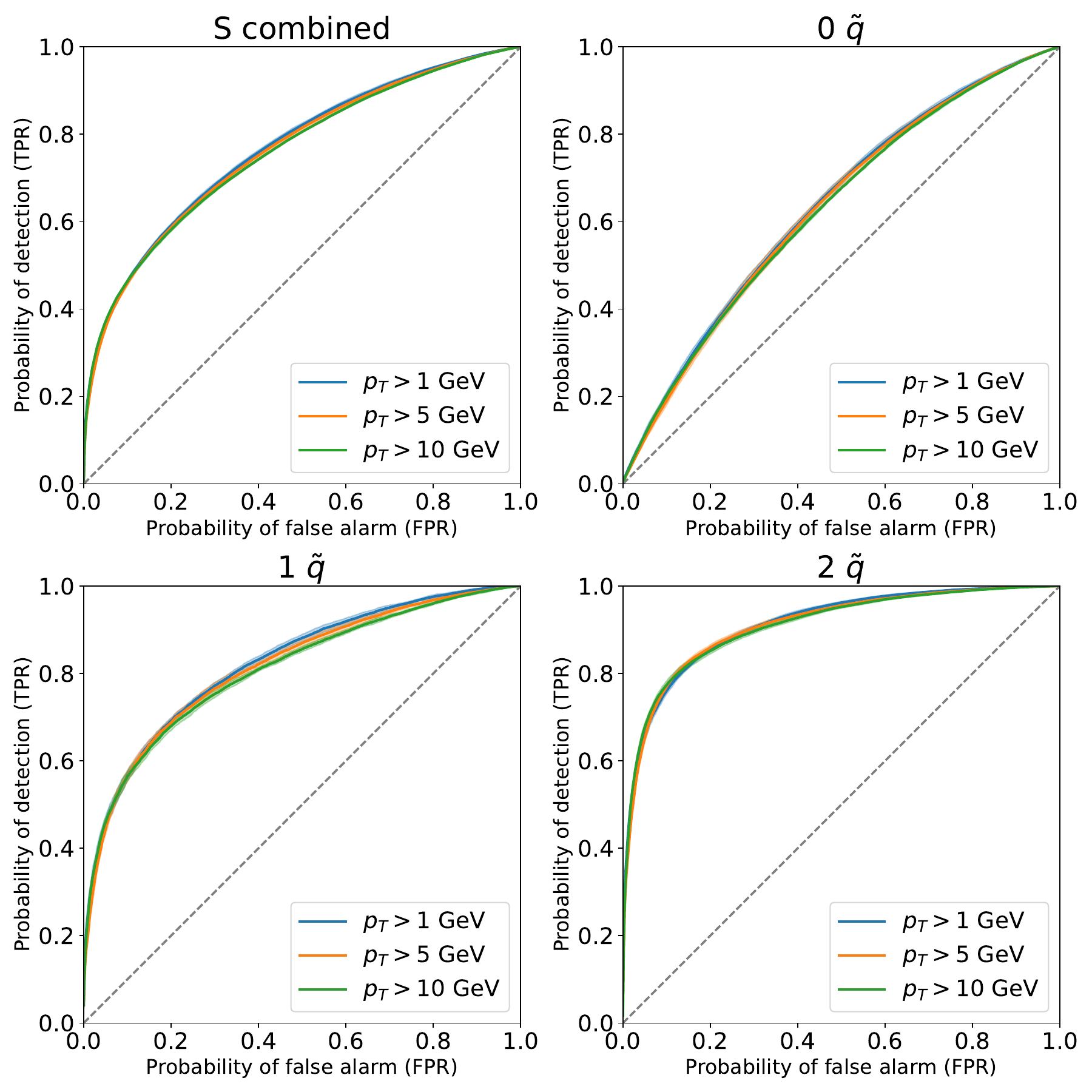}
    \caption{\small 
    The ROC curves for the Higgsino-like scenario with $m_{\tilde h}=300$ GeV and $m_{\tilde q}=2.2$ TeV. 
    The blue, orange and green curves correspond to the \pt threshold for soft particles of 1, 5 and 10 GeV, respectively.  
    The plots are for the combined signal (top left), 
    0 $\tilde q$ (top right), 
    1 $\tilde q$  (bottom left)
    and 2 $\tilde q$ (bottom right). }
    \label{fig:rocpt-higgsino}
\end{figure}

Fig.\ \ref{fig:rocpt-higgsino} depicts the impact of changing the \pt cut value for the Higgsino sample. One can see that, in this case, there is no clear correlation between classification performance and the rejection of soft particles.

This study leads us to the following 
conclusions: 
i) GNN learns the information contained in the soft activity; 
ii) correlations between soft particles are difficult to learn and cause relatively large uncertainty in the prediction; 
(iii) the $\ppt$ cut with the 5 GeV threshold is enough to give reliable results; 
iv) exploitation of soft particles may improve the performance of squark-Wino search.

\subsection{Cross evaluation}\label{sec:cross}

In the previous sections, we have trained and evaluated GNN models with the same signal hypothesis, e.g.\ a model trained with the Wino training sample was evaluated with the Wino test sample. 
In reality, however, we do not know whether the observed signal comes from Wino-, Higgsino- or Bino-like electroweakino. 
In this section, we consider this fact and perform a cross evaluation
of different models.

\begin{figure}[t!]
    \centering
    \includegraphics[width=1\linewidth]{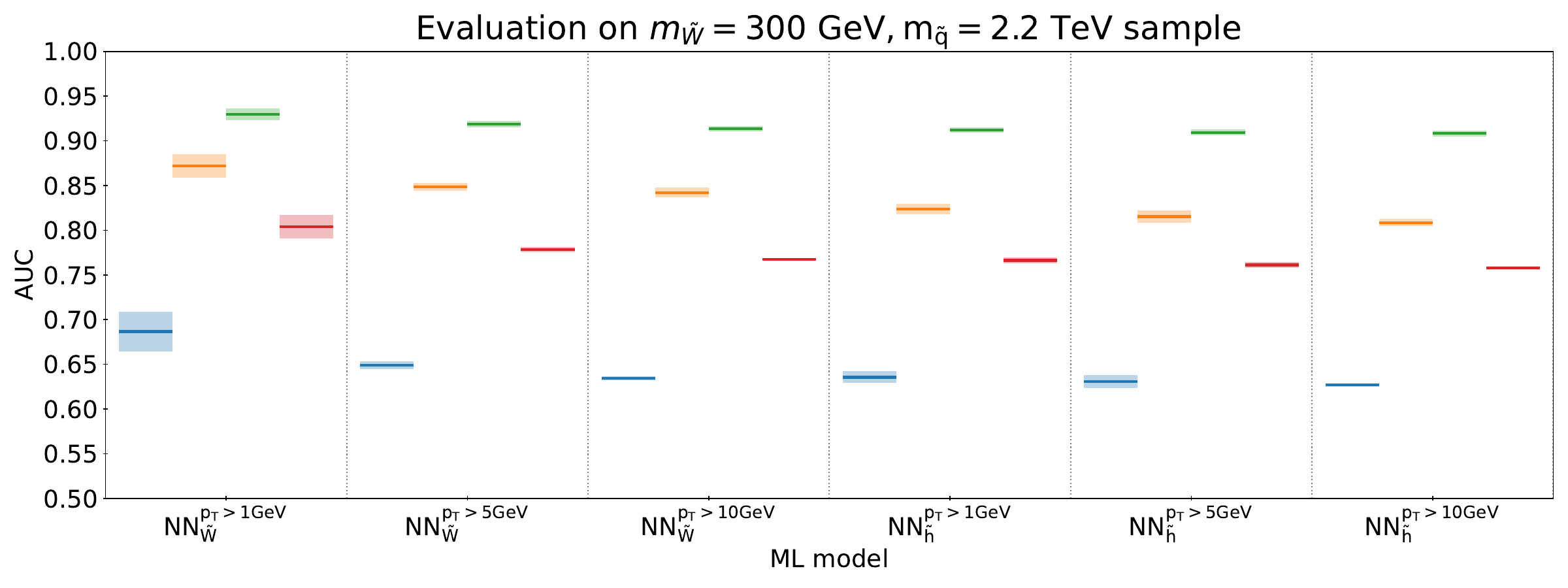}
    \caption{\small 
    The AUC of NN models trained on six samples: (Wino, Higgsino)\,$\times$ $(p_{\rm T} > 1, 5, 10\,{\rm GeV})$. 
    All models are evaluated on the Wino-like neutralino sample with $m_{\tilde W}=300$ GeV and $m_{\tilde q}=2.2$ TeV. 
    The colours indicate the signal classes: combined signal (red), 0 $\tilde q$ (blue), 1 $\tilde q$ (orange) and 2 $\tilde q$ (green).}
    \label{fig:cross-wino}
\end{figure}

Fig.\ \ref{fig:cross-wino} shows the results of the evaluation of six ensembles of GNN models, trained either on Wino-like or Higgsino-like electroweakino, with one of the three \pt cut values: $p_\mathrm{T} > 1$, 5 or 10 GeV. 
All models are evaluated on the benchmark Wino sample, i.e.\ sample with $m_{\tilde W}=300$ GeV and $m_{\tilde q}=2.2$ TeV. 
For each model, the mean AUC and its standard deviation are shown for two signal hypotheses (Wino- or Higgsino-like) and three different $p_T$ cuts. 
We can see that the models trained with Higgsino samples provide worse results than those trained with the Wino sample when evaluated with the Wino sample, as we expected.  
We can also see the trend we discussed in the Sec.\ \ref{sec:ptcut}; lowering the $p_T$ threshold leads to larger AUC values and larger uncertainties.   
We also observe that the 0 $\tilde q$ signal class is the most sensitive to the change of the training sample and the $p_T$ threshold.

\begin{figure}[t!]
    \centering
    \includegraphics[width=1\linewidth]{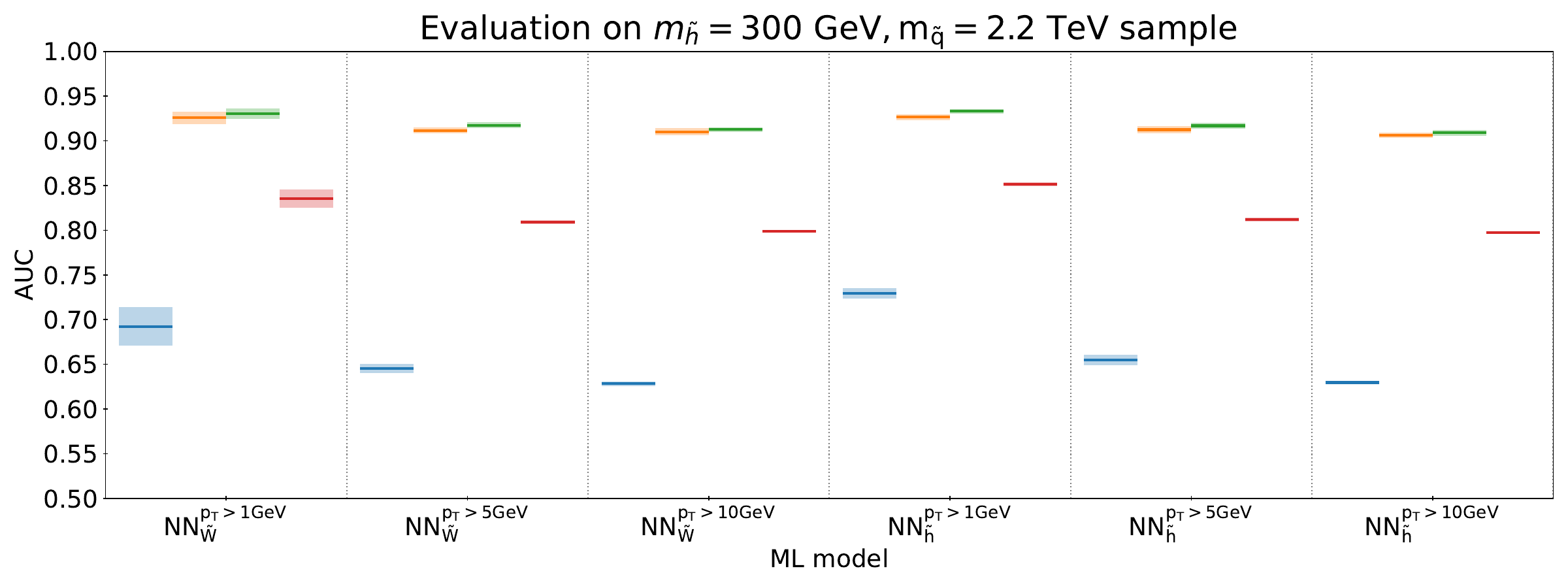}
    \caption{\small 
    The AUC of NN models trained on six samples: (Wino, Higgsino)\,$\times$ $(p_{\rm T} > 1, 5, 10\,{\rm GeV})$. 
    All models are evaluated on the Higgsino-like neutralino sample with $m_{\tilde h}=300$ GeV and $m_{\tilde q}=2.2$ TeV. 
    The colours indicate the signal classes: combined signal (red), 0 $\tilde q$ (blue), 1 $\tilde q$ (orange) and 2 $\tilde q$ (green).    
    }
    \label{fig:cross-higgsino}
\end{figure}

Fig.\ \ref{fig:cross-higgsino} shows results of the cross evaluation on the Higgsino-like sample with $m_{\tilde h}=300$ GeV and $m_{\tilde q}=2.2$ TeV. 
One see that all models report AUC $>$ 0.90 for the 1 $\tilde q$ and 2 $\tilde q$ signal classes.  
The impact of the wrong signal hypothesis is most clearly observed for the 0 $\tilde q$ signal class. 
The classification performance of the models trained on the Wino samples for the 0 $\tilde q$ class is much worse compared with the Higgsino models.
In particular, for the 1 GeV $p_T$ threshold, the result becomes less stable for the Wino model, leading to the large blue band.    
This suggests that soft particles in Higgsino pair production events have some additional correlations that networks were able to learn.

\begin{figure}[t!]
    \centering
    \includegraphics[width=1\linewidth]{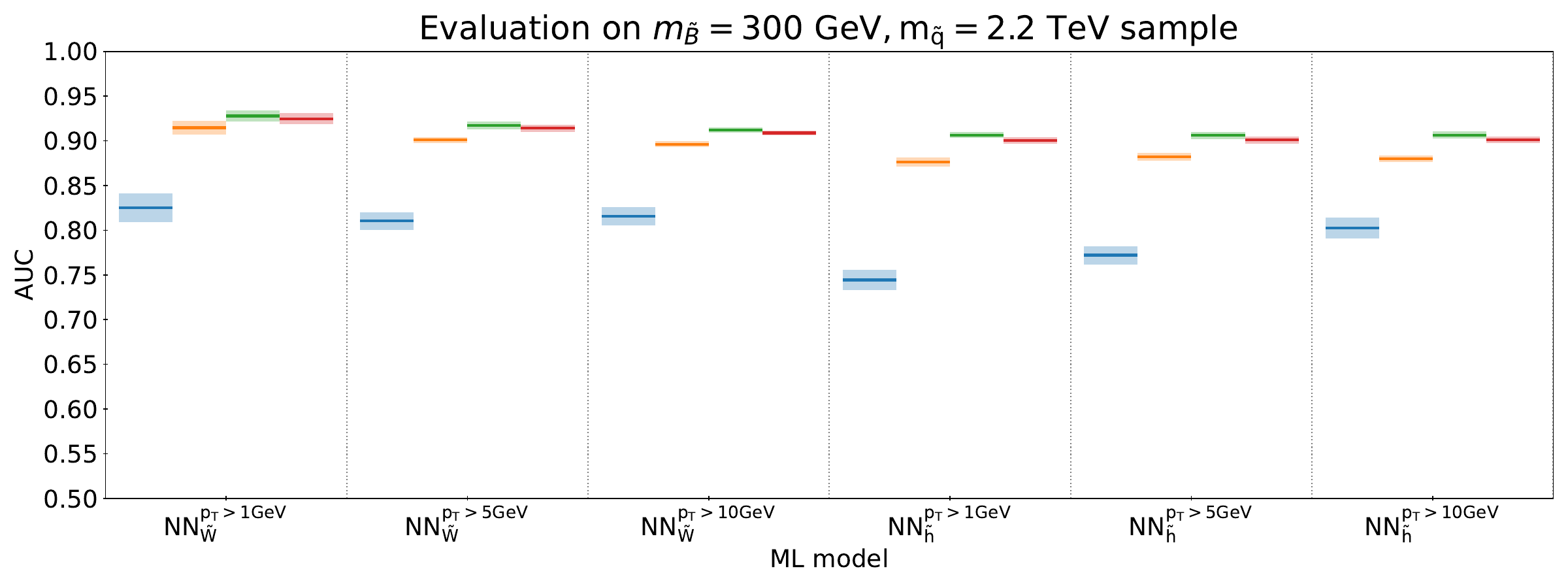}
    \caption{\small 
The AUC of NN models trained on six samples: (Wino, Higgsino)\,$\times$ $(p_{\rm T} > 1, 5, 10\,{\rm GeV})$. 
    All models are evaluated on the Bino-like neutralino sample with $m_{\tilde B}=300$ GeV and $m_{\tilde q}=2.2$ TeV. 
    The colours indicate the signal classes: combined signal (red), 0 $\tilde q$ (blue), 1 $\tilde q$ (orange) and 2 $\tilde q$ (green).
}
    \label{fig:cross-bino}
\end{figure}

We finally evaluate our six models on a sample with Bino-like neutralino, with $m_{\tilde B}=300$ GeV and $m_{\tilde q}=2.2$ TeV in Fig.\ \ref{fig:cross-bino}. 
Surprisingly, all six models perform quite well, reporting AUC $>$ 0.73 for 0 $\tilde q$ and AUC $>$ 0.85 for other classes, including the combined signal. 
The models trained on Wino-like 
neutralino samples perform better for 0 $\tilde 
q$ and 1 $\tilde q$ signal classes when evaluated 
on Binos than Winos. This suggests that Binos are easier to classify than Winos.
Moreover, the classification performance of Binos appears to 
be resilient to the change of the \pt cut, 
suggesting that the soft activity is not very 
important in this case. 
This is likely due to the fact that the 0 $\tilde q$ signal class is practically absent in the Bino-like sample.

\section{Projected sensitivies}\label{sec:limits}
\subsection{Wino-like LSP}

In this subsection, we estimate the projected sensitivity on the ($m_{\tilde W}$, $m_{\tilde q})$ plane that our GNN analysis may achieve at the end of LHC Run-3 or HL-LHC.
When simulating the $pp \to \tilde q \tilde q$ process and calculating its cross section, we fix the gluino mass at 10 TeV, around which the process still has a weak dependency on the gluino mass. 
The sensitivities and the uncertainty bands are derived  
{with GNN models trained on the Wino samples with ($m_{\tilde W}$, $m_{\tilde q}) = (300\,{\rm GeV},\,2.2\,{\rm TeV})$. 
Throughout this section, we include the contribution from the second leading background process $W + {\rm jets}$, followed by $W \to \tau \nu$.}
We estimated that this process increases the background by a factor of 1.3.

\begin{figure}[t!]
    \centering
    \includegraphics[width=0.7\linewidth]{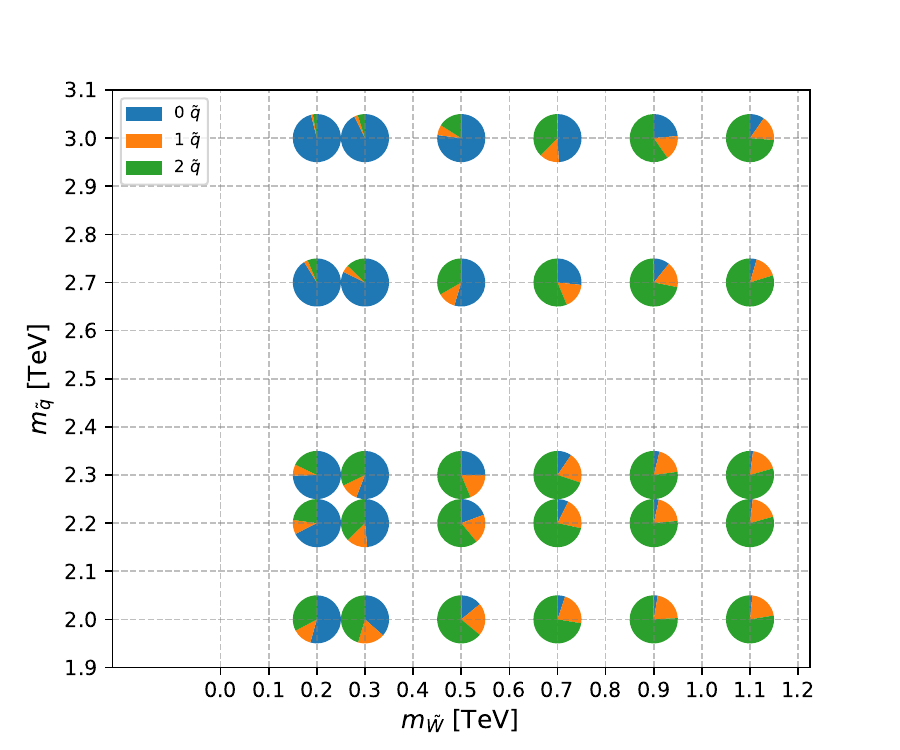}
    \caption{\small The composition of signal samples at mass-scan points in the Wino mass vs squark mass plan. 
    The colours correspond to the signal classes: 0 $\tilde q$ (blue),  1 $\tilde q$ (orange) and 2 $\tilde q$ (green).}
    \label{fig:wino-scan-composition}
\end{figure}

The grid of 30 mass points on the ($m_{\tilde W}$, $m_{\tilde q}$) plane is shown in Fig.\ \ref{fig:wino-scan-composition}, where the composition of the signal (after the preselection) is depicted with pie charts. 
The colours represent different signal classes: 0 $\tilde q$ (blue), 1 $\tilde q$ (orange) and 2 $\tilde q$ (green).
As expected, one can see that the signal is dominated by the 0 $\tilde q$ process for light Winos and heavy squarks, while in the opposite case, i.e.\ for heavy Winos and light squarks, the 2 $\tilde q$ process dominates the signal.
The Wino-squark associated production, 1 $\tilde q$, is never the largest process but is still important. 
Since the classification performance for the 1 $\tilde q$ and 2 $\tilde q$ processes is much better than for the 0 $\tilde q$ process, the signal composition has a large impact on the final result.

\begin{figure}[t!]
    \centering
     \includegraphics[width=0.49\linewidth]{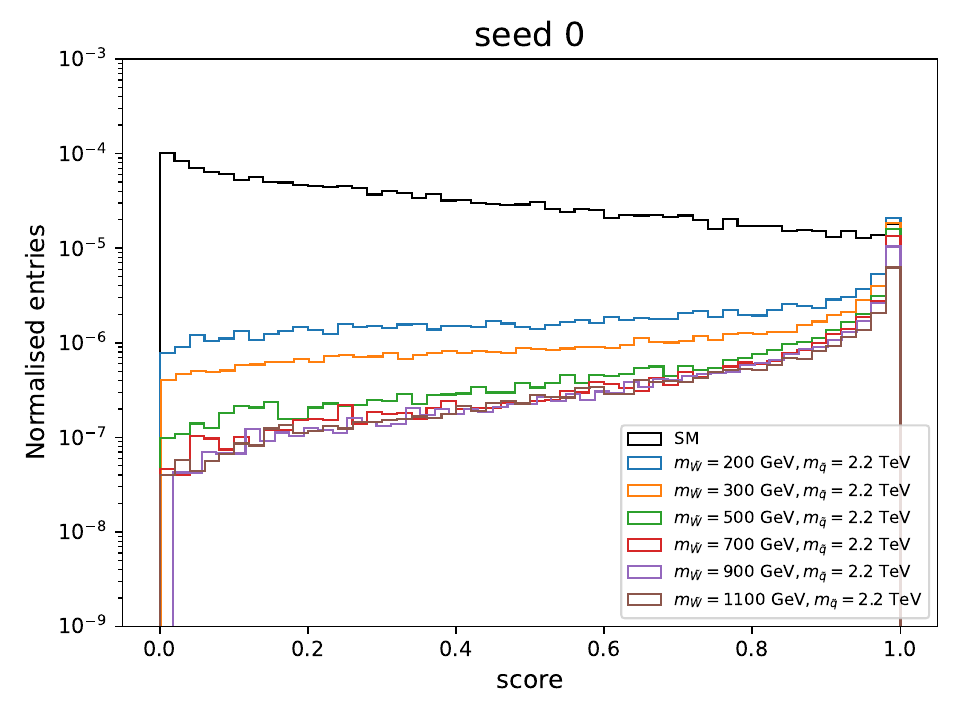}
    \includegraphics[width=0.49\linewidth]{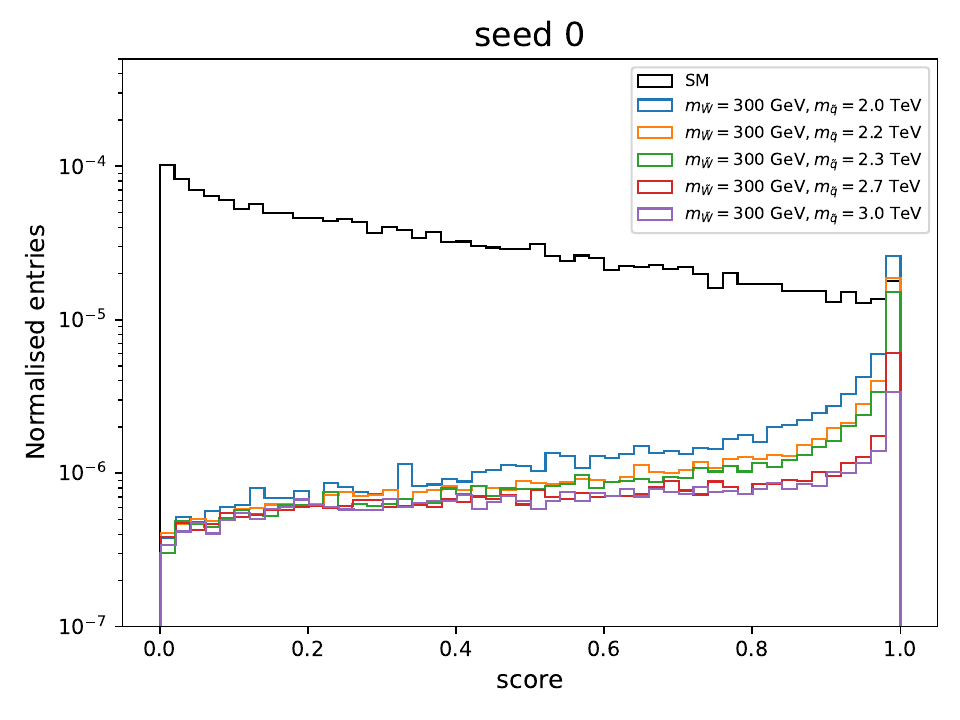}
    \caption{\small 
    The distributions of the NN score for different Wino masses.
    In the left (right) panel, the distributions are normalised to one (the cross section).
    }
    \label{fig:wino-scan-score}
\end{figure}

Fig.\ \ref{fig:wino-scan-score} displays the score distributions for different Wino and squark masses.  
In the left panel, the Wino mass is varied from 200 to 1100 GeV, keeping the squark mass fixed at 2.2 TeV.
As the Wino mass increases, the rate of the 0 $\tilde q$ process decreases, while that of 2 $\tilde q$ does not change.
Since the 0 $\tilde q$ process dominates the signal in a wide score range between 0 and 0.8 (see Fig.\ \ref{fig:wino-score}), the sore distributions are affected mainly in this region.
The right-hand panel presents the analogous plot with Wino mass fixed to 300 GeV and squark mass varied between 2 and 3 TeV. 
An increase of the squark mass results in a decrease of the 2 $\tilde q$ and 1 $\tilde q$ cross sections, while the event rate for the 0 $\tilde q$ is much less sensitive to the squark mass.
Since the 2 $\tilde q$ events dominate the signal in the high score region $s > 0.8$, we can understand the decline of the $s \simeq 1$ peak when the squark mass is varied from 2.2 to 3.0 TeV.

\begin{figure}[t!]
    \centering
     \includegraphics[width=0.49\linewidth]{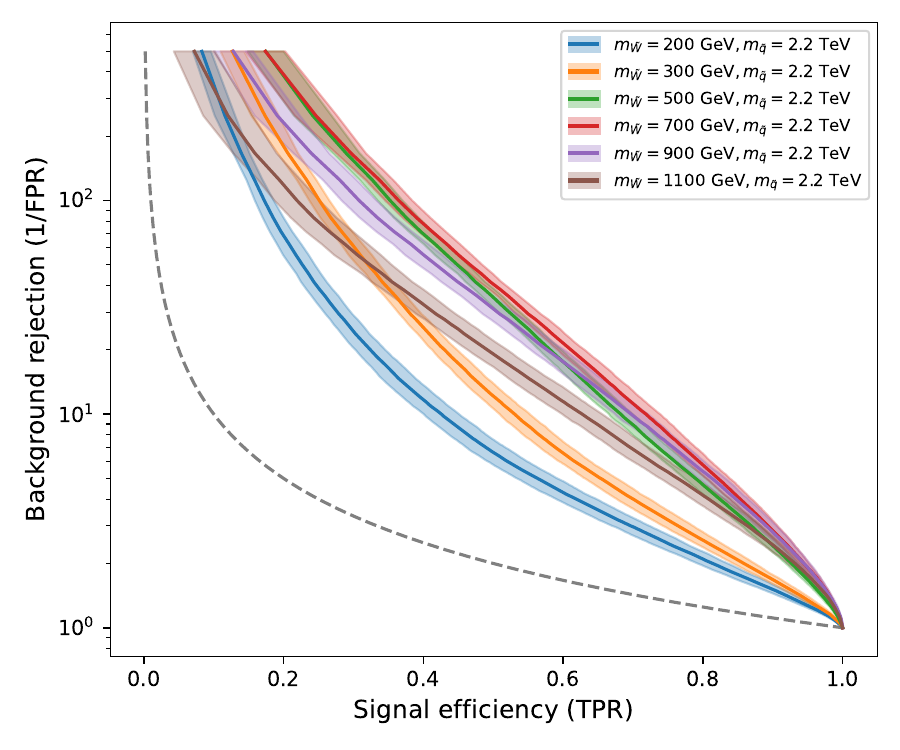}
    \includegraphics[width=0.49\linewidth]{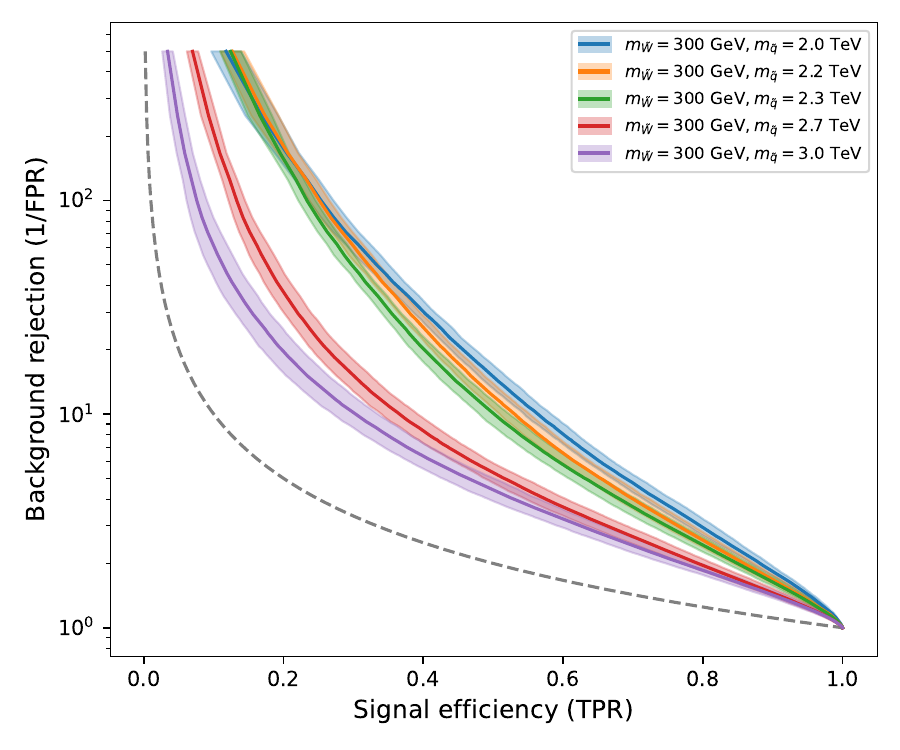}
    \caption{\small 
    The ROC curves for the Wino-like scenario.
    In the left (right) panel, the Wino (squark) mass is varied.
    }
    \label{fig:wino-scan-roc}
\end{figure}
The effects of changing sparticle masses on the ROC curves are displayed in Fig.\ \ref{fig:wino-scan-roc}. 
In the left (right) plot, the Wino (squark) mass is varied while keeping the squark (Wino) mass fixed. 
We observe in the left panel that the increase of the Wino mass results in better discrimination, as long as it is not too heavy.
The better performance is due to the fact that the relative portion of 2 $\tilde q$ events in the signal sample effectively increases because the 0 $\tilde q$ cross section decreases as the Wino mass increases. 
The performance improves in this case because the network can discriminate the signal from the background much more easily for 2 $\tilde q$ events than 0 $\tilde q$ ones.  
For the Wino mass larger than $700~\rm{GeV}$, the 0 $\tilde q$ process is already insignificant, and the classification performance starts to drop, probably due to the large difference between Wino masses used in the training and evaluation.
On the right panel, we see that the classification improves as the squark mass decreases.
This is also due to the fact that the smaller squark masses result in larger compositions of the 2 $\tilde q$ process in the sample. 
These results suggest that the classifier trained on $m_{\tilde W}=300~\mathrm{GeV}$ and $m_{\tilde q}=2.2~\mathrm{TeV}$ sample can be used for other mass points.

\begin{figure}[t!]
    \centering
     \includegraphics[width=0.49\linewidth]{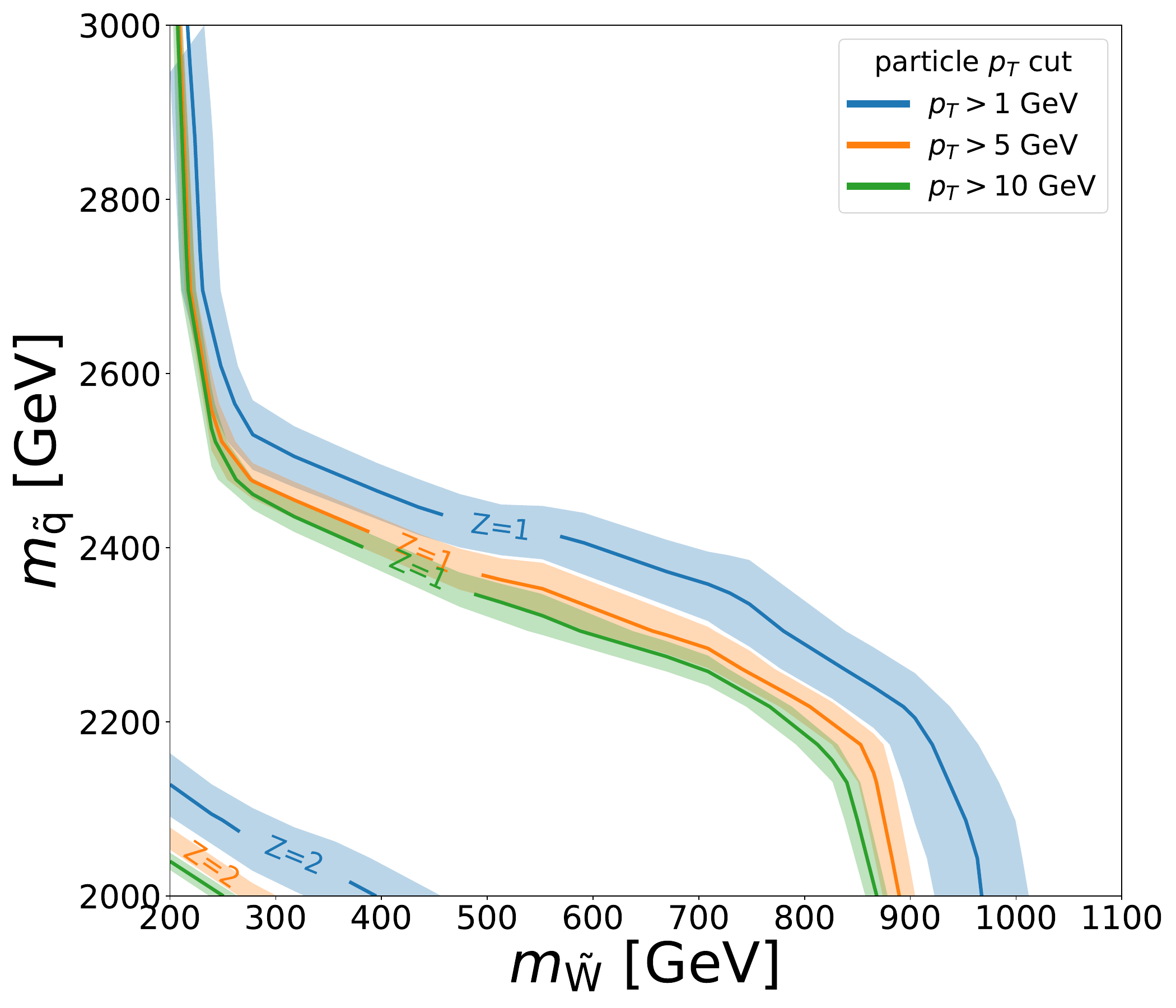}
    \includegraphics[width=0.49\linewidth]{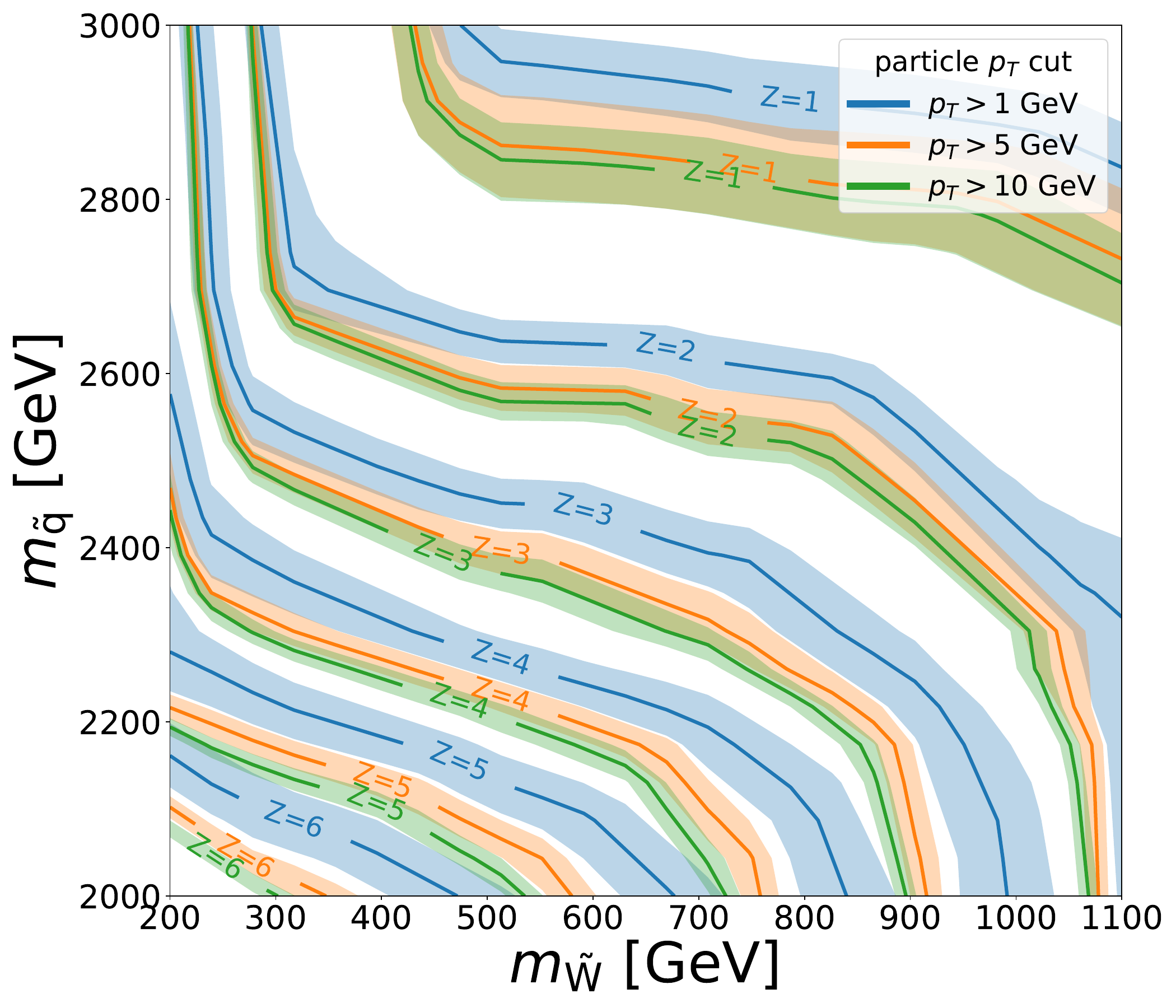}
    \caption{\small 
    The projected sensitivities in the Wino vs squark mass plane for $L=300~\mathrm{fb}^{-1}$ (left) and  $L=3000~\mathrm{fb}^{-1}$ (right). 
    The contours of various naive significance values are shown, with the bands representing the uncertainties.
    The colours correspond to the soft \pt cut: $p_\mathrm{T} > 1$ GeV (blue),  $p_\mathrm{T} > 5$ GeV (orange) and $p_\mathrm{T} > 10$ GeV (green).
    }
    \label{fig:wino-limit}
\end{figure}

Finally, in Fig.\ \ref{fig:wino-limit}, we show the contours of the naive significance $Z=S/\sqrt{S+B}$ on the ($m_{\tilde W}$, $m_{\tilde q}$) mass plane. 
In the left (right) plot, the integrated luminosity of \lhc (\hllhc) is assumed. 
For each value of $Z$, three contours are shown for different \pt cuts on soft particles in the low-level data; 
the blue, orange and green colours correspond to the requirement on the particle flow objects with $p_\mathrm{T} > 1$, 5 and 10 GeV, respectively.
The widths of the curves represent the fluctuation observed in the ten different NN models obtained from different training samples. 
Naively, the contour of $Z = n$ corresponds to the $n$-$\sigma$ sensitivity expected at the given luminosity. 
In the left panel, we see that our GNN analysis is insensitive to the considered Wino-like scenario beyond the 2-$\sigma$ level. 
We also see that the maximum masses that can be probed get smaller for stronger \pt cuts while the uncertainty shrinks.  
When squark is relatively light, i.e.\ $m_{\tilde q} \simeq 2$ TeV, the maximal mass of Wino that can be probed with $Z=2$ is 400 (290) [250] GeV for $p_\mathrm{T}>1$, 5 and 10 GeV, respectively. 
The sensitivities at the HL-LHC shown in the right panel are more promising.
It suggests that the Wino mass below 1100 GeV can be probed at 2-$\sigma$ level if the squark mass is less than 2.4 TeV.
When squarks are heavy enough, their production cross section diminishes, and the signal sample is composed almost exclusively of Wino-pair production processes. Hence, the limit becomes insensitive to further increase of $m_{\tilde q}$.
For light squark mass around 2 TeV, the Wino mass can be probed up to 680 GeV with $Z=5$, enabling a statistically significant discovery of Supersymmetry.

\subsection{Higgsino-like LSP}

\begin{figure}[t!]
    \centering
    \includegraphics[width=0.7\linewidth]{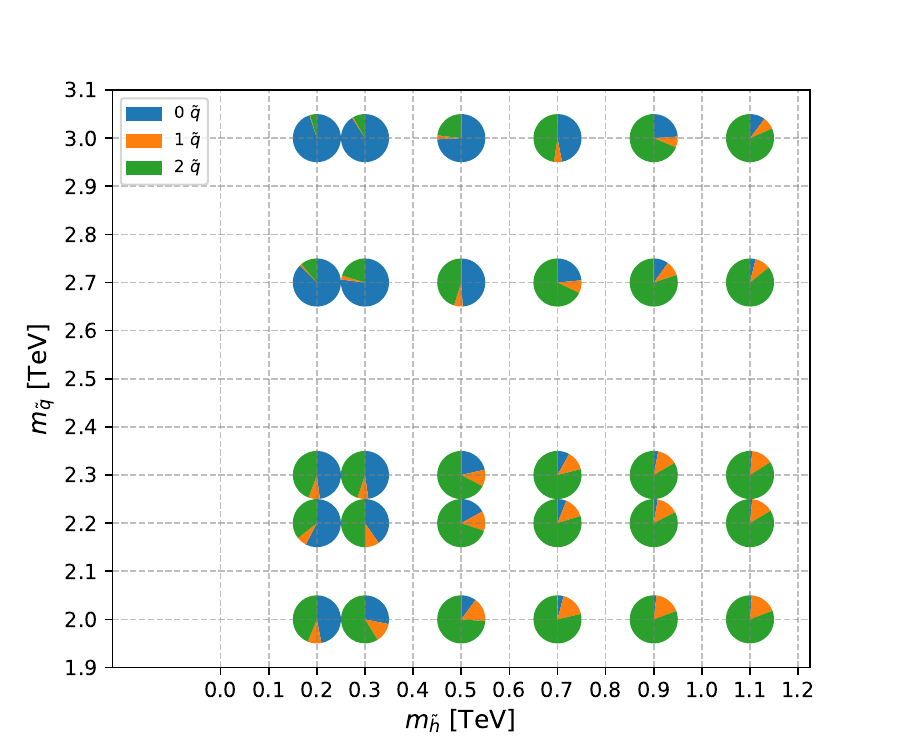}
    \caption{\small The composition of signal samples at mass-scan points in the Higgsino mass vs squark mass plan. 
    The colours correspond to the signal classes: 0 $\tilde q$ (blue),  1 $\tilde q$ (orange) and 2 $\tilde q$ (green).}
    \label{fig:higgsino-scan-composition}
\end{figure}

This subsection discusses the projected sensitivities of the LHC Run-3 and HL-LHC to the Higgsino-like scenario.
The sensitivities are derived using GNN models trained on the Higgsino samples with $(m_{\tilde h}, m_{\tilde q}) = (300\,{\rm GeV},\,2.2\,{\rm TeV})$.
Fig.\ \ref{fig:higgsino-scan-composition} displays the signal sample composition at various mass points. 
As can be seen, when squarks are heavy, the Higgsino pair production process, 0 $\tilde q$, dominates the signal sample, while for heavy Higgsinos and relatively light squarks, the squark pair production, 2 $\tilde q$, occupies most of the signal sample.
As before, the composition of the samples is a crucial factor affecting the final sensitivities.

\begin{figure}[t!]
    \centering
     \includegraphics[width=0.49\linewidth]{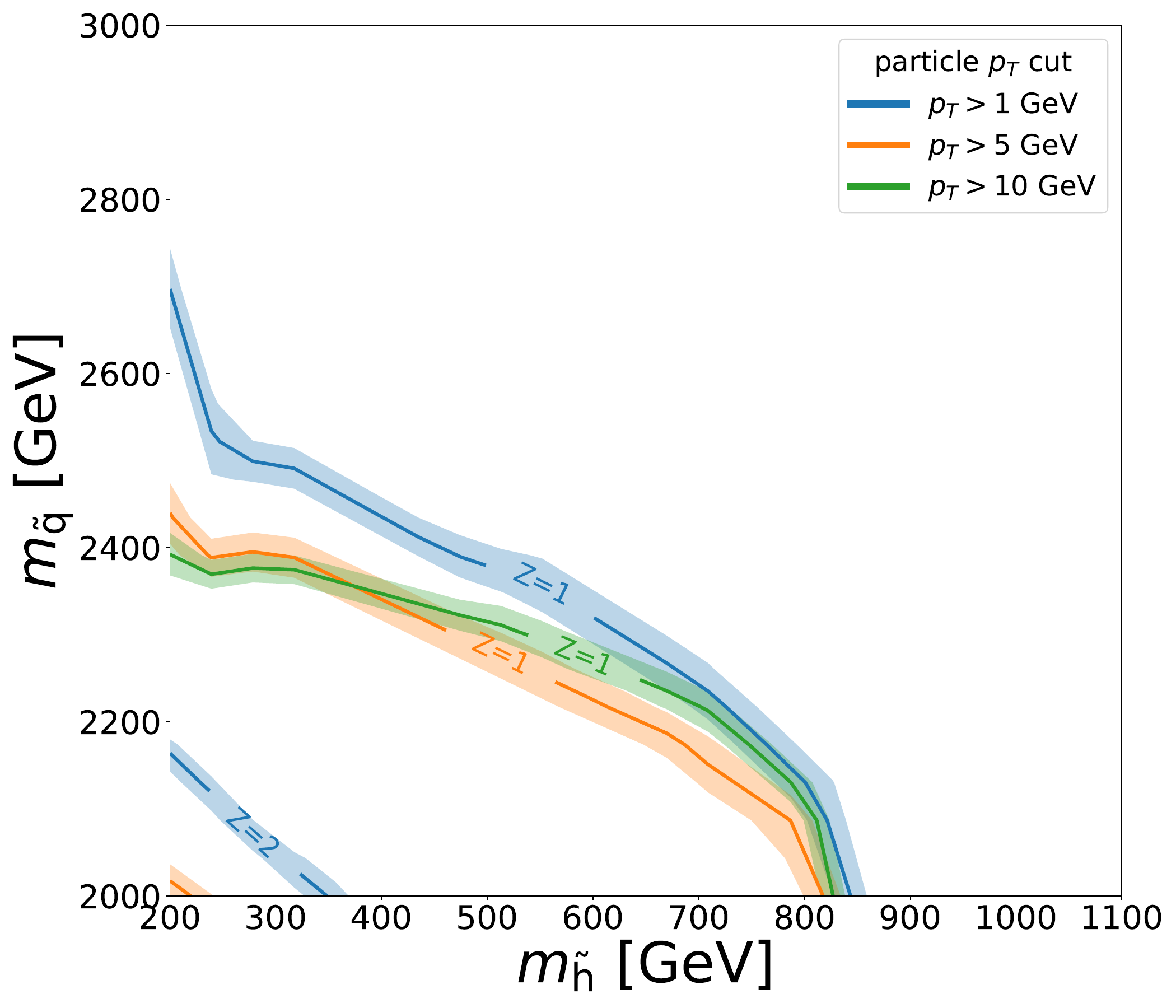}
    \includegraphics[width=0.49\linewidth]{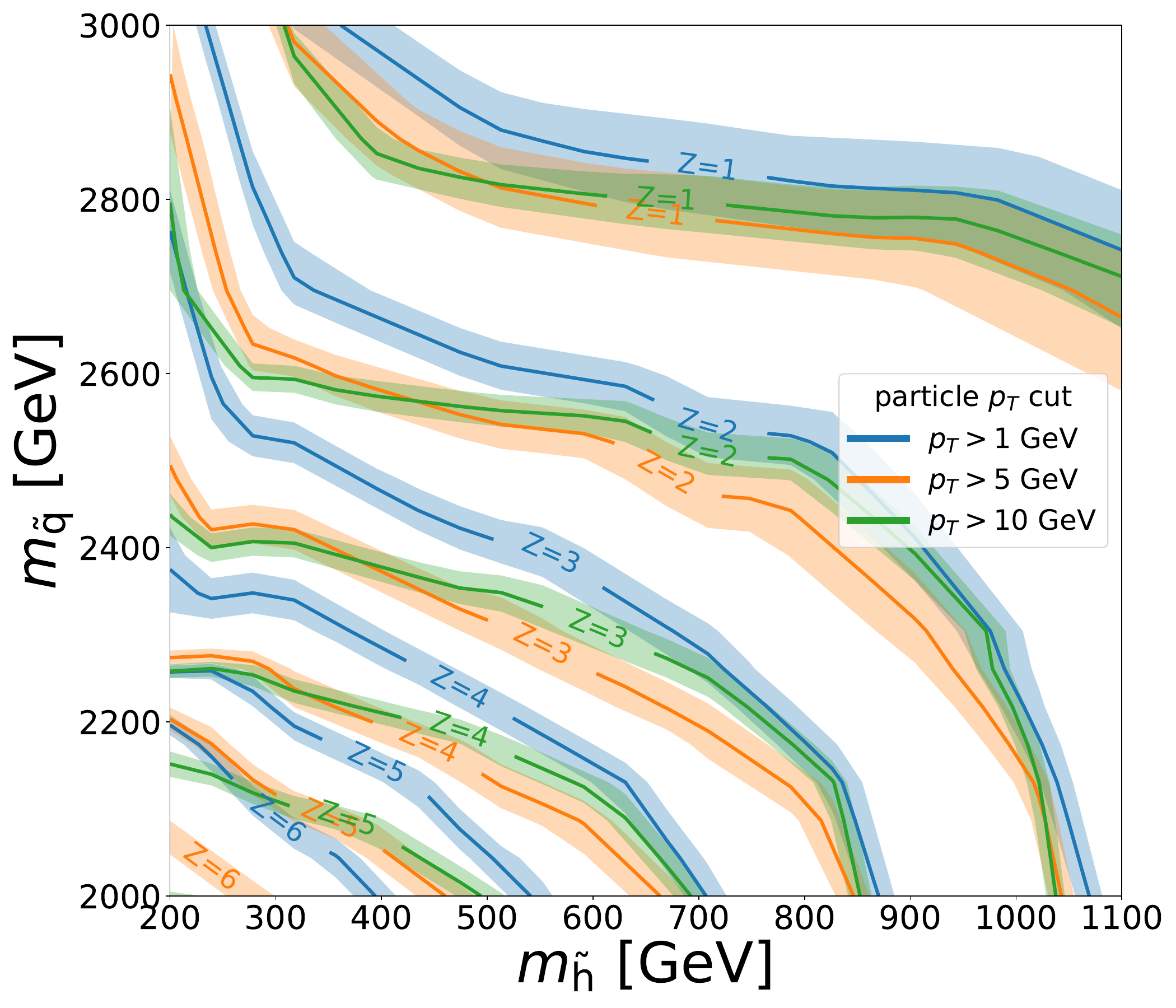}
    \caption{\small 
    The projected sensitivities in the Higgsino vs squark mass plane for $L=300~\mathrm{fb}^{-1}$ (left) and  $L=3000~\mathrm{fb}^{-1}$ (right). 
    The contours of various naive significance values are shown, with the bands representing the uncertainties.
    The colours correspond to the soft \pt cut: $p_\mathrm{T} > 1$ GeV (blue),  $p_\mathrm{T} > 5$ GeV (orange) and $p_\mathrm{T} > 10$ GeV (green).
    }
    \label{fig:higgsino-limit}
\end{figure}

Fig.\ \ref{fig:higgsino-limit} shows the contours of the naive significance $Z$ in the ($m_{\tilde h}$, $m_{\tilde q}$) plane.
The left (right) plot assumes $L = 300$ (3000) fb$^{-1}$ corresponding to the LHC Run-3 and HL-LHC, respectively. 
As in the previous subsection, the blue, orange and green contours represent the results with the $\ppt$ thresholds of 1, 5 and 10 GeV on particle flow objects in the low-level variables, respectively.
The bands of the contours represent the fluctuation observed in our ensemble of NN models.

Comparing Fig.\ \ref{fig:higgsino-limit} with the results for Winos in Fig.\ \ref{fig:wino-limit}, we observe that in the former figure, the uncertainties are smaller and the limits are a little weaker. 
In Fig.\ \ref{fig:rocpt-higgsino}, we observed that
the soft \pt cut has a very little effect on the classification performance in the training mass point, $(m_{\tilde h}, m_{\tilde q}) = (300\,{\rm GeV},\,2.2\,{\rm TeV})$.
{In Fig.\ \ref{fig:higgsino-limit}, we see that this is no longer true if the network is used away from the training point. 
In particular, if the network is used at higher Higgsino mass points, soft particle information around $p_T \sim 5-10$ GeV negatively contributes to the classification.
This is visible in both plots in Fig.\ \ref{fig:higgsino-limit}, where we observe that the projected sensitivity with a soft particle cut of $p_T > 5$ GeV is worse than that with a $p_T > 10$ GeV cut in large Higgsino mass regions.
{This non-trivial correlation between the soft activity around $5-10$ GeV and the squark and Higgsino masses indicates that some of those soft particles originate from the hard process. 
}
As this behaviour is not observed in the Wino case, it implies that the network uses the soft particle information differently between the Wino and Higgsino scenarios, which will be partially confirmed in the next section (Fig.\ \ref{fig:interpretation-corr-low}).
However, the manner in which the soft particle information is utilised within the network is highly complex, and it is difficult to pinpoint the exact cause of this behaviour.
}



The left panel in Fig.\ \ref{fig:higgsino-limit} depicts naive significance contours for \lhc. The studied model can be observed at Run-3 with $\rm Z=2$ significance, allowing to probe neutralino masses up to 350 GeV with 2.0 TeV squarks, or 2.18 TeV squarks if Higgsino is 200 GeV. 
The right plot in Fig.\ \ref{fig:higgsino-limit} shows naive significance contours for \hllhc. 
Generally, the limits that can be derived from this plot are slightly weaker than in the Wino case. 
Higgsino mass can be probed up to 1080 GeV with $Z=2$. If we require $Z=5$ and adopt $p_{\rm T} > 1~{\rm GeV}$ cut, then we can constrain neutralino mass up to 550 GeV and squark mass up to 2250 GeV.

\subsection{Bino-like LSP}\label{sec:bino-scan}

The cross evaluation in Fig.\ \ref{fig:cross-bino} revealed that models trained on Winos or Higgsinos could be effectively used for the classification of Bino-like neutralinos. 
Therefore, in this section, we present the results of a mass scan for Binos. 
Unlike the previous two sections, we do not train a separate ensemble on Bino-like neutralino, but we reuse the ensemble trained on Winos.

\begin{figure}[ht!]
    \centering
    \includegraphics[width=0.7\linewidth]{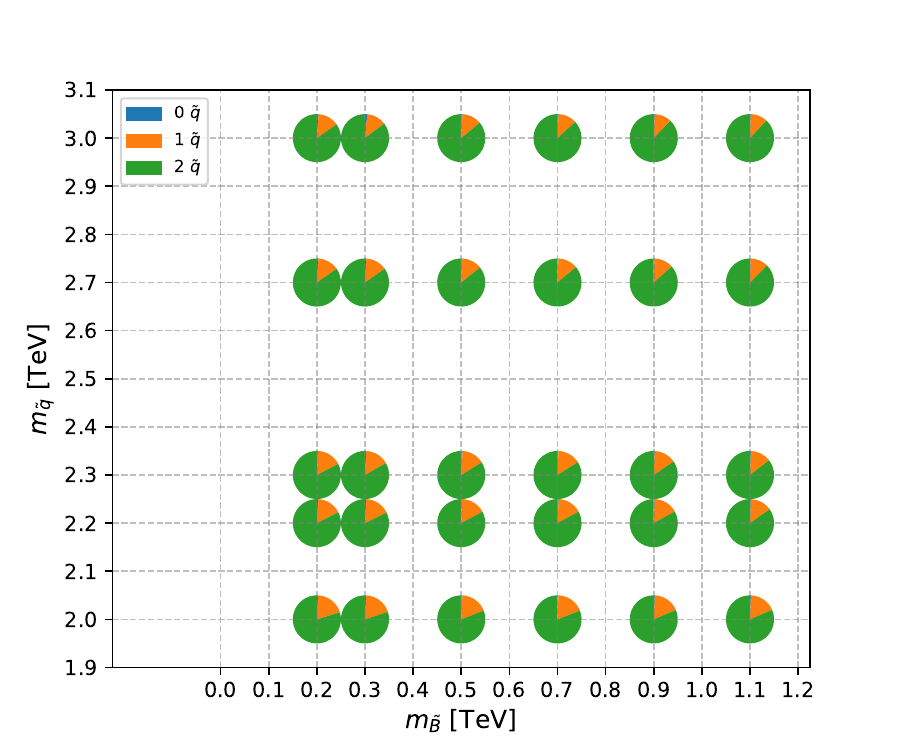}
    \caption{\small 
    The composition of signal samples at mass-scan points in the Bino mass vs squark mass plan. 
    The colours correspond to the signal classes: 0 $\tilde q$ (blue),  1 $\tilde q$ (orange) and 2 $\tilde q$ (green).}
    \label{fig:bino-scan-composition}
\end{figure}

In Fig.\ \ref{fig:bino-scan-composition}, we show the sample composition for different masses of Binos and squarks. One immediate observation is that the 0 $\tilde q$ class of events is almost nonexistent because the particle considered is almost pure Bino, and the pure Bino does not couple to gauge bosons and cannot be produced via $s$-channel. 
The dominant production of Bino-like neutralinos is the production of two squarks that decay to quarks and neutralinos, which explains why ML models trained on Winos and Higgsinos are so good at classifying Binos. 
The Bino-squark associated production constitutes about 12-20\% of produced events.

\begin{figure}[ht!]
    \centering
     \includegraphics[width=0.49\linewidth]{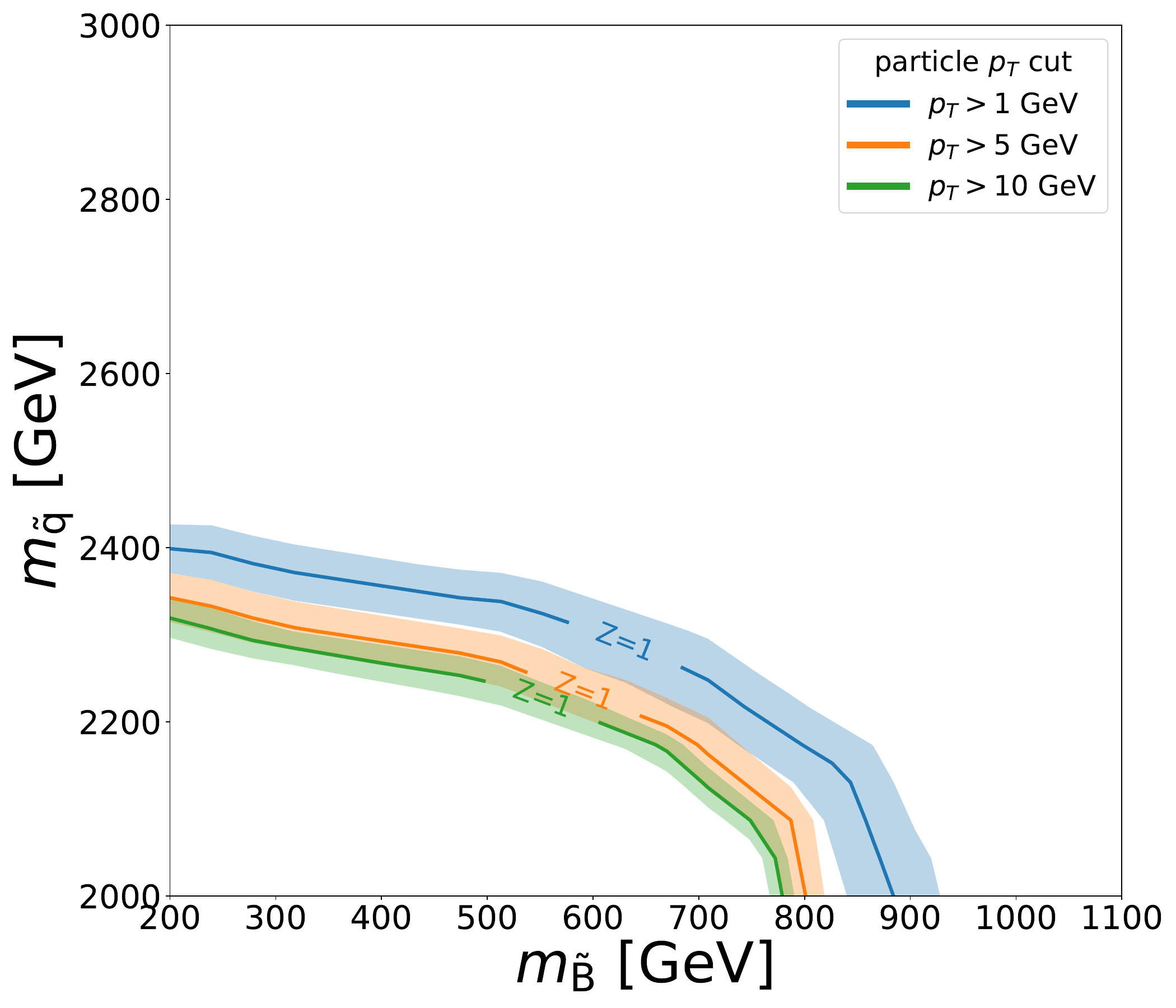}
    \includegraphics[width=0.49\linewidth]{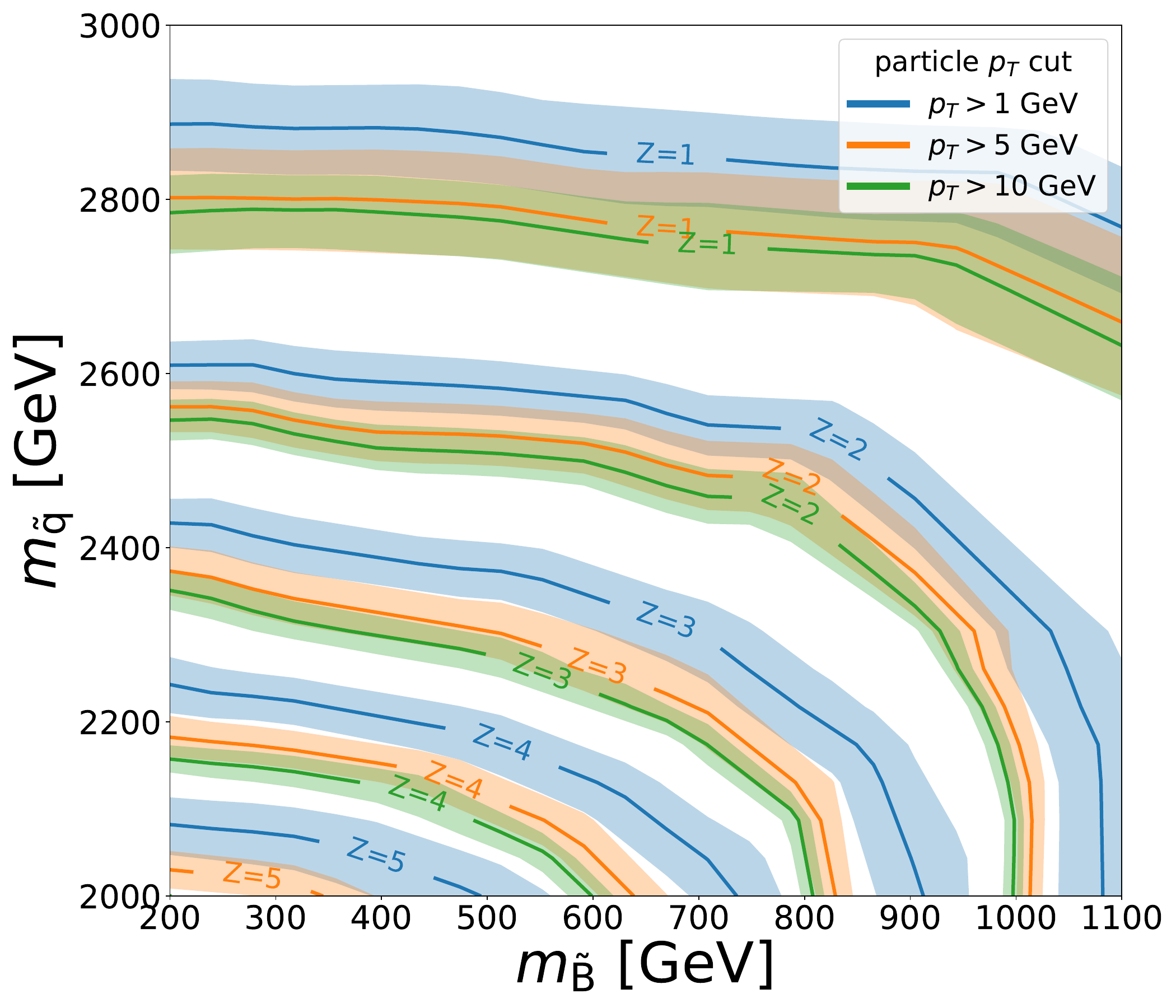}
    \caption{\small 
    The projected sensitivities in the Bino vs squark mass plane for $L=300~\mathrm{fb}^{-1}$ (left) and  $L=3000~\mathrm{fb}^{-1}$ (right). 
    The contours of various naive significance values are shown, with the bands representing the uncertainties.
    The colours correspond to the soft \pt cut: $p_\mathrm{T} > 1$ GeV (blue),  $p_\mathrm{T} > 5$ GeV (orange) and $p_\mathrm{T} > 10$ GeV (green).    
    }
    \label{fig:bino-limit}
\end{figure}

In Fig.\ \ref{fig:bino-limit}, we present naive significance $Z$ contours in the Bino vs squark mass plane. 
The left panel is for \lhc, while the plot on the right-hand side is for \hllhc. 
The colouring scheme is the same as in Figs.\ \ref{fig:wino-limit} and \ref{fig:higgsino-limit}. 
The limits for Bino are a little weaker than for Winos and Higgsinos. 
In the \lhc case, there are only contours for $Z=1$. For $p_\mathrm{T}>1$ GeV cut, Bino masses up to 900 GeV can be tested, and squark mass can be probed up to 2.4 TeV. For \hllhc, one can test all Bino masses for the squark mass up to 2.6 TeV if we require $Z=2$. 
If we demand $Z=5$, we can test Binos with masses up to 500 GeV and squark masses up to 2100 GeV.
All limits exhibit strong dependence on $m_{\tilde q}$ and only a moderate dependence on $m_{\tilde B}$.
Similarly to Winos, a stronger cut on particles' \pt results in slightly weaker limits.

\section{Interpretation}\label{sec:interp}

In this section, we attempt to interpret the neural network, identify the relevant features that most affect the final result, and seek possible improvement directions.
We also check to what extent the soft particle information is used in the classification as they are less reliable due to mismodelling of soft activities and pileup effects.
We work with the $m_{\tilde \chi_1^0}=300~\mathrm{GeV}$ and $m_{\tilde q}=2.2~\mathrm{TeV}$ mass point.

\subsection{Input-output correlation}

The easiest and most natural method of estimating the importance of input features is to calculate the Pearson correlation coefficient between each of the input variables and the output score. Pearson correlation coefficients are defined as:
\begin{equation}
    r_{xs} = \frac{\sum^n_{i=1} (x_i - \bar x) (s_i - \bar s)}{\sqrt{\sum^n_{i=1} (x_i - \bar x)^2}\sqrt{\sum^n_{i=1} (s_i - \bar s)^2}},
\end{equation}
where $n$ is the sample size, $x_i$ and $s_i$ are individual sample points for high-level feature $x$ and score $s$,  $\bar x$ and $\bar s$ denote their means.
A large positive correlation indicates that a high value of a variable 
is likely to occur for signal events. In contrast, a large negative 
correlation marks a feature that is useful for recognising SM events. 
Small, in terms of the 
absolute value, correlation is inconclusive and might be just a result of overtraining or simply a fluctuation.

\begin{figure}[t!]
    \centering
     \includegraphics[width=0.49\linewidth]{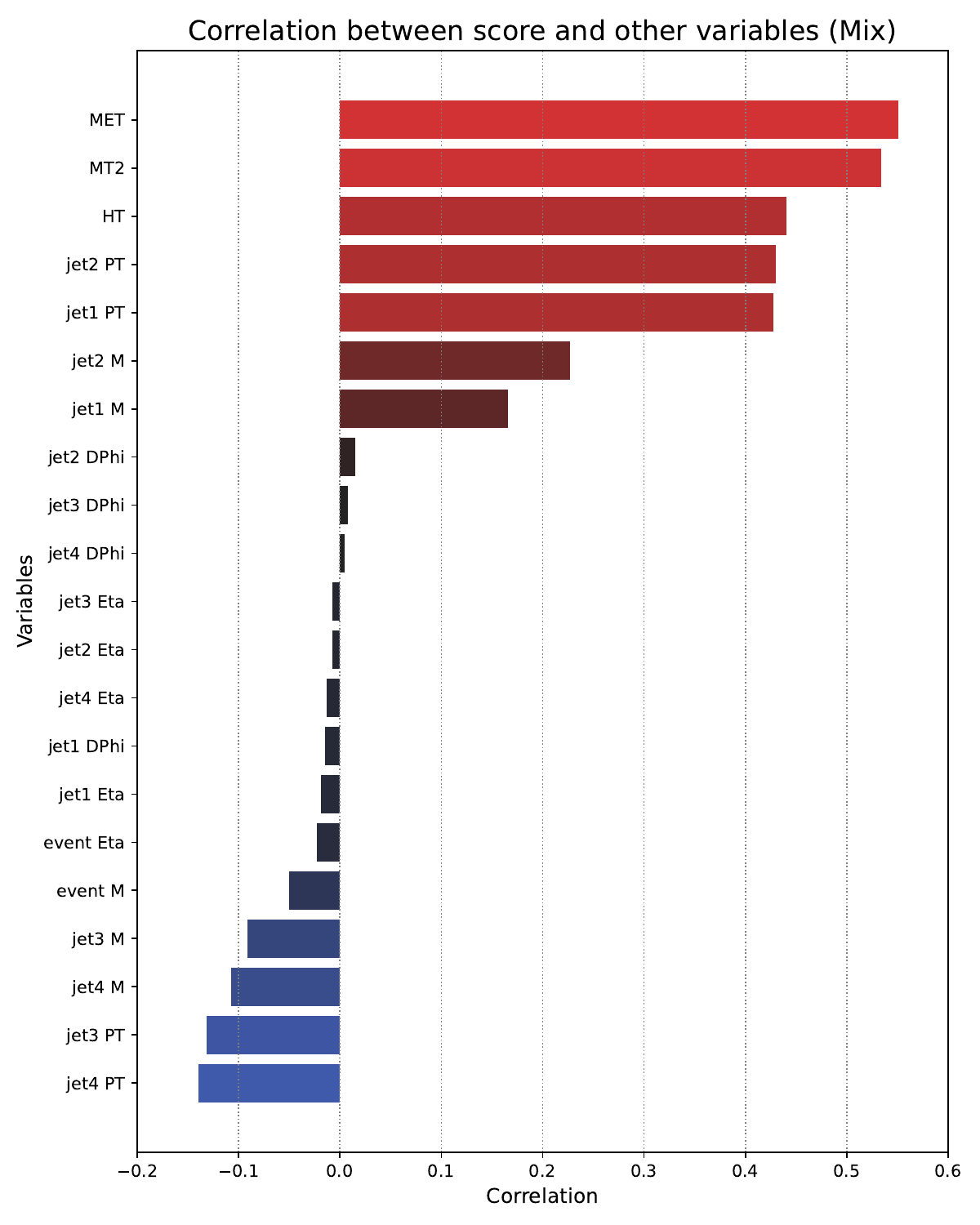}
    \includegraphics[width=0.49\linewidth]{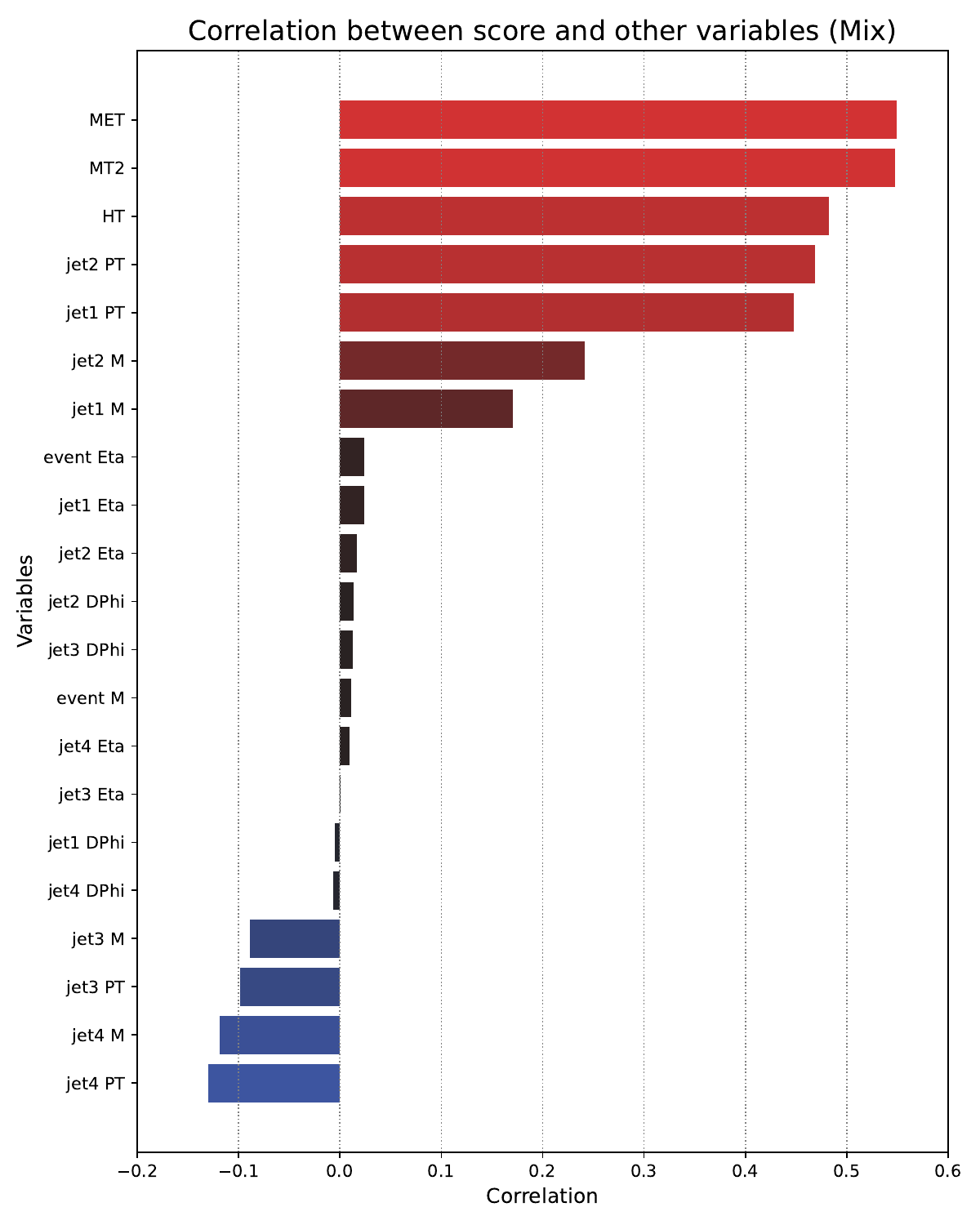}
    \caption{\small 
    The correlation between the output of NN model and high-level input variables for the Wino- (left) and Higgsino-like (right) samples.
    }
    \label{fig:interpretation-corr-high}
\end{figure}

The Pearson correlation coefficients are shown in Fig.\ \ref{fig:interpretation-corr-high} for 21 high-level inputs from Table \ref{tab:high-level-variables}. 
The left panel in Fig.\ \ref{fig:interpretation-corr-high} is for one of the networks trained on the Wino sample, and the plot on the right-hand side is for a model trained on Higgsinos. The correlation is calculated for a sample containing a balanced mixture of both SM background and signal.
The correlation plots in Fig.\ \ref{fig:interpretation-corr-high} are very similar, meaning that the same information from high-level variables is learned for both Higgsino and Wino samples. 
The most important features are $E_T^{\rm miss}$ and $m_{T2}$, which are strongly correlated ($r_{xs}>0.5$) with the signal label.
This is expected since $E_T^{\rm miss}$ is larger for BSM events due to a large mass of neutralinos. The $m_{T2}$ is known to be an effective discriminative variable. 
Large masses and momenta of the leading and second leading jets, as well as large $H_\mathrm{T}$, are characteristic for events with on-shell squarks because the jets from squark decays are highly boosted due to a large mass difference between squarks and neutralinos. 
When it comes to anticorrelated variables, the most important are the momenta and masses of the third and fourth jets. Since we use zero-padding, 
this strongly suggests that events from the SM background are typically characterised by a larger number of jets than BSM events. 
However, these variables are less powerful discriminators because a single parton, e.g.\ a quark from a decay of a squark, might result in more than one reconstructed jet in the detector.

\begin{figure}[t!]
    \centering
     \includegraphics[width=0.49\linewidth]{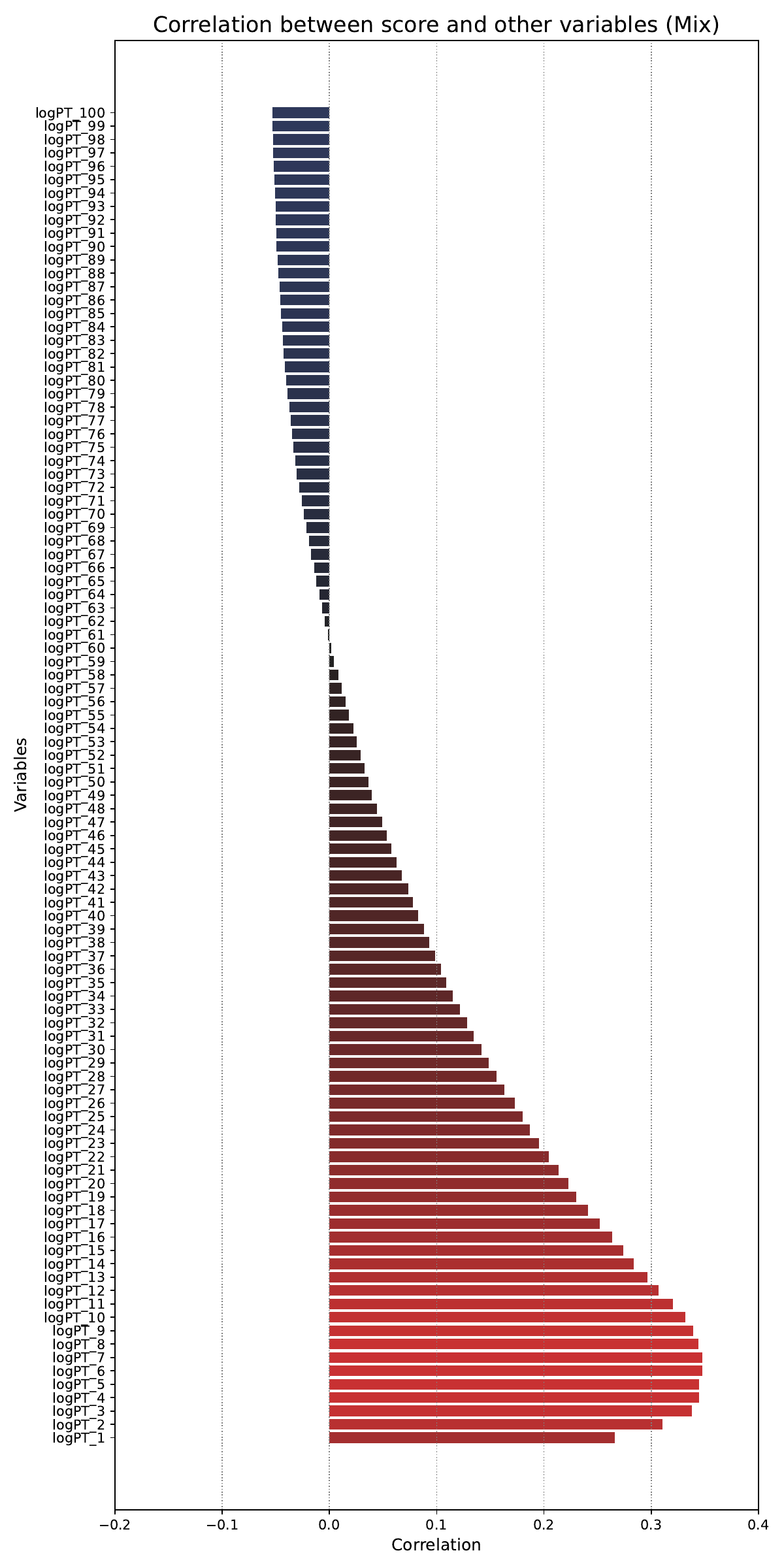}
    \includegraphics[width=0.49\linewidth]{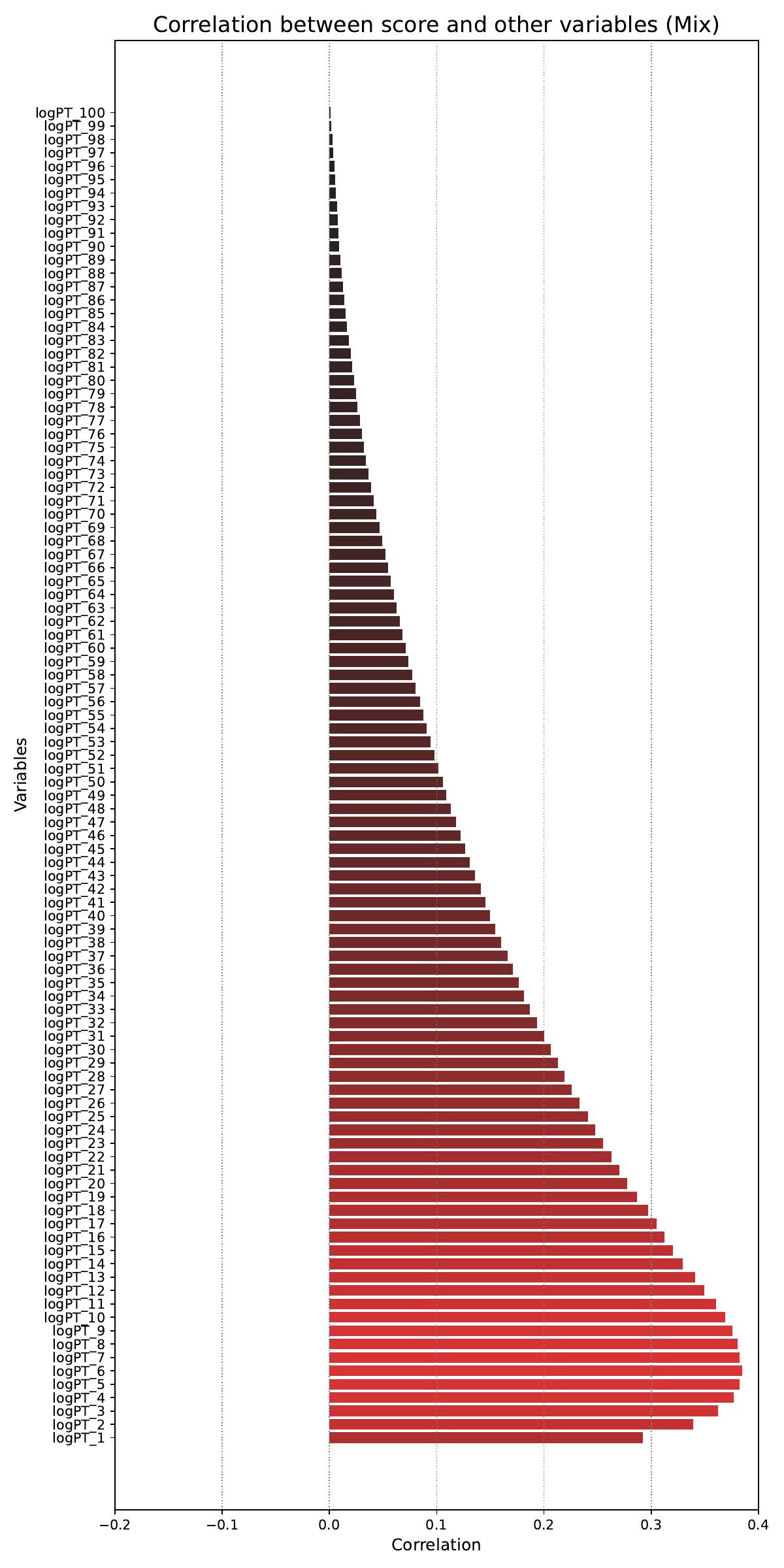}
    \caption{\small 
    The correlation between the output of NN model and particle $p_{\rm T}$s for the Wino- (left) and Higgsino-like (right) samples.
    }
    \label{fig:interpretation-corr-low}
\end{figure}

Fig.\ \ref{fig:interpretation-corr-low} depicts correlation coefficients between the NN output and \logpt of the first 100 particles (ordered by \pt from the largest transverse momentum). 
The left panel of Fig.\ \ref{fig:interpretation-corr-low} is for a sample with Wino signal, and the right panel is for a sample with Higgsinos. In both cases an equal amount of signal and background is mixed.
The comparison between the two plots reveals a clear difference. 
For Winos, $p_{\rm T}$s of the first 60 particles are correlated with the score, while particles further in the list are anti-correlated. 
For Higgsinos, $p_{\rm T}$s of all of the first 100 particles are positively correlated with the NN output.
In both cases, the largest correlations are for the first 10-15 particles, then drops.
This implies the soft particle information gives a subleading contribution to the classification. 
We therefore expect the classification result is not highly sensitive to the exact modelling of soft activities and pileup effects.
Clearly, $p_{\rm T}$s of the first few particles contribute to the \pt of the leading or second leading jet and are therefore related to highly boosted jets from squark decays. 
Anti-correlations visible in the Wino case indicate that the presence (we pad particles with zeros if there are less than 250 in an event) of more soft particles is probably related to the presence of ISR jets and background-origin, which would explain why large \pt cuts deteriorate the classification performance.
Interestingly, the behaviour of the Higgsino model is different.

We have investigated the cause of this difference. For the SM background events, the transverse momenta, $p_{\rm T}$, of the first 100 particles are always positively correlated for 
both Higgsino and Wino samples. In the case of $1~\tilde q$ and $2~\tilde q$ events, a strong positive correlation is observed 
for the hardest 40–60 particles in Wino samples and 50–70 
particles in Higgsino samples. Beyond this range, the softer 
particles are anticorrelated with the neural network score.
This effect is, however, weaker for Higgsino sample.
For signal events containing only ISR jets ($0~\tilde q$), the ~70 hardest particles exhibit a positive correlation with the NN 
score for both types of electroweakinos. However, the softer 
particles show differing behavior: they are positively 
correlated in the case of Higgsinos, while slightly negatively 
correlated for Winos. All these effects combined result in the difference between the two panels in Fig. \ref{fig:interpretation-corr-low}.

\subsection{Importance of high-level inputs}

Another well-established method of investigating the importance of features is checking the robustness of the model prediction against random shuffling of input data. The procedure is simple. 
A model trained on some set of features is evaluated on another data set in which one of the features has been randomly shuffled, resulting in a loss of information relating to that feature. 
The procedure is repeated for all features, and the results are compared with the unshuffled case.

\begin{figure}[p]
\thispagestyle{empty}
    \centering
    \begin{subfigure}[b]{0.38\textwidth}
        \centering
        \includegraphics[width=\textwidth]{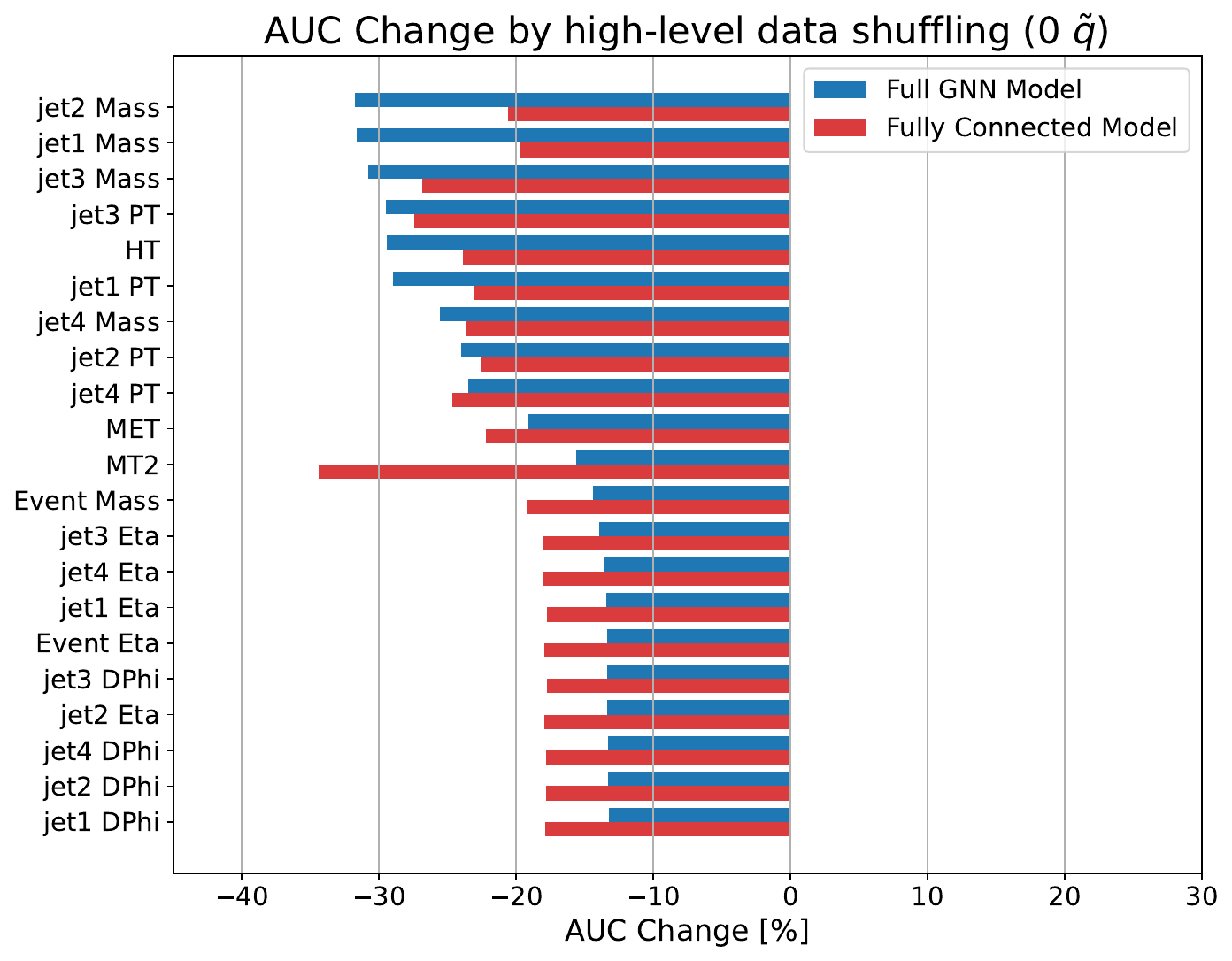}
        \caption{\footnotesize Wino pair production.}
        \label{fig:wino0}
    \end{subfigure}
    \hfill
    \begin{subfigure}[b]{0.38\textwidth}
        \centering
        \includegraphics[width=\textwidth]{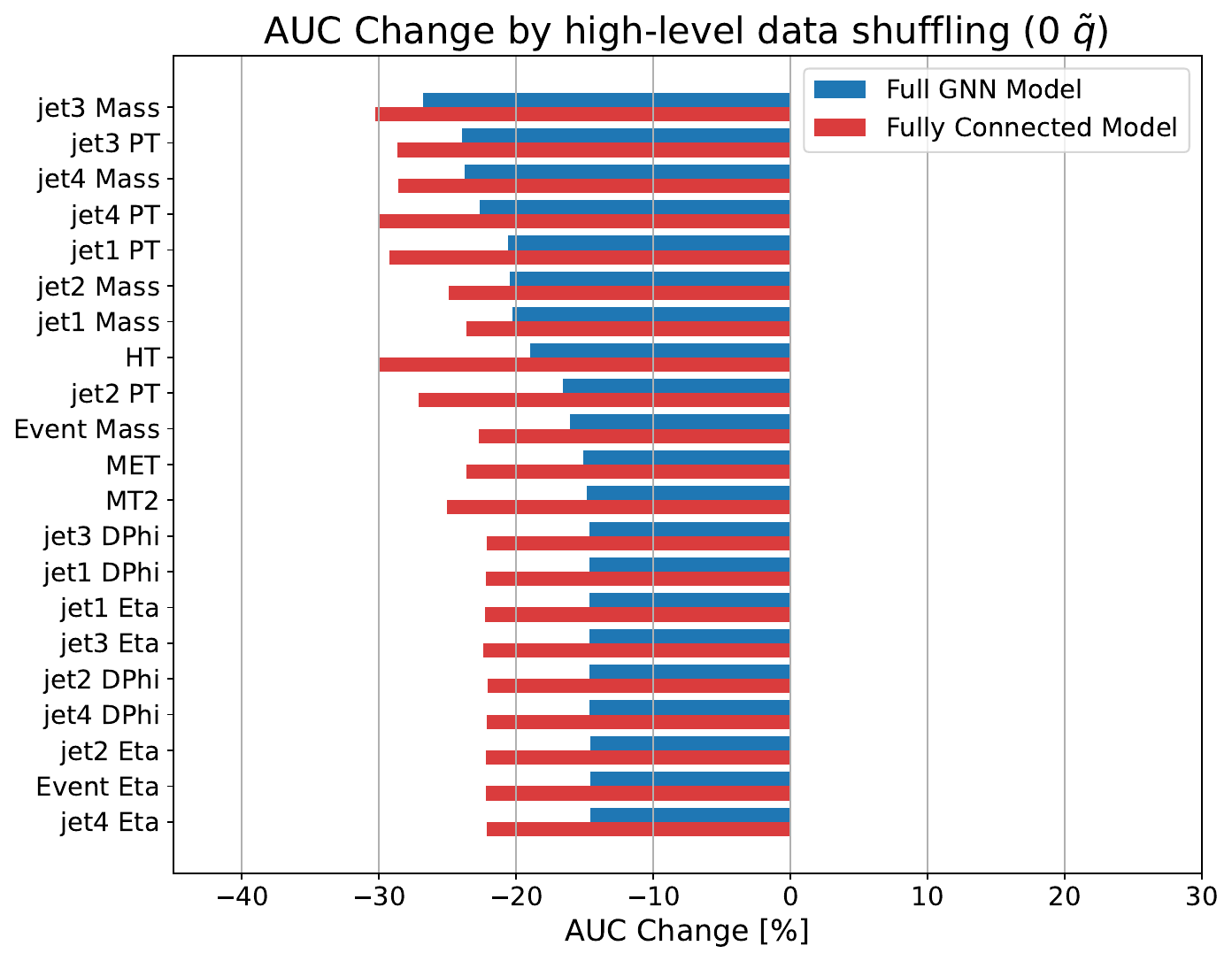}
        \caption{\footnotesize Higgsino pair production.}
        \label{fig:higgsino0}
    \end{subfigure}
    \vfill
    \begin{subfigure}[b]{0.38\textwidth}
        \centering
        \includegraphics[width=\textwidth]{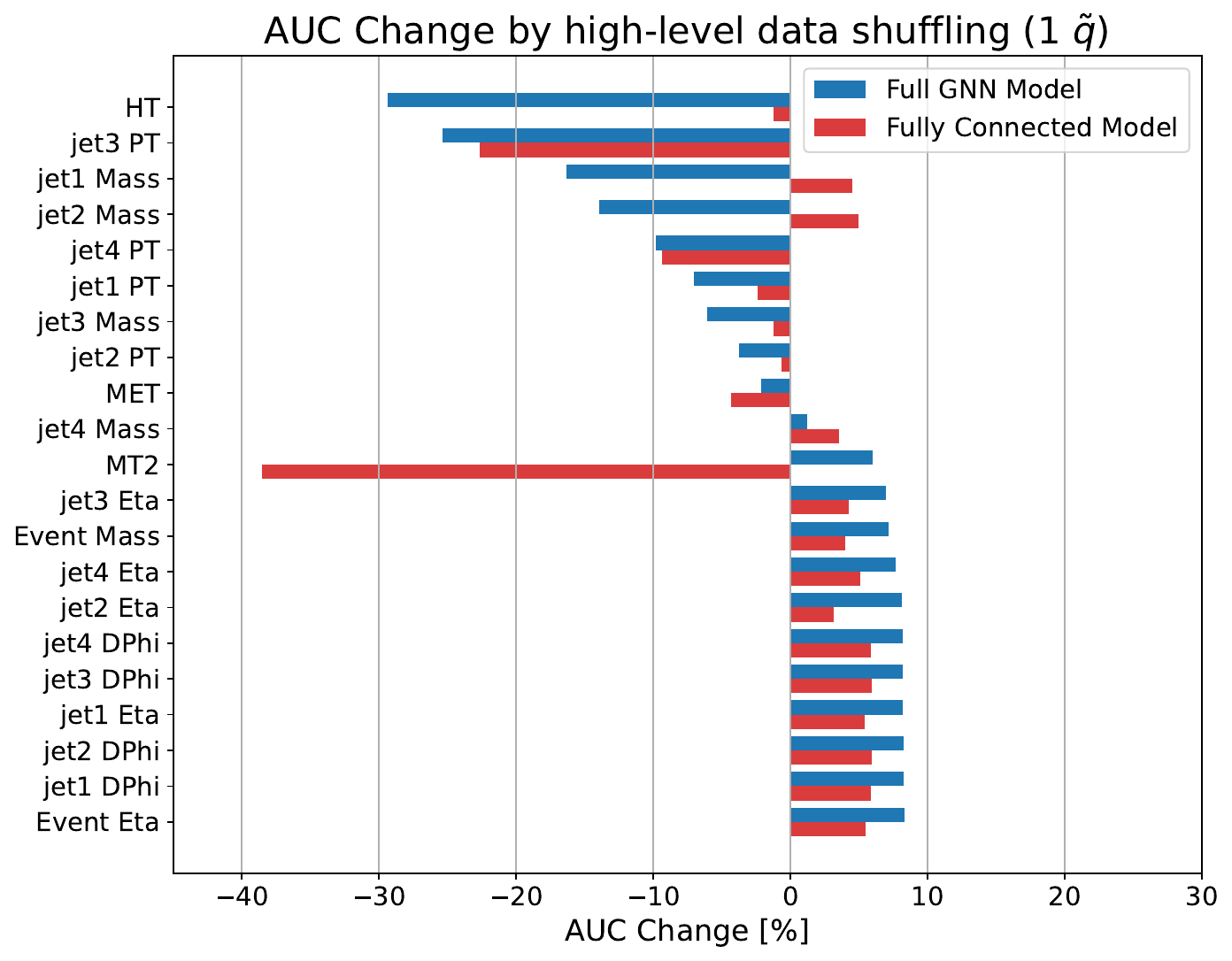}
        \caption{\footnotesize Wino-squark production.}
        \label{fig:wino1}
    \end{subfigure}
    \hfill
    \begin{subfigure}[b]{0.38\textwidth}
        \centering
        \includegraphics[width=\textwidth]{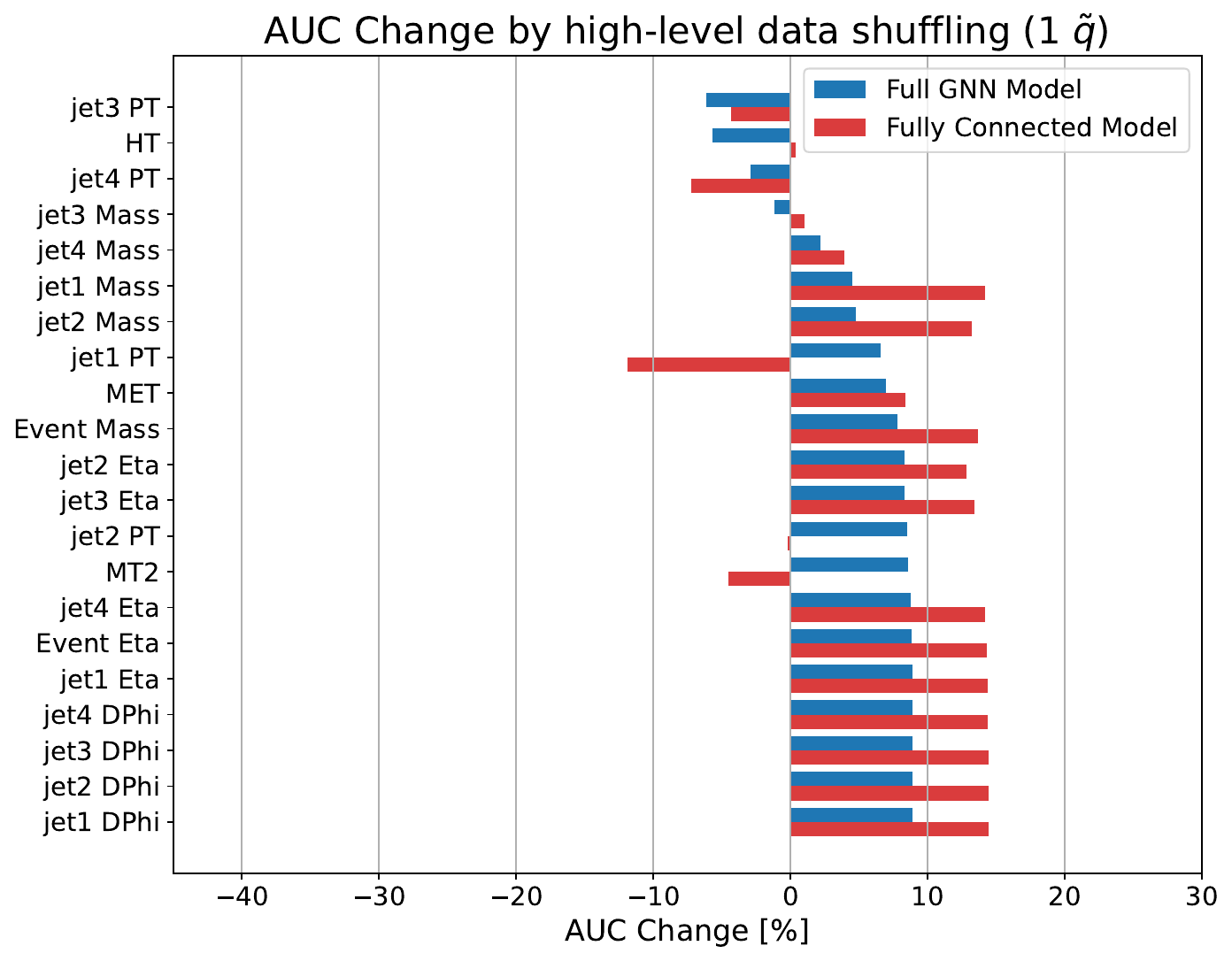}
        \caption{\footnotesize Higgsino-squark production.}
        \label{fig:higgsino1}
    \end{subfigure}
    \vfill
    \begin{subfigure}[b]{0.38\textwidth}
        \centering
        \includegraphics[width=\textwidth]{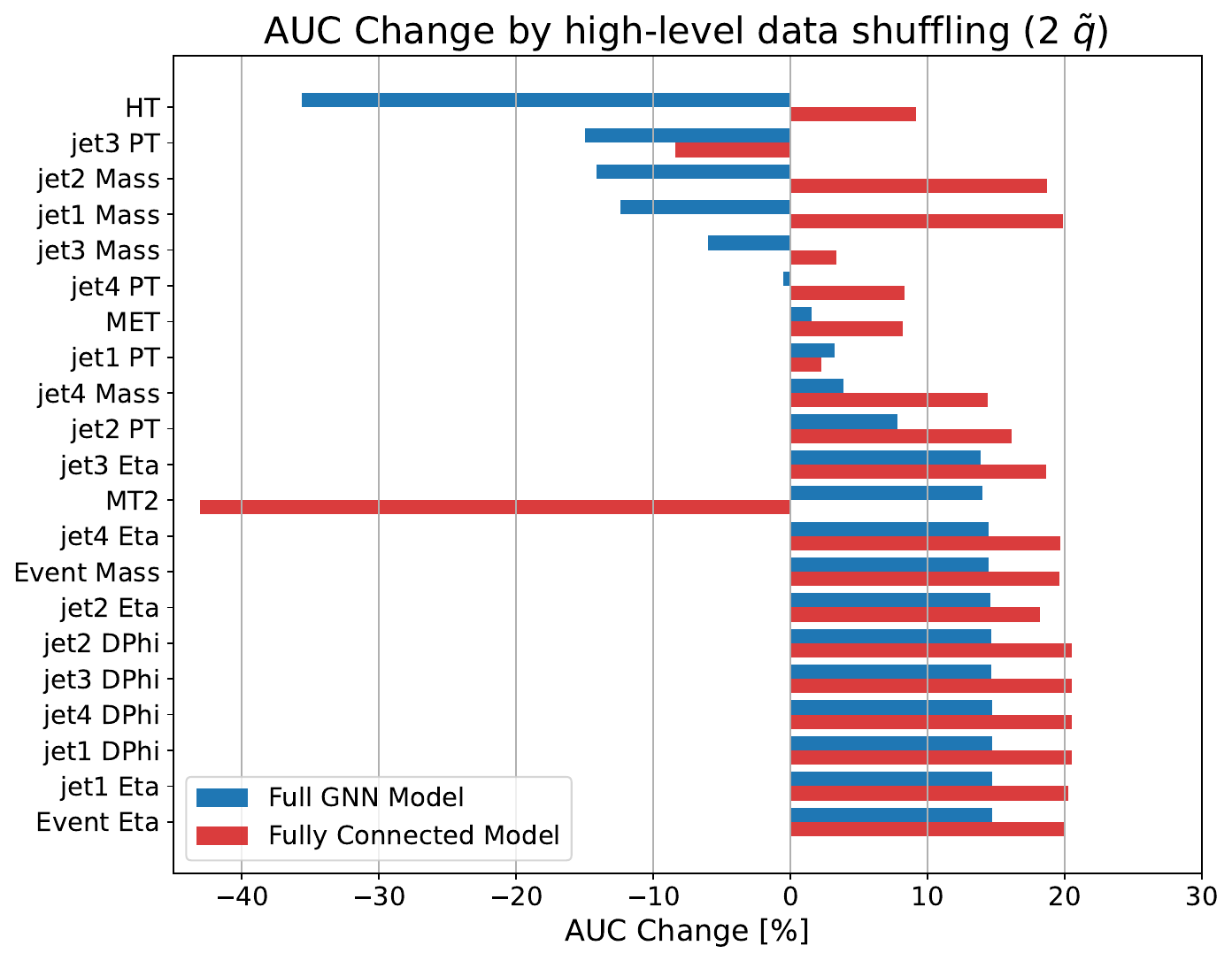}
        \caption{\footnotesize Squark-pair production, model trained on Wino-like sample.}
        \label{fig:wino2}
    \end{subfigure}
    \hfill
    \begin{subfigure}[b]{0.38\textwidth}
        \centering
        \includegraphics[width=\textwidth]{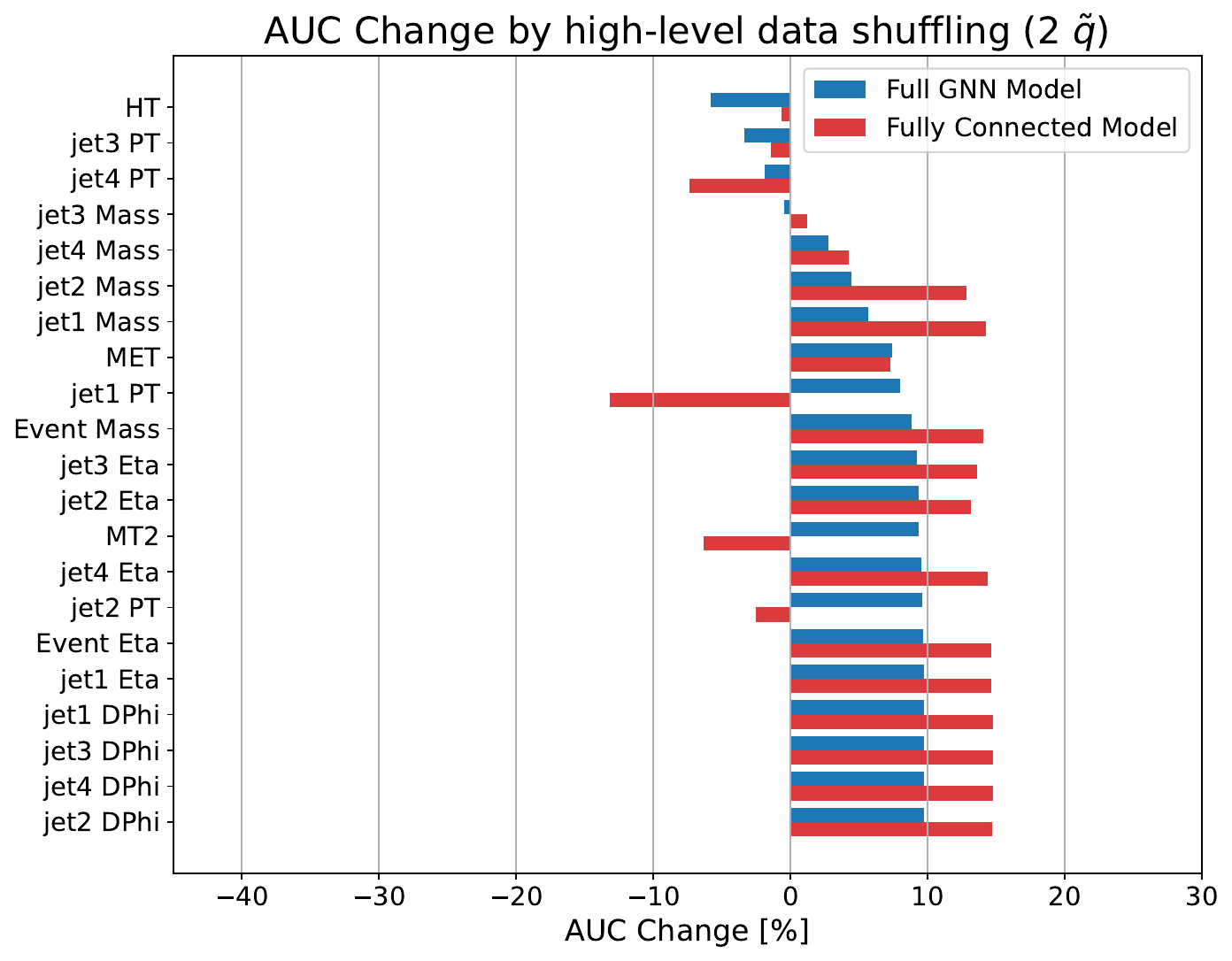}
        \caption{\footnotesize Squark-pair production, model trained on Higgsino-like sample.}
        \label{fig:higgsino2}
    \end{subfigure}
    \vfill
    \begin{subfigure}[b]{0.38\textwidth}
        \centering
        \includegraphics[width=\textwidth]{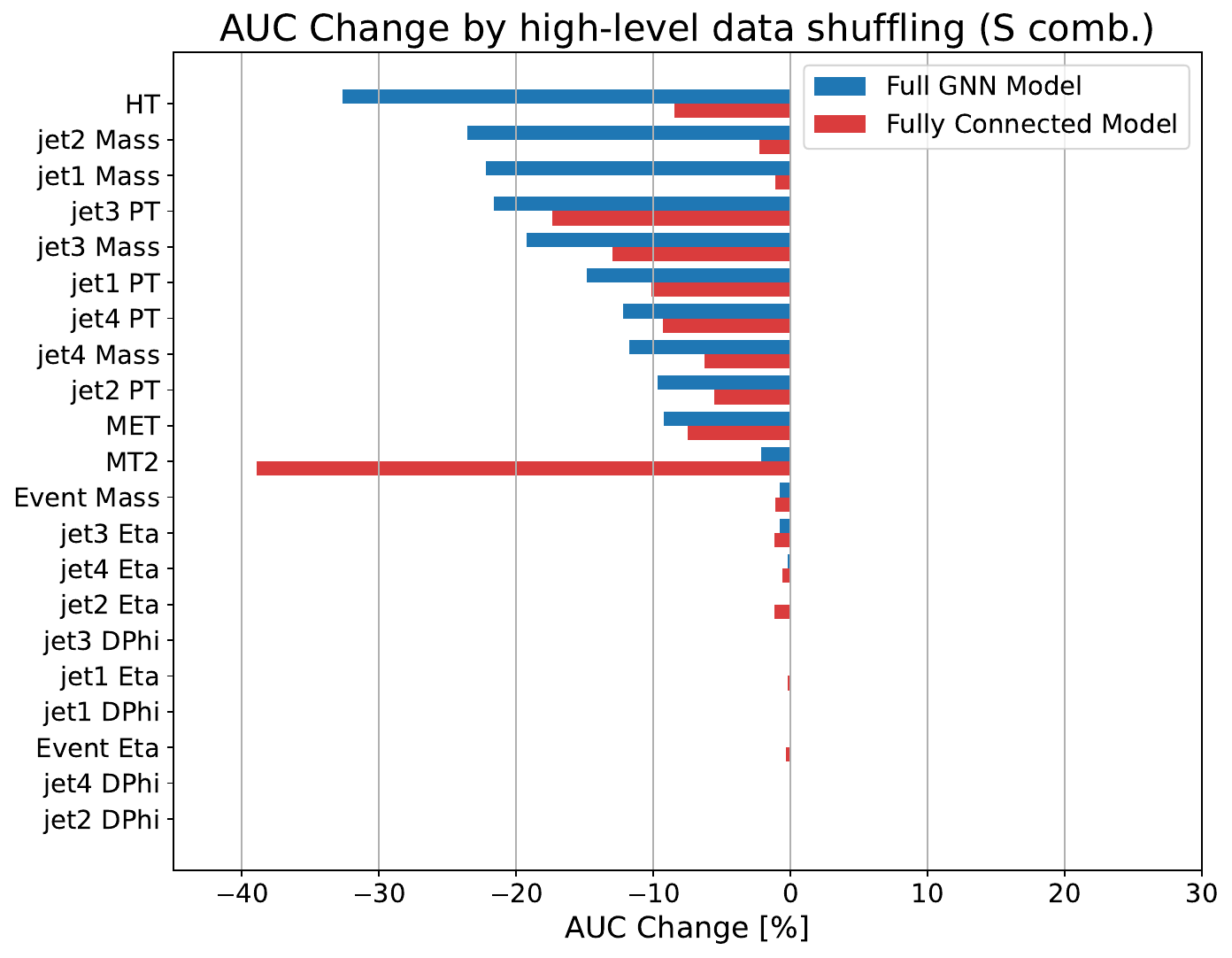}
        \caption{\footnotesize Combined signal with Wino-like neutralino.}
        \label{fig:wino3}
    \end{subfigure}
    \hfill
    \begin{subfigure}[b]{0.38\textwidth}
        \centering
        \includegraphics[width=\textwidth]{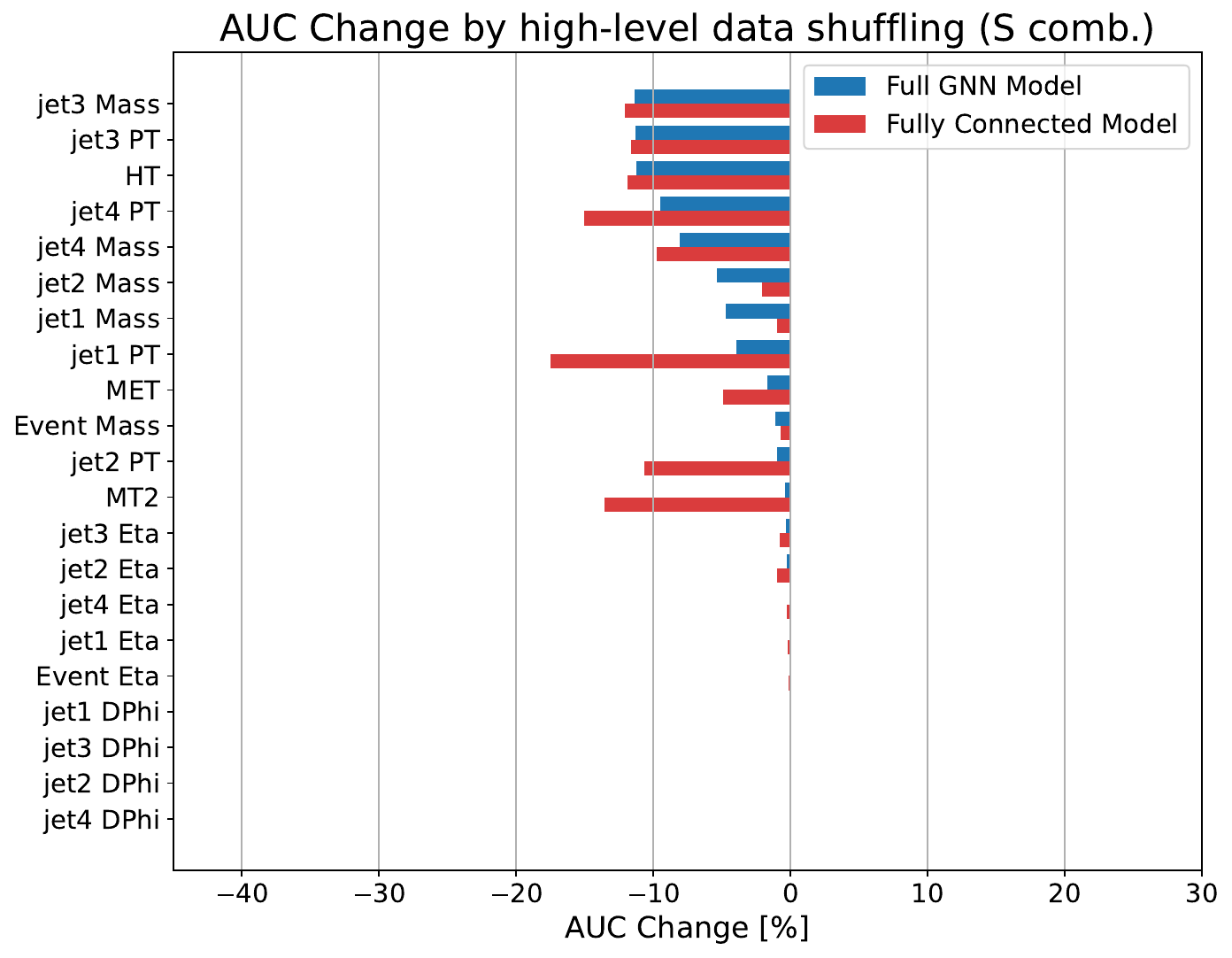}
        \caption{\footnotesize Combined signal with Higgsino-like neutralino.}
        \label{fig:higgsino3}
    \end{subfigure}
    \caption{\footnotesize The relative change of the AUC for the Wino- (left) and Higgsino-like (right) samples evaluated with the full GNN model (blue) and fully connected part of it (red). 
    Both models are trained on $m_{\tilde \chi_1^0}=300$ GeV and $m_{\tilde q}=2.2$ TeV.
    $E_T^{\rm miss}$ is denoted by MET. 
    }
    \label{fig:interpretation-auc}
\end{figure}

Fig.\ \ref{fig:interpretation-auc} depicts the result of the described procedure for high-level inputs for both Winos (left column) and Higgsinos (right column), with different classes of signal specified. 
Both models were evaluated on the test samples for the $m_{\tilde \chi_1^0}=300~\mathrm{GeV}$  $m_{\tilde q}=2.2~\mathrm{TeV}$ mass point.
The modification of the classification performance is quantified with the percentage change of the AUC metric. 
In addition to the AUC change for the full GNN models, depicted in Fig.\ \ref{fig:interpretation-auc} in blue, we also show the AUC change for Fully Connected Models in red. 
These models correspond to the fully connected part of the baseline model described in Sec.\ \ref{sec:architecture}. 
They use only high-level information and are characterised by much worse classification performance. 
Nonetheless, comparing them with full GNN models provides an interesting insight into how the importance of high-level features changes due to the presence of low-level analysis part of the architecture.

Fig.\ \ref{fig:wino0} shows the change of the AUC for Wino sample, only for events without on-shell squarks. One can see that shuffling any of the variables results in worse classification performance for the full GNN model, ranging from over 13 \% for jet angular variables to 32\% for masses of the first three jets. 
This means that the network uses all of the high-level variables when classifying events. 
The most useful are the masses and transverse momenta of jets, but $H_{\rm T}$ and $E_T^{\rm miss}$ are also useful. 
Comparison with the fully connected model reveals several interesting facts. 
Similarly to the full model, all high-level variables are used, but the size and ordering of contributions from individual features are different. 
The largest difference is for the $m_{T2}$, shuffling of which reduces the AUC by 34\%. 
This is not surprising because the $m_{T2}$ is known to be a very effective high-level discriminative variable. 
Nevertheless, the result in Fig.\ \ref{fig:wino0} suggests that a part of the information carried by it can be deduced from the low-level inputs. 
On the other hand, invariant masses of the leading and second jets are more important for the full model. This indicates that the full model has combined the jet information with constituent information to enhance the classification performance. 

An analogous plot for Higgsino is shown in Fig.\ \ref{fig:higgsino0}. 
One can see that for both models, all high-level variables are used and shuffling always results in a
decrease in the AUC. 
The effect is consistently larger for a fully connected model, indicating its higher dependence on high-level features.

Fig.\ \ref{fig:wino1} shows changes in the AUC for the classification of 1 $\tilde{q}$ events. 
We observe that the most impactful high-level input, leading to a drop of about 30\% in AUC for the GNN model, is the $H_\mathrm{T}$. 
Other useful variables are masses and transverse momenta of jets. 
Shuffling the rest of the high-level variables results in a slightly higher AUC.
This can be explained in the following way.
The network is trained for signal events containing 0 $\tilde q$, 1 
$\tilde q$, and 2 $\tilde q$ processes, proportional to their cross sections. It is trained to obtain the best classification 
performance for the mixed signal event, which allows for a situation 
in which the network tunes more to one of the dominant signal classes, 
i.e. $0 ~\tilde q$, by the expense of the others. 
Another explanation stems from the fact that using the 
Mean Squared Error loss function and weight regularisation prevents 
the network from mode collapse and forces it to learn all classes of 
signal events. Since $0~\tilde q$ class is much harder to discriminate 
from the background than 1 $\tilde q$ and 2 $\tilde q$, 
it requires using all available variables. At the same 
time events with decaying squarks can be  
classified with good accuracy based only on a subset of high-level 
variables.

Interestingly, results for a fully connected model are very different. 
The most useful quantity is, again, $m_{T2}$, shuffling of which reduces the AUC of the fully connected model by almost 40\%, while the GNN model does not use this variable at all. 
Another notable difference is that transverse momenta are used only for subleading jets in the fully connected model. 

This might indicate the structure of the leading jet is less important for leading jets. 
However, it is hard to identify the actual cause of the behaviours as 
the classification is optimized for  0 $\tilde{q}$ events.

Fig.\ \ref{fig:higgsino1} shows the results for 1 $\tilde{q}$ classification in the Higgsino sample. 
In this case, the situation differs from the Wino results shown in Fig.\ \ref{fig:wino1}. 
Only $H_\mathrm{T}$, $p_\mathrm{T}$s of the third and fourth jets and the mass of the third jet contribute to the classification performance of the full GNN model. 
For the fully connected model, the relevant variables are jet $p_\mathrm{T}$s and $m_{T2}$. 
The most striking difference between the models is for the \pt of the leading jet. 
The fully connected model relies on its information, but the GNN-based model does not, as it can deduce the relevant information from the low-level inputs. 
These differences might arise from the different compositions of the signal modes. For the $m_{\tilde \chi_1^0}=300~\mathrm{GeV}$
$m_{\tilde q}=2.2~\mathrm{TeV}$ mass point,
the fractions of $0~\tilde q$, $1~\tilde q$, and $2~\tilde q$ signal classes for Winos are 49\%, 13\%, and 37\%, respectively. For Higgsinos the corresponding fractions are 40\%, 10\%, and 50\%.


Fig.\ \ref{fig:wino2} shows the results for the 2 $\tilde{q}$ signal class for the Wino-like neutralino. In this case, the most useful variable for the GNN model is $H_\mathrm{T}$, shuffling of which reduces the AUC by 36\%. 
Other relevant variables are \pt of the third jet and invariant masses of the first three jets. 
When it comes to the fully connected model, it relies almost entirely on $m_{T2}$ (43\%) with a little help from the \pt of the third jet (8\%).

Fig.\ \ref{fig:higgsino2} presents results for the 2 $\tilde{q}$ signal class for the Higgsino-like neutralino. The result is very similar to that in Fig.\ \ref{fig:higgsino1}, and similar conclusions are drawn.

Figs.\ \ref{fig:wino3} and \ref{fig:higgsino3} show results for the combined signal for Wino and Higgsino samples, respectively. 
One can see that shuffling any of the high-level features either drops the AUC or has no effect, which is correct behaviour as the network is 
trained for a full signal sample. It also confirms 
our interpretation of Figs. \ref{fig:wino2},\ref{fig:higgsino2},\ref{fig:wino3},\ref{fig:higgsino3}
For Winos, as shown in Fig.\ \ref{fig:wino3}, both models use jet $p_\mathrm{T}$s, masses of the third and fourth jets, and $E_T^{\rm miss}$. 
However, the fully connected model relies heavily on the $m_{T2}$, while the GNN model uses $H_\mathrm{T}$ and invariant masses of the first and second jets instead, which indicates correlation of the high-level jet information and jet constituent information is important.  
For Higgsino in Fig.\ \ref{fig:higgsino3}, both models rely on masses and transverse momenta of the third and fourth jets and $H_\mathrm{T}$. 
However, unlike the GNN model, the fully connected network relies significantly on the leading and second jet \pt and $m_{T2}$. 
The full model is able to retrieve this information from the low-level inputs.

An overall conclusion from Fig.\ \ref{fig:interpretation-auc} is that high-level inputs are optimized for 
classification of signal events without on-shell squarks. 
Classification of events with decaying squarks is much easier, and it requires only a subset of high-level variables. 
The fully connected network relies greatly on the $m_{T2}$, while the full model uses the jet information to analyze the correlation with jet constituents. 
Since the fully connected network does not heavily rely on the jet mass variables, its predictions are more robust against mismodelling of soft activities and the pileup effect.

\subsection{Jet images}

\begin{figure}[t!]
    \centering
    \includegraphics[width=\linewidth]{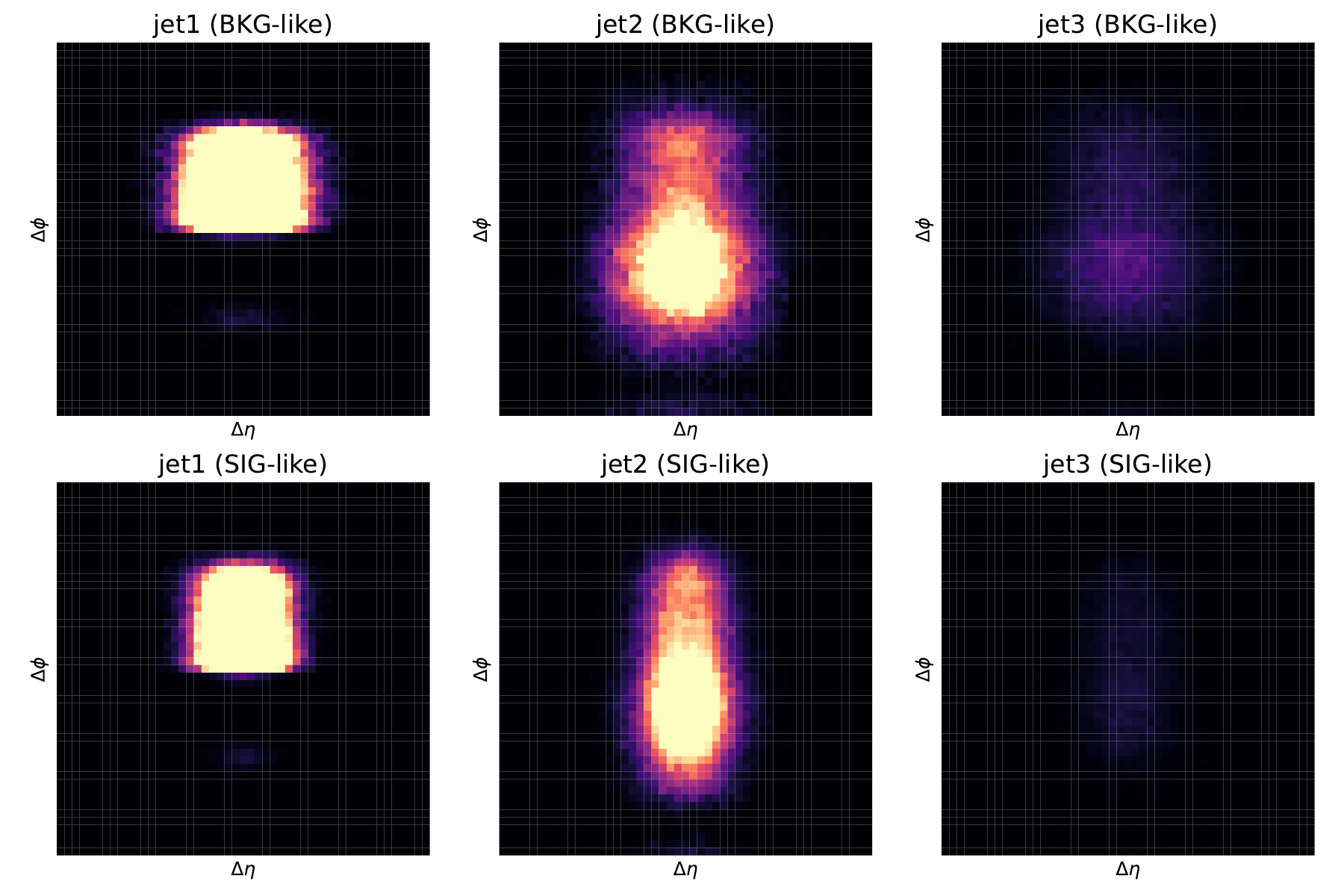}
    \caption{\small Images of the first three jets for the background-like (top) and signal-like (bottom) events for Winos. 
    Normalisation is consistent for all images, allowing us to compare the energy deposited by each of the jets.}
    \label{fig:interpretation-jet-wino}
\end{figure}

In Fig.\ \ref{fig:interpretation-jet-wino}, we show the images of the first three jets. 
To make these pictures, events from the combined SM background and Wino-like signal samples are split into two subsets: background-like and signal-like events. 
The background-like events are events for which the network returned score values less than 0.1, while the signal-like events correspond to score values greater than 0.9. 
It is important to stress that both signal and background events may fall into any of the two subsets because we are interested in the classifier's judgement rather than the truth labels. 
The plots in the top row in Fig.\ \ref{fig:interpretation-jet-wino} are for the background-like events, while plots in the bottom row are for signal-like. 
The columns in Fig.\ \ref{fig:interpretation-jet-wino} correspond to the first (left), second (centre) and third (right) jets in terms of \pt. 
For each of the jets, the average transverse energy carried by its constituents is depicted in the $\Delta \eta$ vs $\Delta \phi$ plane, where $\Delta \eta$ and $\Delta \phi$ are the pseudorapidity and the azimuthal angle coordinates relative to the position of the event centroid, respectively. 
The plane has been divided into 100 bins in both dimensions. 
For each bin, the transverse energy of corresponding particles has been aggregated over all particles in all events and divided by the number of events. In addition, for events in which $\Delta \phi(j_1)<0$, $\phi$ coordinates of all particles have been flipped, $\phi \to - \phi$. 
A common normalisation has been used for all plots in Fig.\ \ref{fig:interpretation-jet-wino}, allowing us to compare the average energy depositions between different jets.

The images in Fig.\ \ref{fig:interpretation-jet-wino} indicate that jets in events classified as background-like are, on average, wider than for the signal-like event.
It makes sense since jets from squark decays are typically highly boosted due to the large mass difference between squarks and Winos. 
Due to $\Delta \phi$ flipping, the activity is always in the upper half of the plot. 
The activities of the second leading jets are split. 
The main activity lies in the lower half of the plane, but there is also a significant contribution in the upper half. 
The contributions to $\Delta \phi < 0$ mean that the event topology is similar to a boosted di-jet, where most of the event energy is distributed between the leading and the second leading jets. 
The activity in the upper half of the plane suggests that constituents of both jets are moving in the same direction. 
For the background and the 0 $\tilde q$ signal events, it may be that the showering split the leading jet into two, and the ``real'' second jet is moving in the opposite direction. 

\begin{figure}[t!]
    \centering
    \includegraphics[width=1\linewidth]{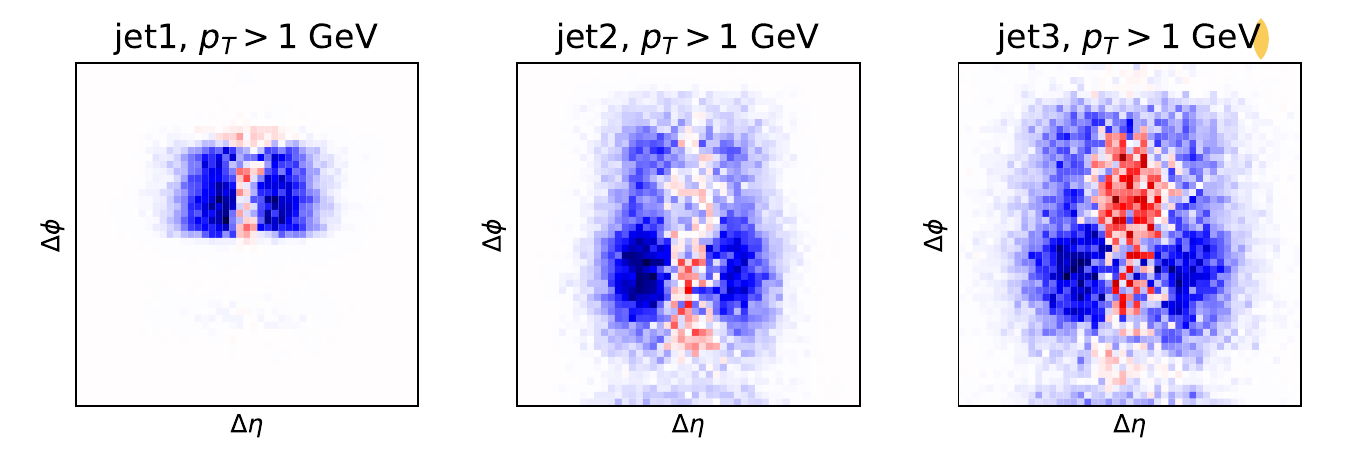}
    \caption{\small The differences between the background-like and signal-like jets (the signal-like distribution minus the background-like distribution) for the Wino-like scenario. The red and blue colours indicate the positive and negative values, respectively.}
    \label{fig:interpretation-jet-diff-wino}
\end{figure}

Fig.\ \ref{fig:interpretation-jet-diff-wino} depicts differences in shapes of the first three jets between background-like and signal-like samples.
The blue pixels correspond to higher activity in the background-like case, while the red pixels are for signal-like events. 
The plots on the left, middle and right correspond to the first, second and third jets, respectively. 
We clearly observe that jets in the background-like events are typically much wider.
Interestingly, the third jet for signal-like events is more likely to follow the leading jet in a similar direction, while for background-like events, it is more likely to follow the second leading jet.

\begin{figure}[t!]
    \centering
    \includegraphics[width=\linewidth]{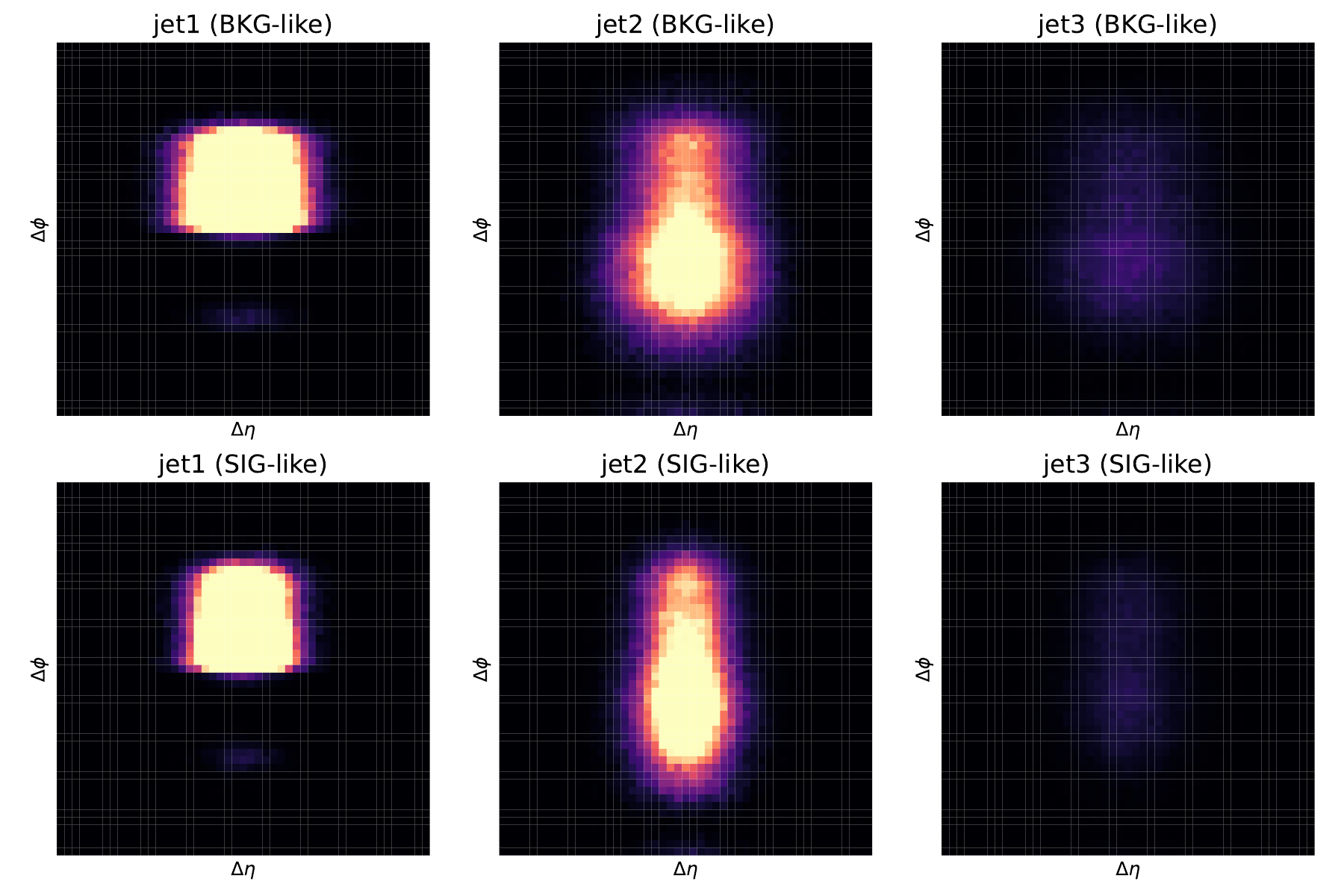}
    \caption{\small 
    Images of the first three jets for the background-like (top) and signal-like (bottom) events for Higgsinos. 
    Normalisation is consistent for all images, allowing us to compare the energy deposited by each of the jets.}
    \label{fig:interpretation-jet-higgsino}
\end{figure}

\begin{figure}[t!]
    \centering
    \includegraphics[width=1\linewidth]{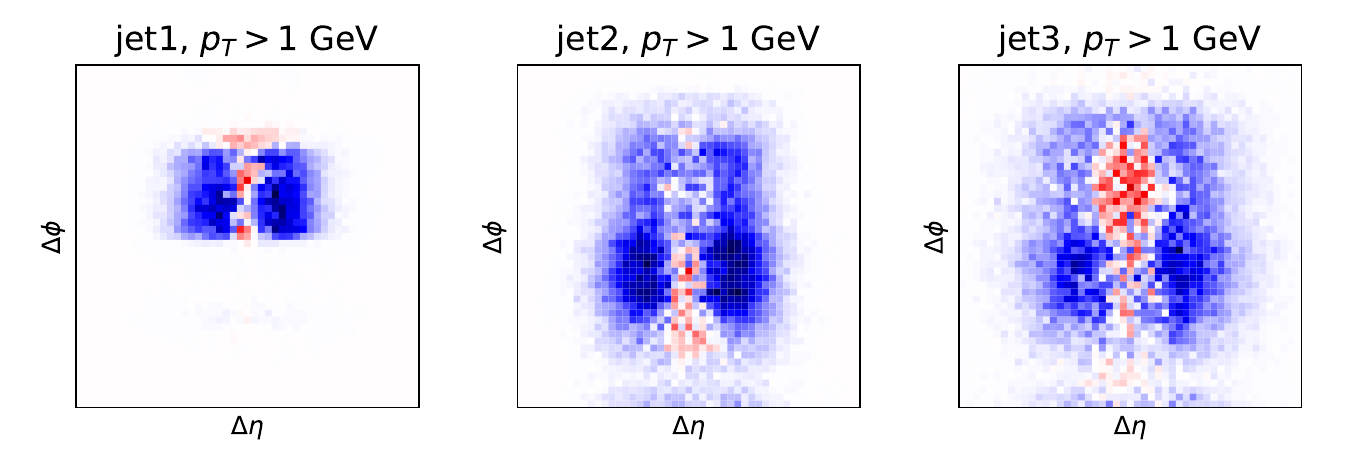}
    \caption{\small The differences between the background-like and signal-like jets (the signal-like distribution minus the background-like distribution) for the Higgsino-like scenario. The red and blue colours indicate the positive and negative values, respectively.}
    \label{fig:interpretation-jet-diff-higgsino}
\end{figure}

The jet images for a model trained and evaluated on a sample with Higgsino-like neutralinos are depicted in Fig.\ \ref{fig:interpretation-jet-higgsino}, and differences in shapes between signal-like and background-like events are shown in Fig.\ \ref{fig:interpretation-jet-diff-higgsino}. The results are very similar to Winos, and similar conclusions apply.


\section{Conclusions}\label{sec:concl}

In this study, we have proposed a new ML-assisted analysis of squark-neutralino searches in the monojet channel at the Run-3 and High Luminosity LHC.
Our analysis consisted of two parts. 
First, we started with a cut-and-count preselection in order to remove most of the SM background. 
Next, we used state-of-the-art Machine Learning tools to analyse data with high- and low-level objects. 

We found that events with decaying squarks are much easier to classify because of significant differences between ISR jets and jets from squark decays. 
Nevertheless, even for processes without squarks, networks can enhance the signal/background discrimination power. 
Moreover, we have demonstrated that our approach is significantly better at solving the posed classification problem than the well-established Boosted Decision Trees.  
We have shown the importance of low-level information for classification results, especially for discrimination between ISR jets in the background and signal processes. 
Finally, we have performed a cross evaluation to determine the robustness of our algorithm against changes in the electroweakino mixing, and we found that the ML 
models trained on Wino or Higgsino 
samples can be effectively used also to 
search for Binos.

We have performed a mass scan for Winos, Higgsinos and Binos
to derive the discovery and exclusion prospects for LHC at the end of Run-3 and High Luminosity phases. 
 The robustness of our approach against changes in sparticle masses enabled us 
 to derive limits on squark and 
 electroweakino masses without 
 requiring retraining for each mass 
 point. 
We have shown that Wino and Higgsino can be probed up to 2-$\sigma$ level 
at the end of Run-3 if their masses are smaller than 400 and 350 GeV, respectively, and the squark mass is around 2 TeV.
The prospects for HL-LHC are much more promising. 
For Winos, Higgsino or Binos with 
masses up to 680 GeV, 550 GeV and 500 
GeV, respectively, 5-$\sigma$ discovery 
is possible as long as squarks are not 
too heavy and soft particles can be 
used.

Last but not least, we have attempted to interpret our ML algorithm.
We found when classifying events from the electroweakino pair production process, high-level variables are very useful, especially for Winos, but for events with squark decaying, low-level variables seem to be more relevant. 
It seems that the network has learned to discriminate between the background and signal events based on the sizes and shapes of the jets, i.e.\ jets from squark decays are more collimated. 
We have also found signs of overfitting, which indicates that the results could improve if a larger data set were available.

Despite the excellent performance of the recent pileup mitigation techniques, some effects of pileup and underlying events are expected at the Run-3 and High Luminosity phases of LHC.
Since the accurate simulation of those soft activities is challenging, we instead study how the result changes when the soft $p_T$ cut is varied between 1 and 10 GeV.
We also checked in Figs.\ \ref{fig:interpretation-corr-low} and \ref{fig:interpretation-auc} that the network uses soft particles and the jet mass information as supplementary information, suggesting the classification is not highly sensitive to those less reliable information.

The final result of this study is a new analysis based on Graph Neural Network 
architecture, which combines data on multiple levels, from individual particles to whole events. 
Our approach was carefully evaluated and proven robust to changes in the underlying parameters of the signal model, which allows it to be efficiently used for real data after a more thorough optimisation.

\section*{Acknowledgements}
{
R.M. would like to acknowledge the financial support provided by:
Polish National Science Centre grants 2021/41/N/ST2/00972
and partially 2020/38/E/ST2/00243; and Polish National Agency For Academic exchange grant 
BPN/BEK/2022/1/00253/DEC/1. 
M.M.N is supported by Grant-in-Aid for Transformative Research Area (A) 22H05113 and Grant-in-Aid for Scientific Research(C) JSPS KAKENHI Grant Number 22K03629. 
This work was performed in part at the Aspen Center for Physics, which is supported by National Science Foundation grant PHY-2210452.
The work of KS was performed in part at the workshop on “Effective Theories for Nonperturbative Physics" at the Mainz Institute for Theoretical Physics
(MITP) of the DFG Cluster of Excellence PRISMA+ (Project ID
390831469).
}

\bibliographystyle{utphys28mod}
\bibliography{refs} 

\newpage

\begin{appendices}
\section{Sofware version}\label{app:a}
In order to ensure reproducibility of the study, we provide in Table \ref{tab:software} the versions of the software that have been used.

\begin{table}[ht!]
    \centering
    \begin{tabular}{l|l}
    \hline
         SuSpect  & 3.1.1 \\
         SUSY-HIT & 1.5a \\
         MadGraph5 & 2.7.3 \\
         Pythia & 8.309 \\
         Delphes & 3.4.3pre12\\
         FastJet & 3.4.0\\
         ExRootAnalysis & 1.1.2\\
         CERN ROOT & 6.24/02 \\
         python & 3.8.10\\
         numpy & 1.24.4\\
         awkward0 & 0.15.5\\
         tensorflow & 2.7.0\\
         pandas & 1.4.1\\\hline
    \end{tabular}
        \caption{\label{tab:software}
        \small Versions of the software used in the study}    
\end{table}
\section{Optimal hyperparameters for XGBoost}\label{sec:hyperparameters}
\begin{table}[hb!]
    \centering
    \begin{tabular}{r|l|l}        
         \hline
        1 & n\_estimators & 99\\
        2 & max\_depth & 5\\
        3 & subsample & 0.82 \\
        4 & colsample\_bytree & 0.79\\
        5 & learning\_rate & 0.091\\
        6 & reg\_lambda& 0.052\\
        7 & alpha& 0.040\\
        8 & min\_child\_weight & 6\\
        9 & eta& 0.13\\
        10 & gamma& 1.26e-4\\
        11 & grow\_policy&lossguide\\ \hline
    \end{tabular}
        \caption{\label{tab:bdt-hyper}
        \small Optimal hyperparameters found for the BDT models.}
\end{table}

\end{appendices}

\end{document}